\documentclass[aps,pre,onecolumn]{revtex4}
\usepackage{graphicx, bm, bbm,amsmath, amssymb, epsfig}

\newcommand{\nn}{\noindent}

\newcommand{\bq}{\begin{align}}
\newcommand{\eq}{\end{align}}

\begin{document}
\title{Collective locomotion of two-dimensional lattices of flapping plates}
\author{Silas Alben$^*$}
\affiliation{Department of Mathematics, University of Michigan,
Ann Arbor, MI 48109, USA}
\email{alben@umich.edu}

\date{\today}

\begin{abstract}
We study the propulsive properties of rectangular and rhombic lattices of flapping plates at O(10--100)
Reynolds numbers in incompressible flow. Here the fluid dynamics often converge to time-periodic in 5-30 flapping periods, facilitating accurate computations of time-averaged thrust force and input power. We classify the
propulsive performances of the lattices and the periodicity of the flows in a space of five parameters: flapping amplitude, frequency (or Reynolds number), horizontal and vertical spacings between plates, and oncoming fluid stream velocity. Nonperiodic states are most common at small streamwise spacing, large lateral spacing, and large Reynolds number. 

Lattices that are closely spaced in the streamwise direction produce intense vortex dipoles between
adjacent plates. The lattices transition sharply from drag- to thrust-producing as these dipoles switch from upstream to downstream orientations at critical flow speeds. Near these transitions the flows pass through a variety of periodic and nonperiodic states, with and without up-down symmetry, and multiple stable self-propelled speeds can occur. As the streamwise spacing increases (and with large lateral spacing), the plates may shed typical vortex wakes (e.g.~reverse von K\'{a}rm\'{a}n streets) that impinge on downstream neighbors. The most efficient streamwise spacing increases with flapping amplitude.
With small lateral spacing, the rectangular lattices have Poiseuille-type flows that yield net drag, while the rhombic lattices may shed vortices and generate net thrust, sometimes with relatively high efficiency. As lateral spacing increases to the vicinity of a plate length and beyond, the rectangular lattices begin to shed vortices and generate thrust, eventually with slightly higher efficiencies than the rhombic lattices, as the two types of lattice flows converge. 
At Re = 70, the lattices' maximum Froude efficiencies are about twice those of an isolated plate (only considering nearly periodic lattice flows). As Re decreases, the lattices' efficiency advantage increases; the lattices 
obtain net thrust (and self-propulsion) at lower Re than an isolated flapping plate does. 
    
The mean input power needed to generate the lattice flows can be estimated in the limits of small and large streamwise spacings, with small-gap and Poiseuille-like flows between the plates respectively in the two cases. For both lattices, the mean input power saturates as the lateral spacing becomes large (and thrust occurs). At small lateral spacings, the rhombic lattices' input power may be much larger when the plates overlap, leading to a decrease in Froude efficiency.

\end{abstract}


\maketitle

\section{Introduction \label{sec:Introduction}}

Propulsion by flapping foils has garnered considerable interest in recent years, as a bio-inspired alternative to traditional designs for aquatic and aerial vehicles. Flapping propulsion has potential advantages in efficiency, maneuverability, and stealth, particularly at small and medium scales \cite{lighthill1960nss,wu1971hsp,sparenberg1995hpa,
ASBT_JFluidMech_1998,SBL_PAeroSci_1999,TTY_AnnRevFluidMech_2000,LH_JFluidMech_2003,triantafyllou2004review,heathcote2005ffa,fish2006passive,miller2012using,smits2019undulatory}.
Some of the different types of flapping bodies and motions considered are: rigid or flexible foils \cite{lighthill1960nss,wu1971hsp,ASBT_JFluidMech_1998,heathcote2005ffa} undergoing heaving and/or pitching motions \cite{Fre88,LH_JFluidMech_2003,Von03,triantafyllou2004review,buchholz2005evolution,smits2019undulatory};
flexible foils oscillated at one point and otherwise bending passively \cite{Alben2008JFM1,michelin2009resonance,AlbenJCP2009,yeh2014free,hoover2018swimming,hess2020cfd}, or with an internal driving force distributed all along the foil \cite{tytell2016role,ming20193d}; foils oscillated transversely to an imposed oncoming flow \cite{ASBT_JFluidMech_1998,LH_JFluidMech_2003}, or
swimming (translating/rotating) freely under a force balance law \cite{vandenberghe2004symmetry,alben2005coherent,spagnolie2010surprising,alben2012dynamics,yeh2014free}. Another large body of work has considered the stability and dynamics
of passive flexible flags, plates, and membranes in fluid flows \cite{shelley2011flapping}. A common way to understand the physics of force generation by flapping foils is to relate the forces on the foil to the vorticity shedding patterns, often von K\'{a}rm\'{a}n vortex streets or other ordered arrays. Given a certain body motion, the formation of such vorticity distributions depends on unsteady large-scale boundary layer separation and is difficult to describe
with a simple analytical approach. Computational and experimental approaches are more commonly used to describe the phenomena. Several works have found that Froude efficiency is maximized when a reverse von K\'{a}rm\'{a}n street is formed, typically near Strouhal numbers of 0.2--0.5 for biological and biomimetic swimmers \cite{TTG_JFluidsStruct_1993,JDP_AIAAJ_1998,ASBT_JFluidMech_1998,TNR_Nature_2003,rohr2004strouhal,triantafyllou2004review,dabiri2009optimal,eloy2012optimal}.
Outside this range, other ordered and disordered vortex wakes are observed \cite{triantafyllou2004review,godoy2008transitions,schnipper2009vortex}.

A number of works have extended the study to the case of multiple flapping foils interacting in a fluid \cite{sparenberg1975efficiency,akhtar2007hydrodynamics,wang2007effect,boschitsch2014propulsive,kurt2018flow,lin2019phase}. Key parameters are the phase differences between the foils' oscillations, and the spacings (in-line and/or transverse) between the foils. 
If one body interacts with a typical vortex wake of another (e.g.~an inverse von K\'{a}rm\'{a}n street), the spacings and phasings will largely determine the types of vortex-body collisions that occur and the resulting forces. Vortices impinging on foils alter the pressure distribution and vortex shedding at the leading and trailing edges \cite{akhtar2007hydrodynamics,rival2011recovery}. The vortex wakes may be strengthened or weakened through the interactions, with the possibility of increased thrust or efficiency in some cases \cite{triantafyllou2004review}. Related lines of work have addressed interactions between a single body and ambient vorticity (e.g.~shed from a static obstacle) \cite{streitlien1996efp,liao2003fish,beal2006ppv,fish2006passive,liao2007review,alben2009swimming,alben2009passive}, vortex-wall collisions \cite{doligalski1994vortex,rockwell1998vortex,alben2011interactions,alben2012attraction,flammang2013functional,quinn2014unsteady}, and interactions between multiple passive flapping flags and plates \cite{ristroph2008ahd,zhu2009interaction,alben2009wake,kim2010constructive,uddin2013interaction}. Although much is known, the complicated physics of vortex shedding remains an obstacle to a simple quantitative description of multiple-body/body-vortex interaction problems \cite{lentink2010vortex,li2019energetics}. Even the apparently simpler case of collective interactions in the zero Reynolds number limit \cite{dombrowski2004self,saintillan2008instabilities,lauga2009hydrodynamics,koch2011collective,elgeti2015physics,saintillan2018rheology}, with linear flow equations but geometrical complexities, has many open issues, among them close interactions between bodies \cite{paalsson2020integral,wu2020solution}.

When multiple bodies are considered, the number of degrees of freedom increases enormously even with many simplifying assumptions. We now have to choose a particular geometry and kinematics for each body (including relative phases for periodic motions). We need to resolve the flow on a wide range of scales simultaneously, from the size of a large group of bodies and their vortex wakes to the scale of thin, time-dependent boundary layers and separation regions on each body surface.  For prescribed spatial configurations of the bodies, there are many possible choices. A potential way to simplify the problem is to allow a group of bodies to evolve dynamically and look for 
configurations that are attracting states of various initial conditions
\cite{becker2015hydrodynamic,ramananarivo2016flow,dai2018stable,mirazimi2018meso,peng2018hydrodynamic,peng2018collective,im2018schooling,oza2019lattices,newbolt2019flow,li2019energetics}. Many of these involve quantized spacings that are related
to the natural spacings of vortex streets.
If the spatial configuration evolves dynamically according to the forces on the bodies, the nonlinear dynamics are generally sensitive to initial conditions as well as the details of close interactions and/or collisions between bodies. It is very difficult to classify the whole range of possibilities in such systems. Many studies have instead focused on configurations seen in groups of biological organisms \cite{weihs1975some,partridge1979evidence,svendsen2003intra,liao2003fish,akhtar2007hydrodynamics,portugal2014upwash,daghooghi2015hydrodynamic,marras2015fish,hemelrijk2015increased,gravish2015collective,li2016hydrodynamic, khalid2016hydrodynamics,ashraf2017simple,park2018hydrodynamics}. Other recent studies have used machine learning to determine optimal motions of groups of swimmers \cite{gazzola2016learning,novati2017synchronisation}. Another large body of work concerns the use of simplified laws of interaction in place of detailed fluid dynamics, to model schools and flocks of bodies \cite{katz2011inferring,filella2018model}.



Following previous models \cite{childress2004transition,oza2019lattices,newbolt2019flow}, experiments \cite{vandenberghe2006unidirectional,becker2015hydrodynamic,ramananarivo2016flow}, and simulations \cite{wang2000vortex,alben2005coherent,huang2007mechanism,alben2008implicit,deng2016dependence,deng2018horizontal} inspired by biology
\cite{borrell2005aquatic}, we consider a particular version of the multiple flapping foil problem, with simple body geometries and kinematics, that is amenable to a wide (though by no means exhaustive) exploration of parameter space: thin plates that are oscillated vertically and moved horizontally together through a viscous fluid.  The plates and motions considered here are fore-aft symmetric for simplicity; adding a pitching motion \cite{spagnolie2010surprising}, an asymmetric body thickness profile \cite{huang2007mechanism}, and/or active and/or passive deformations \cite{novati2017synchronisation,dai2018stable} can enhance the thrust generated and the propulsive efficiency. The main quantities of interest are the time-averaged horizontal force (i.e.~thrust or drag) and the input power needed to oscillate the plates vertically in the fluid. We study perhaps the most common measure of efficiency,
the Froude (propeller) efficiency, a ratio of average propulsive power to average input power required to oscillate the foils \cite{lighthill1960nss,ASBT_JFluidMech_1998}. We also
study an alternative output quantity, the self-propelled speed(s) of the foils \cite{vandenberghe2006unidirectional,alben2005coherent,alben2012dynamics,novati2017synchronisation,dai2018stable}.

\section{Model \label{sec:Model}}


\begin{figure} [h]
           \begin{center}
           \begin{tabular}{c}
               \includegraphics[width=7in]{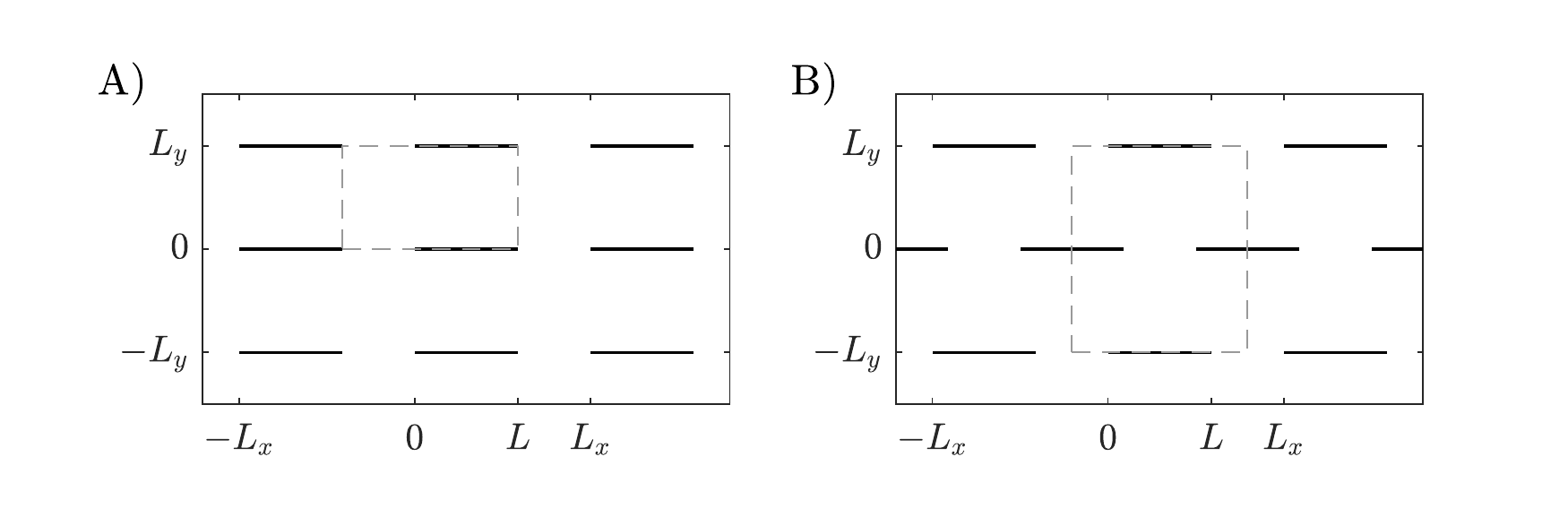}\\
           \vspace{-.25in}
           \end{tabular}
          \caption{\footnotesize A) A rectangular lattice of plates. An $L_x$-by-$L_y$ unit cell is shown with a dashed outline. B) A rhombic lattice of plates.  An $L_x$-by-$2 L_y$ double unit cell is
shown with a dashed outline.
 \label{fig:SchematicPlatesFig}}
           \end{center}
         \vspace{-.10in}
        \end{figure}

We consider a lattice of plates (rectangular or rhombic, shown in figure \ref{fig:SchematicPlatesFig}), each
plate moving with the same velocity $\mathbf{U}(t) = (U, V(t))$, constant in the horizontal direction, and sinusoidal in the
vertical direction:
\begin{align}
\frac{V(t)}{fL} = 2\pi \frac{A}{L}\sin(2\pi t)\, \left(1-e^{-(t/t_0)^2}\right),
\end{align}
\nn with $A$ the amplitude and $f$ the frequency of the vertical displacement corresponding to $V(t)$. The exponential term allows smooth start-up from rest with
time constant $t_0 = 0.2$ (the particular value is not important, as our focus is on the eventual steady state behavior). We nondimensionalize quantities using the plate length $L$ as the characteristic length, the flapping period $1/f$ as the characteristic
time, and the fluid mass density $\rho_f$ as the characteristic 
mass density.

We solve the incompressible Navier-Stokes equations, nondimensionalized, in the rest frame of the lattice \cite{Batchelor1967}:
\begin{align}
\partial_t\mathbf{u} + \mathbf{u}\cdot\nabla\mathbf{u} &= -\nabla p
+\frac{1}{\mbox{Re}_f} \nabla^2 \mathbf{u} - \frac{d\mathbf{U}}{dt}(t),  \label{NS1} \\
\nabla \cdot \mathbf{u} &= 0. \label{NS2}
\end{align}
\nn The basic dimensionless parameters are
\begin{align}
\frac{A}{L}, \,\mbox{Re}_f = \frac{f L^2}{\nu}, \,l_x = \frac{L_x}{L}, \,l_y = \frac{L_y}{L},
\,U_L = \frac{U}{fL}, \label{params}
\end{align}
\nn where $\nu$ is the kinematic viscosity of the fluid and $L_x$ and $L_y$ are the lattice spacings in the $x$ and $y$ directions, respectively.
Other important dimensionless parameters, combinations of those above, are:
\begin{align}
\mbox{Re} = \frac{4A f L}{\nu}, \,\mbox{Re}_U = \frac{U L}{\nu},
\,U_A = \frac{U}{fA}, \,\mbox{St} = \frac{2}{U_A}. \label{params2}
\end{align}
Re is the Reynolds number based on the mean vertical velocity of the foil on each half-stroke, and is therefore a better measure of the ratio of inertial to viscous forces than Re$_f$, which we think of as a dimensionless frequency. It is convenient for computations to nondimensionalize time by the flapping period, but the flapping frequency is one of the kinematic parameters we vary as we search for optimal flapping kinematics as well as plate spacings. Therefore, for comparison across kinematic parameters, Re$_U$ gives a more uniform measure of the horizontal speed of the foil than $U_L$ (since $L$ and $\nu$ are considered fixed in all cases, while $f$ varies). To find the horizontal velocities that yield efficient thrust-generating states and self-propelled states, previous work has shown that we should search in certain ranges of St (or $U_A$, twice its reciprocal) \cite{TTG_JFluidsStruct_1993,JDP_AIAAJ_1998,ASBT_JFluidMech_1998,TNR_Nature_2003,rohr2004strouhal,triantafyllou2004review,dabiri2009optimal,eloy2012optimal}.

Instead of prescribing the horizontal velocity $U_L$, one can allow it to evolve dynamically
according to Newton's second law, setting the plates' rate of change of horizontal momentum equal to the horizontal component of the net fluid forces on them \cite{alben2005coherent,alben2008implicit,spagnolie2010surprising,alben2012dynamics,dai2018stable}. In this case, we have the plates' dimensionless mass $M$ as a control parameter instead of $U_L$. We have simulated this case, with $U_L(t)$ ``free" and $M$ fixed, and the case of fixed $U_L$, and in both cases, periodic and nonperiodic flow dynamics can arise generically at different parameters. The coupling of body and fluid motion seems to add some additional
complexity to the problem, so here we focus on the case with fixed $U_L$, which is also the focus of most previous flapping foil studies, including those that investigated Froude efficiency \cite{lighthill1960nss,TTG_JFluidsStruct_1993,ASBT_JFluidMech_1998,TTY_AnnRevFluidMech_2000}. The case
with fixed $U_L$ and zero time-averaged thrust corresponds to the large-mass limit of cases with time-varying $U_L$---those with initial conditions such that the fixed value of $U_L$ is an attracting state. 

The flow starts at rest, and evolves until it converges to a periodic steady state, or remains nonperiodic up to a chosen end time of a simulation (typically $t = 15$ or 30). Some of these nonperiodic states may eventually converge to periodic in longer simulations. However, most have irregular oscillatory behaviors and seem likely to remain
nonperiodic. These cases seem to require much longer simulations to precisely compute the long-time averages of fluid forces and input power. Thus we mostly focus on the parameters that yield a periodic state, generally those at lower Reynolds numbers,
but give information about nonperiodic results in some cases. 

For a plate with zero thickness in a viscous flow, the pressure and viscous shear stress diverge near the plate tips as the inverse square root of distance
\cite{hasimoto1958flow,ingham1991steady}. In the limit of zero plate thickness, the contribution of the pressure on the plate edges to the net horizontal force is zero. The net horizontal force on the plate is due only to the viscous shear stress on the two sides of the plate,
\begin{align}
F_x = 
\frac{1}{\mbox{Re}_f} \int_0^1 [\partial_y u(x,0,t)]^+_- dx.
\end{align}
\nn The bracket notation denotes the jump in $\partial_y u$ along the plate (the value at
the top minus the value at the bottom).  The
vertical force is due to the pressure difference across the plate:
\begin{align}
F_y = \int_0^1 -[p(x,0,t)]^+_- dx .
\end{align}
\nn Important related quantities are the input power $P_{in}(t)$ and the Froude efficiency $\eta_{Fr}$:
\begin{align}
P_{in}(t) = \frac{V(t)}{fL} F_y ; \; {\tilde P}_{in}(t) = \mbox{Re}_f^3 P_{in}(t); \; \eta_{Fr} = \frac{U\langle F_x(t) \rangle}{\langle P_{in}(t) \rangle}.
\end{align}
\nn Here ${\tilde P}_{in}(t)$ is the input power nondimensionalized with $\nu/L^2$ in place of $f$, for comparison across cases with different $f$ (since $L$ and $\nu$ are assumed fixed). The numerator and denominator of $\eta_{Fr}$ both acquire factors of 
Re$_f^3$ with the same change in nondimensionalization, resulting in no change for 
$\eta_{Fr}$.

Since $\nabla p$ in (\ref{NS1}) is doubly periodic, it has a Fourier decomposition in which the mean (or constant part) has components we denote $\Delta p_x/l_x$ and $\Delta p_y/l_y$, and $\nabla p_1$ is the remainder (the mean-zero part). Thus $p$ is decomposed into mean-zero and linear terms:
\begin{align}
p = p_1 + \frac{\Delta p_x}{l_x}x + \frac{\Delta p_y}{l_y}y.
\end{align}
\nn $p_1$ is determined (up to a constant) by the incompressibility condition, $\nabla\cdot \mathbf{u} = 0$. The constant is fixed by setting $p_1$ to zero at an arbitrary point (e.g.~the lower left corner of the flow domain). To fix the unknowns $\Delta p_x$ and $\Delta p_y$, we impose a condition on the net fluid flow in the vertical and horizontal directions. We assume that the lattice of flapping plates is situated in a larger flow domain that ends
at solid boundaries, where the flow is zero. We therefore assume that the spatially periodic flow approximates the flow away from the boundaries, but take the spatial average of the flow in the lattice to be zero at all times, to match that at the boundaries. The same assumption has been used in theoretical and computational studies of sedimenting suspensions at zero \cite{hasimoto1959periodic,batchelor1972sedimentation,brady1988dynamic,phillips1988hydrodynamic,hinch1988sedimentation,swan2010particle,swan2011hydrodynamics,guazzelli2011fluctuations,af2014fast} and nonzero \cite{ladd1994numerical,mucha2004model,yin2008velocity,fornari2018clustering} Reynolds numbers, and with background turbulence \cite{fornari2016sedimentation}. In these studies the flow is typically solved in a periodic lattice or periodic cell domain, and the velocity of the sedimenting particles relative to zero-volume-flux axes is interpreted as the velocity in the physical or lab frame. In our case, the plates have zero volume, so the volume flux is that of the fluid alone. For periodic domain models of sedimentation and in the present work on flapping locomotion, there is assumed to be a transition region near the boundary where the flow deviates from spatially periodic, to obtain zero flow at the boundary. In a sedimentation simulation, Mucha {\it et al.} found that including the boundary region in the simulation had a negligible effect on particle velocity statistics far from the boundary \cite{mucha2004model}.


\section{Numerical method \label{sec:Numerics}}

\begin{figure} [h]
           \begin{center}
           \begin{tabular}{c}
               \includegraphics[width=6in]{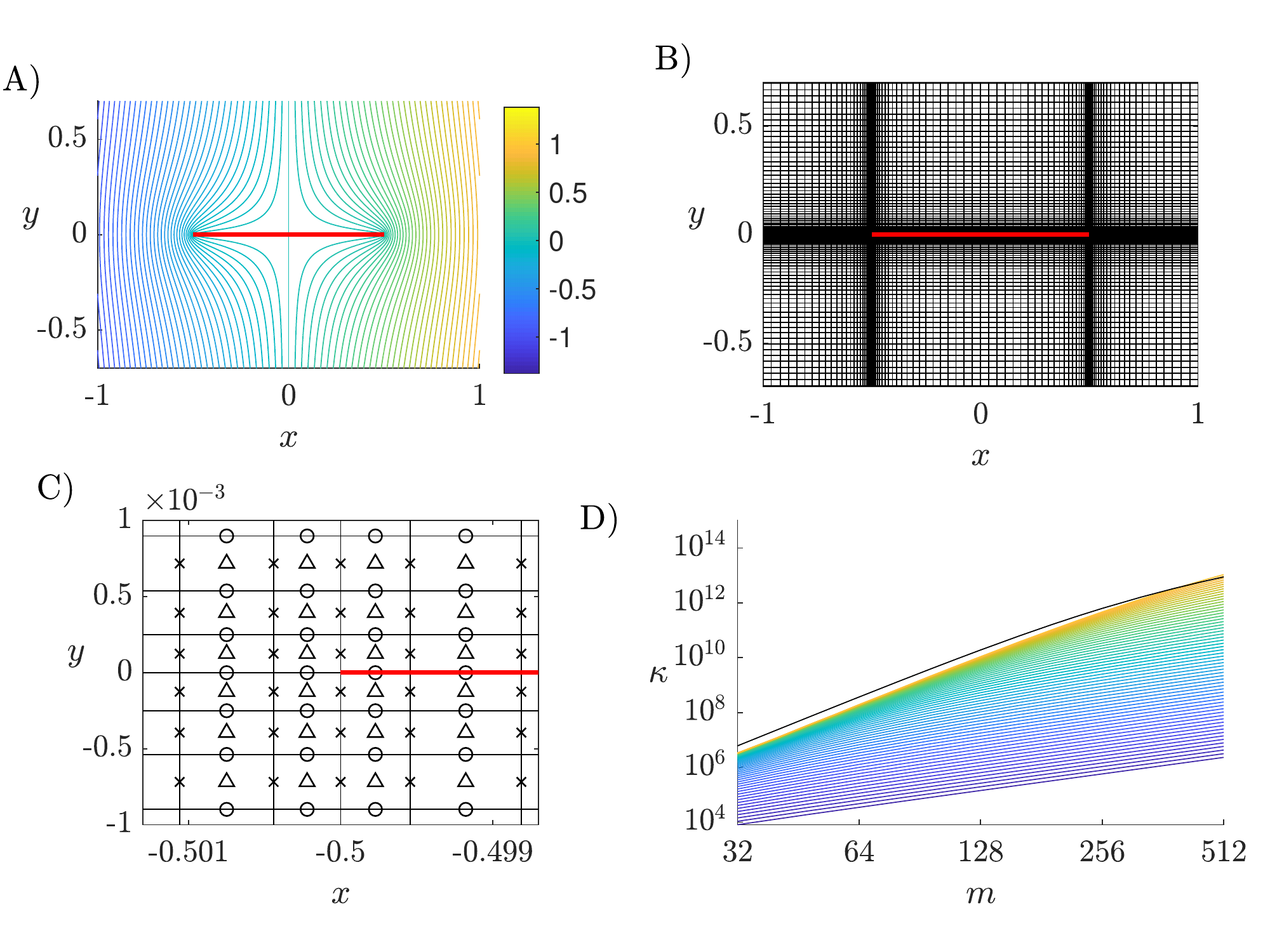}\\
           \vspace{-.25in}
           \end{tabular}
          \caption{\footnotesize Test problem and numerical grids. 
A) Streamlines for potential flow past a flat plate. B) Example of a grid with refinement near the plate (red). C) A close-up of
the grid near the left plate edge. The values of the velocity components and pressure ($u$, $v$, and $p$) are
solved at the locations of the crosses, circles, and triangles, respectively. D) The growth in the 2-norm condition
number of the discrete Laplacian matrix. Each colored line plots the condition number versus $m$ for a given $\eta$, ranging from
$1-2^{-2}$ (darkest blue line) to $1-2^{-14}$ (lightest yellow line); $1-\eta$ decreases by a factor of $2^{-0.2}$ from each line to the one above it, giving an increased concentration of points near the plate edges.
 \label{fig:AnalyticalLaplacianTestFig}}
           \end{center}
         \vspace{-.10in}
        \end{figure}

We choose a flat plate geometry instead of a thin curved body (e.g.~an ellipse) because it fits a periodic rectilinear grid, at the expense of creating flow singularities at the plates' edges. To study the effect of the singularity on a
finite-difference solution of (\ref{NS1})--(\ref{NS2}), we study a simpler model problem with the same type of 
singularity: potential flow past a flat plate, shown in figure \ref{fig:AnalyticalLaplacianTestFig}A. The plate is the red line segment---extending along $(-1/2 \leq x \leq 1/2\;; y = 0)$---and the complex potential is $w(z) = \sqrt{1/4 - z^2}$, with branch cut lying
along the plate. We solve Laplace's equation for the stream function $\psi = \mbox{Im}\{w\}$ in a rectangle $R$
centered at the origin (the plate center), with lengths 3 and 2 in the $x$ and $y$ directions, respectively: 
\begin{align}
\nabla^2\psi &= 0 \quad, \quad   (x,y) \in R: -3/2 \leq x \leq 3/2,\, -1 \leq y \leq 1; \label{Laplace}\\ 
\psi &= 0 \quad, \quad -1/2 \leq x \leq 1/2,\, y = 0 ;\\ 
\psi &= \mbox{Im}\{\sqrt{1/4 - (x+i y)^2}\} \quad , \quad  (x,y) \in \partial R. \label{PsiBdy}
\end{align} 
$\psi$ is continuous but its first derivatives diverge as inverse square roots of distance from
the plate edges. Based on Stokes-flow solutions and local asymptotics of Navier-Stokes solutions \cite{hasimoto1958flow,ingham1991steady}, $\psi$ has the same type of singularity as the velocity components in (\ref{NS1})--(\ref{NS2}) (i.e.~the {\it viscous} flows, not the potential flow defined by $\psi$). Both $\nabla^2 \psi$ and the highest-order derivatives in 
(\ref{NS1})--(\ref{NS2}), i.e.~$\nabla^2 u, \nabla^2 v$, and $\nabla p$, 
diverge as distance from the plate edges to the -3/2-power. We will use second-order finite
differences to discretize both the test problem (\ref{Laplace}) and the viscous problem (\ref{NS1})-(\ref{NS2}), even though the Taylor series expansions underlying the
finite difference approximations diverge at the plate edges. Our goal with the test problem is to 
measure the error in a case with a simple analytical solution, given by (\ref{PsiBdy}) in all of $R$. 

To mitigate the errors, we will employ nonuniform rectilinear (tensor-product) grids, and concentrate grid points near the plate edges, and along the plate surfaces, 
as shown in figure \ref{fig:AnalyticalLaplacianTestFig}B. For the viscous problem, we use the MAC (marker-and-cell) scheme for
incompressible flows \cite{harlow1965numerical} with the grid aligned with the plate as shown
in the sample grid in figure \ref{fig:AnalyticalLaplacianTestFig}C. The velocity components $u$ and $v$ are solved at the crosses and
circles respectively, and the pressure $p$ is solved at the triangles. The $x$ and $y$ locations of the symbols are either on the grid lines, or at midpoints between grid lines. For the test problem, we
solve $\psi$ on the $u$-grid in panel C (i.e.~at the crosses).

The grid lines are defined from uniform grids using a grid-stretching parameter $\eta$. For the $x$-grid on the plate in panels B and C, we first define a uniform grid from -1/2 to 1/2 in $X$, and then the $x$ coordinates of the grid are defined by:
\begin{align}
x = X + \eta \frac{1}{2\pi} \sin{2\pi X}, -1/2 \leq X \leq 1/2. \label{xgrid}
\end{align}
\nn If the uniform grid spacing in $X$ is $\Delta X$, the spacing $\Delta x$ on the stretched grid $x$ 
increases from approximately $1-\eta$ at the plate edges to $1+\eta$ at the plate center. As $\eta$ increases from 0 to 1, the stretched grid transitions from uniform to highly concentrated at the plate
edges. We choose a number of grid points for the plate, and then for the $x$-grid to the left and right of the plate in panel B, we set the number of grid points to be approximately that in the grid along the plate, scaled by
the ratio of the outer region length to the plate length (unity), raised to the 1/2 power. The
functional form of the grids is the same as that in (\ref{xgrid}), with a stretching factor chosen
so that the grid density is approximately continuous at the plate edges.
The $y$ grid is defined similarly to (\ref{xgrid}),
\begin{align}
y = Y - \eta \frac{l_y}{2\pi} \sin{2\pi Y/l_y}, -l_y/2 \leq Y \leq l_y/2, \label{ygrid}
\end{align}
\nn given a uniform grid in $Y$. In the viscous computations that follow, the total numbers of grid points in $x$ and $y$ are similar (within a factor of 2), and in the potential flow test problem here they are equal (and denoted $m$).
For the test problem, we solve (\ref{Laplace}) on the grid shown by crosses in figure \ref{fig:AnalyticalLaplacianTestFig}B, for various values of
$m$ and $\eta$. Due to the discontinuity in flow quantities (e.g.~velocity derivatives and pressure) across the plate, we use one-sided finite differences near the plate for all derivatives in both the test problem and the viscous solver, to maintain their accuracy away from the plate edges. To describe when the accuracy becomes hampered by ill-conditioning, we present the 2-norm condition number of the discrete Laplacian matrix for
$\psi$, for various $m$ and $\eta$ in figure \ref{fig:AnalyticalLaplacianTestFig}D. Each colored line plots the condition number versus $m$ for a given $\eta$, ranging from
$1-2^{-2}$ (i.e.~0.75, darkest blue line) to $1-2^{-14}$ (lightest yellow line), in order of increasing concentration of points near the plate edges. For each $\eta$, the condition number initially grows faster than $m^{-2}$, then asymptotes to this scaling for sufficiently large $m$. The $m$ at which the transition occurs depends on
$\eta$. For $\eta$ = 0.75 (darkest blue line), the line scales as $m^{-2}$ for all $m \geq 32$, while for $\eta = 1-2^{-14}$ (lightest yellow line), the transition is only beginning to occur at $m = 512$. For a given $\eta$, when $m$ is relatively small, increasing $m$ increases the density of points near
the plate edges more than in the rest of the domain. When $m$ is sufficiently large, further increases in $m$ increase the density of points by the same percentage everywhere. At this point we obtain the usual $m^{-2}$ condition number scaling of the discrete Laplacian, albeit with a non-uniform grid. For $m \leq 512$ and $\eta \leq 1-2^{-14}$, the condition number indicates a round-off error at least a few orders of magnitude below double precision (10$^{-16}$). In the viscous simulations, we set $\eta = 0.95$, corresponding to a line in the bottom fifth of those in panel D, and the round-off error is several
orders of magnitude away from double precision.

\begin{figure} [h]
           \begin{center}
           \begin{tabular}{c}
               \includegraphics[width=6.7in]{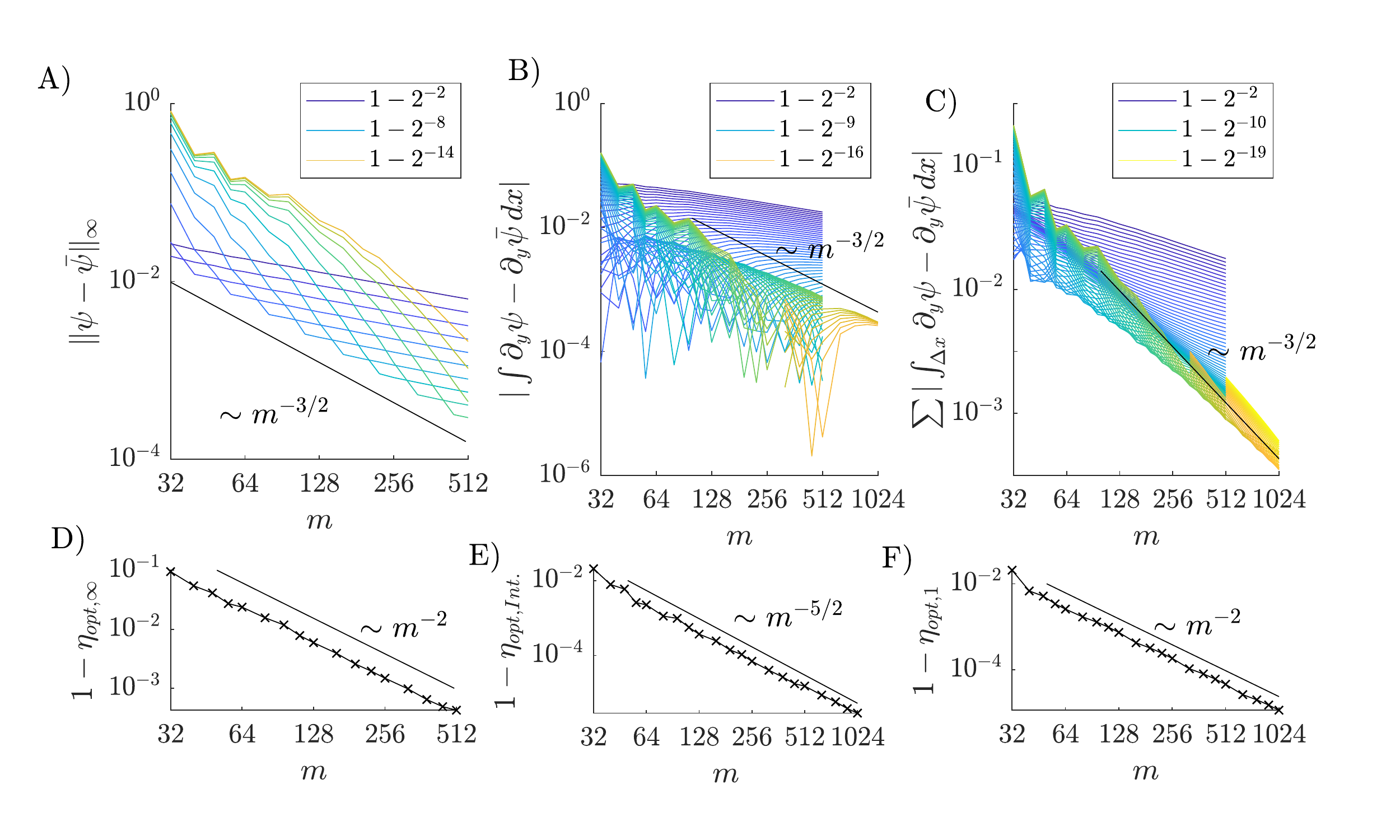}\\
           \vspace{-.25in}
           \end{tabular}
          \caption{\footnotesize Errors in the computed potential flow stream function relative to the exact solution. 
A) Infinity (sup) norm error over the 3-by-2 rectangular domain. B) Error in integral of $\partial_y\psi$ over
the top left half of the plate. C) Sum of the absolute values of the errors in $\partial_y\psi$, at the midpoints of each grid subinterval on the top left half of the plate. D) and F) For each $m$, the 
values of $\eta$ at which the minimum errors occur in panels A and C. E) the values of $\eta$ corresponding to the envelope below the black fit line in panel B.
 \label{fig:AnalyticalLaplacianErrorFig}}
           \end{center}
         \vspace{-.10in}
        \end{figure}

We now study the effects of $m$ and $\eta$ on errors for the test problem, where the analytical solution is known. In figure \ref{fig:AnalyticalLaplacianErrorFig}A-C, we plot a few different measures of
error in $\psi$. Panel A shows the infinity (sup) norm of the error in the computed $\psi$ over the full grid relative
to the exact solution $\bar{\psi}$, given by (\ref{PsiBdy}) in all of $R$. Each colored line again corresponds to a particular $\eta$ value, a few of which are labeled in the legend (the remaining values lie between these values, and are of the form $1-2^{-k/5}$ for $k = 10, 11, \dots, 69, 70$).
For each line, the error initially decreases rapidly with increasing $m$, then much more
slowly (as $m^{-1/2}$) for further increases in $m$. This transition again reflects the transition in
where the density of points is being increased most, near the tip at smaller $m$, and uniformly at
large $m$. The singularity at the plate edges reduces the scaling from $m^{-2}$ for a smooth
problem with second-order finite differences to $m^{-1/2}$. However, by choosing the best $\eta$ for a given $m$, we obtain the lower envelope of the lines in panel A, which has scaling $m^{-3/2}$, closer to the smooth case. In the viscous simulations, we need to compute the forces on the plate, which
require integrating the pressure and velocity gradient, each with inverse-square-root singularities
near the plate edges. For the test problem, $\partial_y\psi$ is the analog of the velocity gradient, with the same singularity strength.
 In panel B, we plot the error in its integral over the top left half of the plate,  
computed with second-order finite differences and then integrated using the trapezoidal rule.
The exact integral is 1/2. Each curve again corresponds to a given $\eta$, which we take to 
$1-2^{-16}$ now, closer to 1, to see the asymptotic behavior at large $m$ better. Each curve eventually
scales as $m^{-1/2}$, but by taking the minimum error over $\eta$ at a given $m$, we can do much better. In fact, for each $m$ there is apparently an $\eta$ for which the error passes through zero, as shown by the downward spikes of the curves on this log scale. The typical error magnitude in the vicinity of this $\eta$ is shown by the upper envelope of the curves at somewhat larger $m$. The black fit line
shows that this envelope scales as $m^{-3/2}$. Therefore, the error in the integral of $\partial_y \psi$
up to the plate edge behaves similarly to the maximum error in $\psi$ over the domain.  The error according to a somewhat more stringent criterion is shown in panel C. We again consider the integrated error in $\partial_y \psi$, but now integrate the absolute value of the error over
the top left half of the plate. This avoids the cancellation of errors over different portions of the plate, which led to the error passing through zero in panel B. Therefore this measure of error avoids the possibility of errors being hidden by cancellation. If we compute it using the trapezoidal rule, the error is $+\infty$ in each case, because of the infinite value of $\partial_y \bar{\psi}$ at the plate edge. Therefore we use a version of the midpoint rule. On each grid interval on the top left half of the plate, we take the average of the computed values at the endpoint minus the exact value at the midpoint (which is always finite), and sum the absolute value of the difference, weighted by the length of each interval. The behavior is similar to that of panel A: for a fixed curve (fixed $\eta$), the error $\sim m^{-1/2}$ at large enough $m$. But 
the lower envelope of the curves $\sim m^{-3/2}$. Panels D--F show, for each $m$, the $\eta$ values
corresponding to the aforementioned envelopes. In panels A and C, these are the error-minimizing $\eta$ at each $m$. Their distance from 1 is seen to decay as $m^{-2}$. In panel B, these are the $\eta$ corresponding to the envelope, and here $1-\eta$ decays slightly faster, approximately as $m^{-5/2}$.

Figure \ref{fig:AnalyticalLaplacianErrorFig} shows that even with the plate edge
singularities, errors can be decreased below 1\% with $m$ not very large $\approx 100$ and $\eta$ close to 1.  
For the viscous simulations, we have the additional need to resolve vorticity throughout the flow domain, though it is strongest near the plate edges. We set $\eta$ to 0.95, close enough to 1 to greatly diminish errors at the plate edges, but far enough to avoid the possibility of ill-conditioning in the viscous system of equations. We take $m$ between 256 to 512, and find that these choices are sufficient to resolve the flows and fluid forces on the plate to within a few percent in relative error for the ranges of parameters (e.g.~domain size, Reynolds number, etc.) studied.

\section{Single flapping plate \label{sec:SinglePlate}}

Before studying lattices of flapping plates, we examine a single flapping plate in flows of various speeds to establish a baseline of performance to which the lattice configurations can be compared. 
We solve the second-order finite difference discretization of (\ref{NS1})--(\ref{NS2}) on the MAC grid
(e.g.~figure \ref{fig:AnalyticalLaplacianTestFig}B and C) as a fully coupled system for $u$, $v$, and
$p$. To simulate an isolated flapping plate in an unbounded fluid, we employ upstream, downstream, and sidewall boundary conditions in the rectangular domain, and take the boundaries sufficiently far from the body that they affect the results by less than a few percent in relative error. The upstream and downstream sides are 3 and 8 plate lengths from the plate's leading edge, and the sidewalls are 5 plate lengths from the plate. At the upstream boundary, $u = -U$ and $v = -V(t)$ are imposed. On the sidewalls, free-slip conditions are imposed ($\partial_y u = 0$, $v = -V(t)$) to avoid vorticity generation. At the downstream boundary, advective outflow conditions are used ($\partial_t u -\mathbf{U}(t)\cdot\nabla u$ = $\partial_t v -\mathbf{U}(t)\cdot\nabla v$ = 0). Similar boundary conditions have
been used in other recent simulations of flows past bodies \cite{tamaddon1994unsteady,sen2009steady,peng2012criticality,yang2012vortex,cid2018numerical}.

\begin{figure} [h]
           \begin{center}
           \begin{tabular}{c}
               \includegraphics[width=7in]{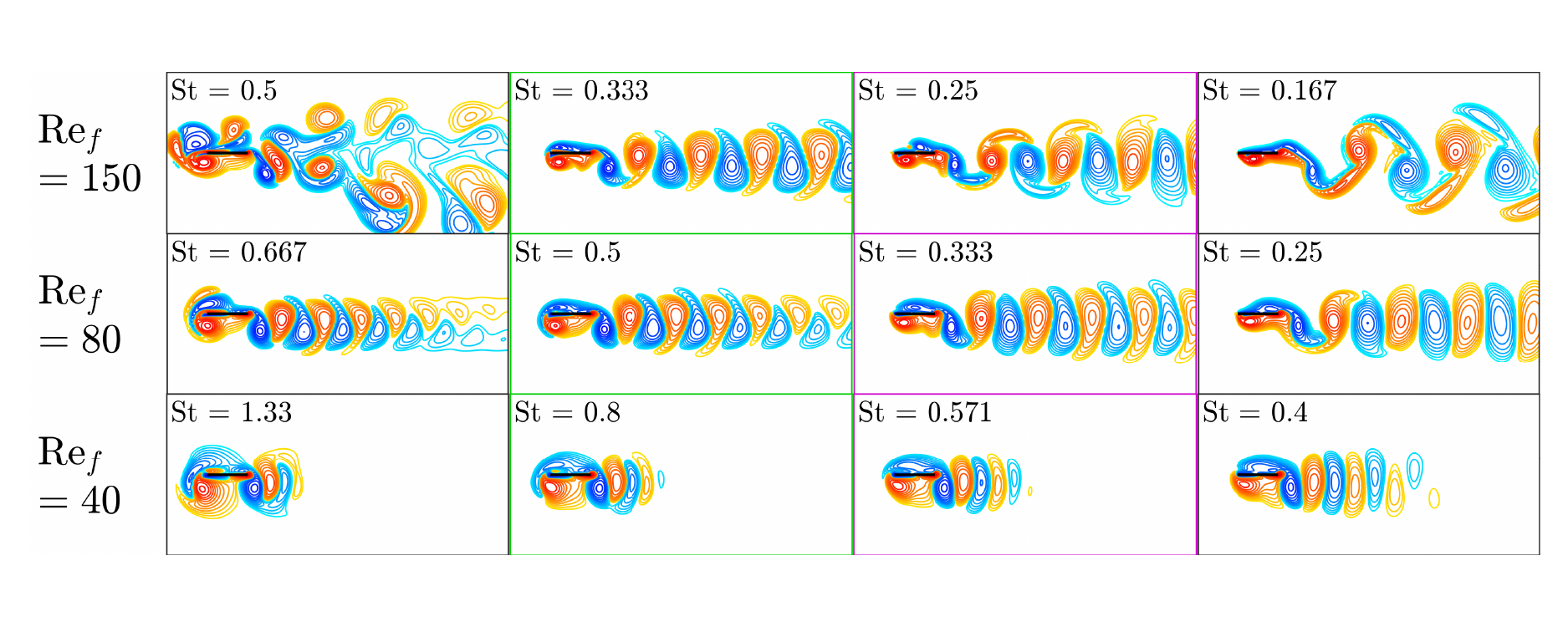}\\
           \vspace{-.25in}
           \end{tabular}
          \caption{\footnotesize Snapshots of vorticity fields with normalized amplitude $A/L$ = 0.2 and different flapping frequencies (Re$_f$, labeled at left) and Strouhal numbers St = $2Af/U$, labeled in each box, corresponding to increasing oncoming flow speed from left to right.
 \label{fig:OneBodyVorticity2BMPFig}}
           \end{center}
         \vspace{-.10in}
        \end{figure}

For an isolated body we have $l_x, l_y \to \infty$ in
(\ref{params}), and we are left with three parameters, $A/L$, Re$_f$, and St.  
Examples of flows with normalized flapping amplitude $A/L$ = 0.2 and various Re$_f$ and St are shown in figure \ref{fig:OneBodyVorticity2BMPFig}. At zero oncoming flow speed, or infinite St (not shown), the flow has a left-right symmetric equilibrium state (with equal and opposite vorticity at the two plate edges). This state becomes unstable to asymmetric motions above a critical value of Re = 4$A$Re$_f/L$ (\cite{vandenberghe2004symmetry,alben2008implicit}). For small but nonzero oncoming flow speed, and sufficiently large Re$_f$, the
vortices shed from the plate edges collide with the body and may travel to the sidewall or upstream boundaries (violating the boundary conditions there). In the upper left panel of figure \ref{fig:OneBodyVorticity2BMPFig} (Re$_f$ = 150, St = 0.5), however, the flow speed is sufficiently large that the vortex wake is generally advected downstream, and is a somewhat disordered array of dipoles, one shed per half-cycle. 
Moving one panel to the right (green box in the second column) is a smaller St value, close to where
the Froude efficiency is maximized for this Re$_f$, and the larger oncoming flow allows the vortex wake to organize into the familiar reverse von K\'{a}rm\'{a}n street \cite{triantafyllou2004review}. One panel further to the right (purple box in the third column of the top row) is St close to the self-propelled state, where $\langle F_x \rangle = 0$. In the last column, the flow speed is much larger and the
body experiences drag although the wake still resembles a reverse von K\'{a}rm\'{a}n street, but with more widely spaced vortices. The second
and third rows show the same sequences of flows as oncoming flow speed is increased, but at successively smaller Re$_f$. Viscous diffusion of vorticity is more apparent, particularly in the bottom row. In the bottom two rows, the negative (blue) vortices move upwards relative to the positive (red) vortices at larger oncoming flow speeds, indicating the transition from reverse towards regular von K\'{a}rm\'{a}n streets \cite{godoy2008transitions}. For fish and birds at Re = $10^3-10^5$, the optimally efficient St are generally in the range 0.2--0.4 \cite{triantafyllou2004review}, while here Re = $10 - 10^2$, and the most efficient St are higher. The St that are close to optimal for Froude efficiency (green boxes) increase as Re decreases (from top to bottom rows), which is also seen in organisms as Re decreases \cite{eloy2012optimal}.
 
\begin{figure} [h]
           \begin{center}
           \begin{tabular}{c}
               \includegraphics[width=7in]{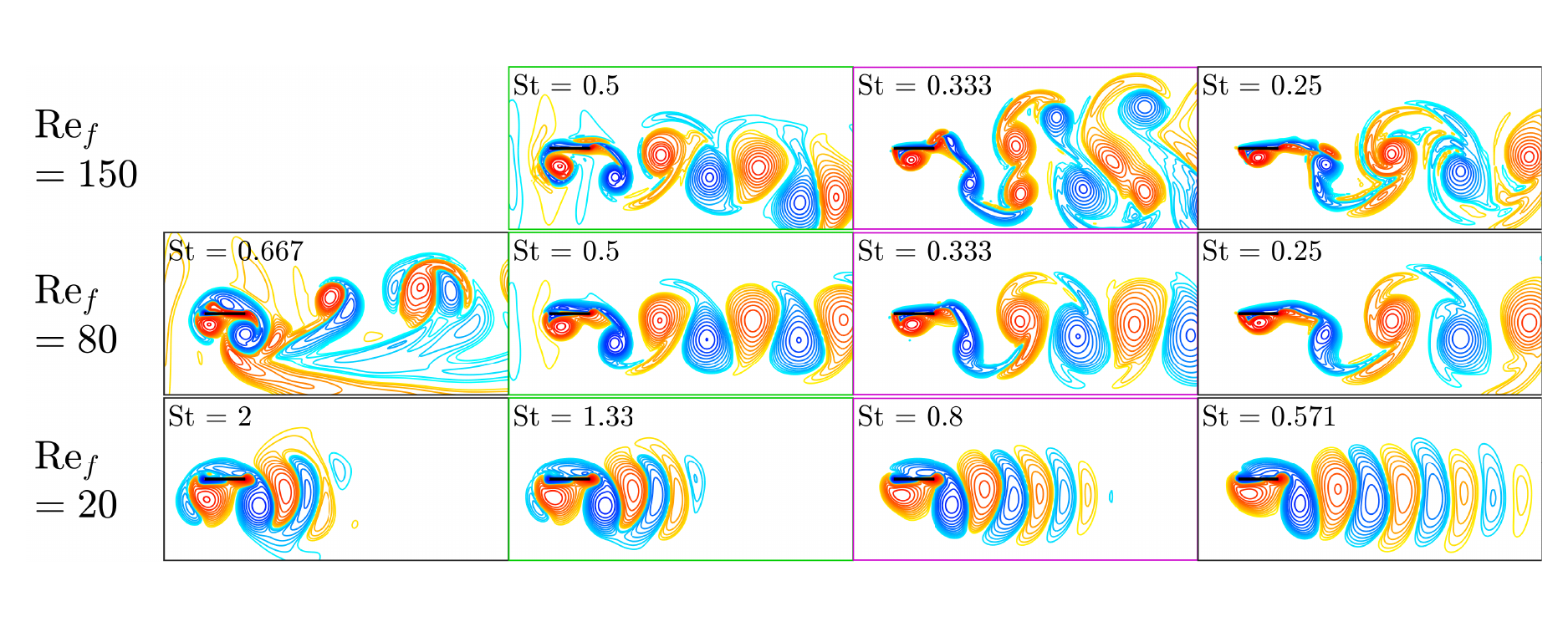}\\
           \vspace{-.25in}
           \end{tabular}
          \caption{\footnotesize Snapshots of vorticity fields with normalized amplitude $A/L$ = 0.4 and
other quantities as described in figure \ref{fig:OneBodyVorticity2BMPFig}.
 \label{fig:OneBodyVorticity4BMPFig}}
           \end{center}
         \vspace{-.10in}
        \end{figure}

Figure \ref{fig:OneBodyVorticity4BMPFig} shows the same transitions with increases in oncoming flow speed, but with $A/L$ increased to 0.4. At the upper left, no flow is shown, because at small flow speeds, vortices collide with the sidewalls. Moving rightward to the green box in the top row, we obtain an up-down asymmetric vortex street and vortex pairing, corresponding generally to nonzero average vertical force on the plate. The approximate self-propelled state (purple box) has an irregular vortex street with multiple vortices shed per half stroke, akin to the 2P wake \cite{schnipper2009vortex}. At Re$_f$ = 80 (second row), at the slowest speed (St = 0.667), the vortex wake again has a complicated structure. At St near the maximum efficiency state (0.5, green box), the wake is a reverse von K\'{a}rm\'{a}n street, which is maintained but dilated downstream at higher oncoming flow speeds. In the third row (Re$_f$ = 20), the vortex street has the reverse von K\'{a}rm\'{a}n structure at all St shown. In general, the effect of increasing $A/L$ to 0.4 is to increase the lateral spacing of the vortex street in the most efficient and self-propelled states (green and purple boxes). The horizontal spacing is influenced most directly by the oncoming flow speed, but $A/L$ also plays a role in the timing of vortex formation and shedding.

\begin{figure} [h]
           \begin{center}
           \begin{tabular}{c}
               \includegraphics[width=6in]{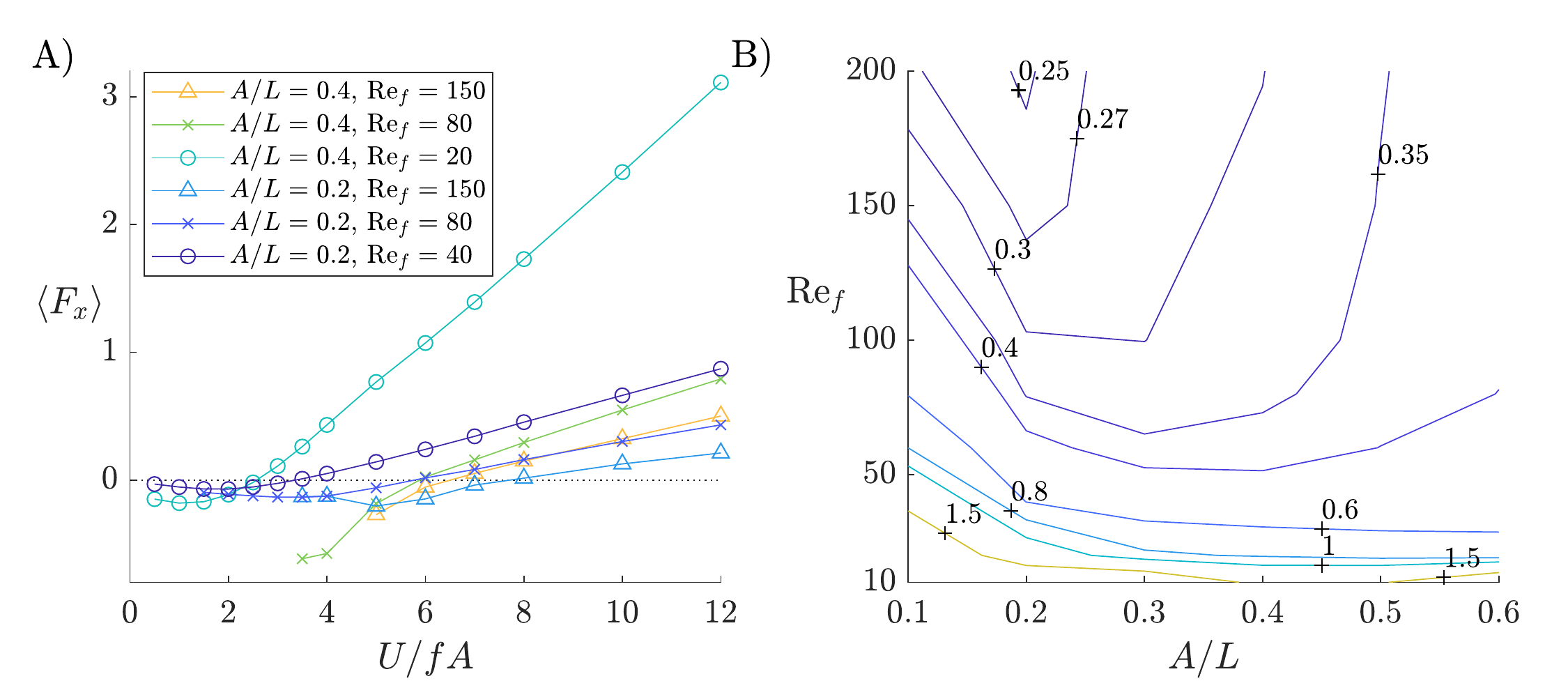}\\
           \vspace{-.25in}
           \end{tabular}
          \caption{\footnotesize A) Average horizontal force $\langle F_x \rangle$ versus normalized
flow speed $U/fA$ = 2/St. B) Contour map of Strouhal numbers corresponding to the self-propelled state ($\langle F_x \rangle$ = 0) of a single flapping plate,
in the space of dimensionless frequency (Re$_f$) and amplitude ($A/L$).
 \label{fig:StSPSFig}}
           \end{center}
         \vspace{-.10in}
        \end{figure}

In figure \ref{fig:StSPSFig}A, we plot the average horizontal force versus normalized oncoming flow speed for the six cases shown in figures \ref{fig:OneBodyVorticity2BMPFig} and \ref{fig:OneBodyVorticity4BMPFig}. Values are omitted where the dynamics are non-periodic, which
occurs over an interval of flow speeds extending from zero; this interval becomes larger as Re$_f$ increases. The curves at the lowest Re$_f$ have a U-shape to the right of zero velocity, indicating that zero velocity is an unstable equilibrium and the self-propelled state is the single stable equilibrium.

To quantify the general features of the self-propelled state, and at smaller oncoming flow speeds, the efficiency-maximizing state,
we compute $\langle F_x \rangle$ and $\eta_{Fr}$ across a wide range of dimensionless frequencies (Re$_f$) and amplitudes ($A/L$).
Figure \ref{fig:StSPSFig}B shows St$_{SPS}$, the Strouhal numbers of the self-propelled states, where $\langle F_x \rangle$ = 0. The numbers grow
rapidly as Re$_f$ decreases to zero, and we expect divergence at a certain Re$_f$, as there is apparently no self-propelled state below a critical Re$_f$ in experiments with rectangular plates \cite{vandenberghe2004symmetry} and simulations of thin ellipses \cite{alben2005coherent}. For a given Re$_f$, the Strouhal number is fairly uniform as $A/L$ varies, indicating that, like steady flows past cylinders \cite{Williamson_AnnuRevFluidMech_1996}
the self-propelled state corresponds approximately to a certain vortex street aspect ratio (roughly St)  that is only slightly modified by $A/L$. Also, St$_{SPS}$ varies smoothly in this region of parameter space, reflecting fairly uniform properties of the reverse von K\'{a}rm\'{a}n street and higher vortex street modes (i.e.~the purple box in the top row of figure \ref{fig:OneBodyVorticity4BMPFig}).

\begin{figure} [h]
           \begin{center}
           \begin{tabular}{c}
               \includegraphics[width=6in]{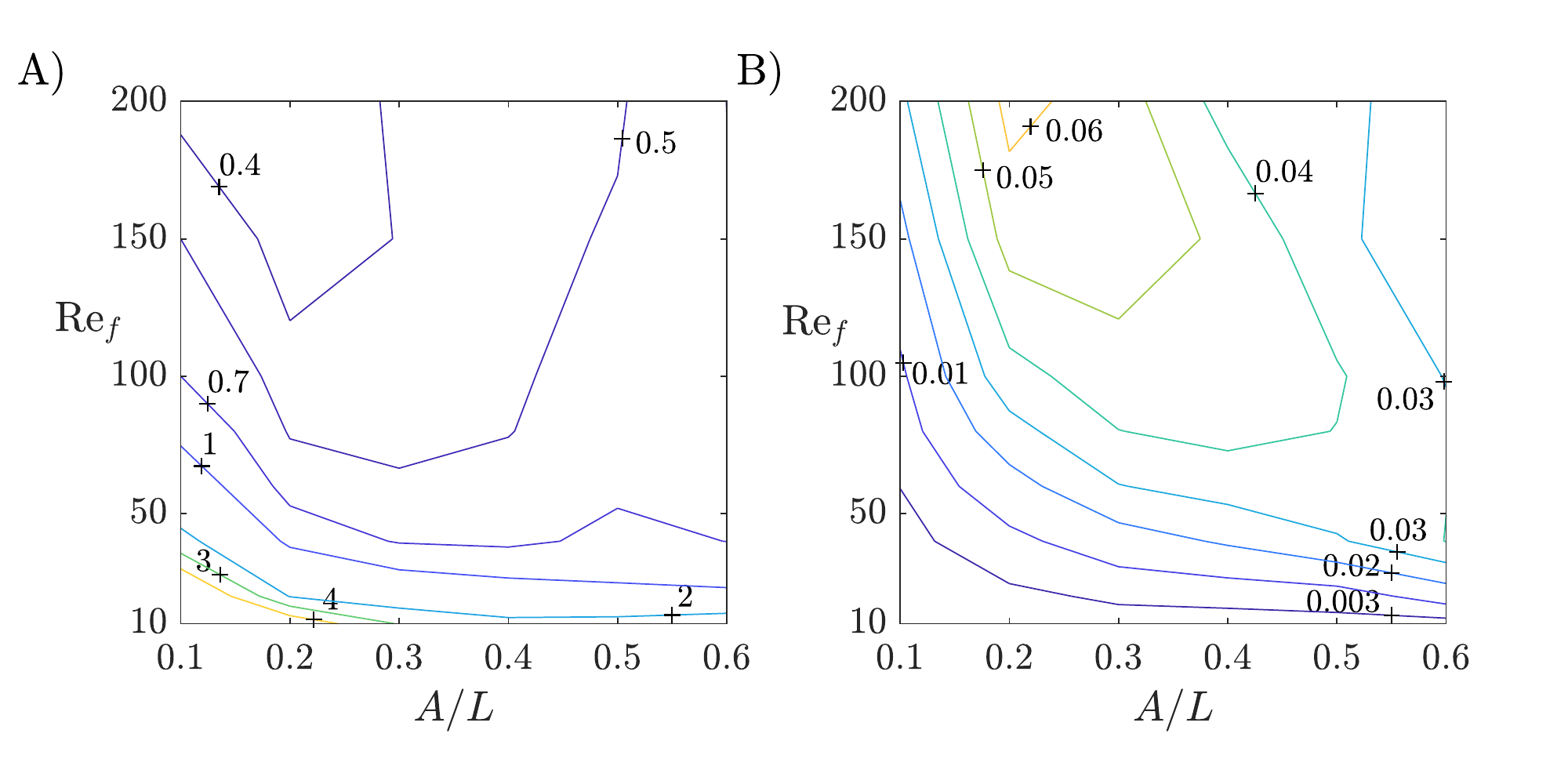}\\
           \vspace{-.25in}
           \end{tabular}
          \caption{\footnotesize A) Strouhal numbers corresponding to maximum Froude efficiency state of a single flapping plate, in the space of dimensionless frequency (Re$_f$) and amplitude ($A/L$). B) Values of Froude efficiency maxima. 
 \label{fig:StFrFig}}
           \end{center}
         \vspace{-.10in}
        \end{figure}

We have seen the maximum Froude efficiency state (approximately the green boxes in figures \ref{fig:OneBodyVorticity2BMPFig} and
\ref{fig:OneBodyVorticity4BMPFig}) occurs at a somewhat lower speed (higher St) than the self-propelled state (purple boxes).
In figure \ref{fig:StFrFig}A we plot contours of maximum-efficiency St and find a pattern of the contours is very similar to
that in figure \ref{fig:StSPSFig}B, but with St roughly 50\% higher in most of the plot. Panel B shows the values of the Froude
efficiency maxima. Efficiency can only be positive above the critical Re$_f$ at which self-propelled locomotion is possible. Not surprisingly, efficiency generally increases with Re$_f$, as vortex shedding becomes more significant. The efficiency reaches a maximum of 0.06 as Re$_f$ increases to 200. Other experimental and computational studies have found the Froude efficiency is nearly unity at much higher Reynolds numbers \cite{ASBT_JFluidMech_1998,floryan2019large}. Panel B also shows that at small Re$_f$, the efficiency-maximizing $A/L > 0.6$, and gradually decreases to 0.2 as Re$_f$ increases to 200. 

\begin{figure} [h]
           \begin{center}
           \begin{tabular}{c}
               \includegraphics[width=4in]{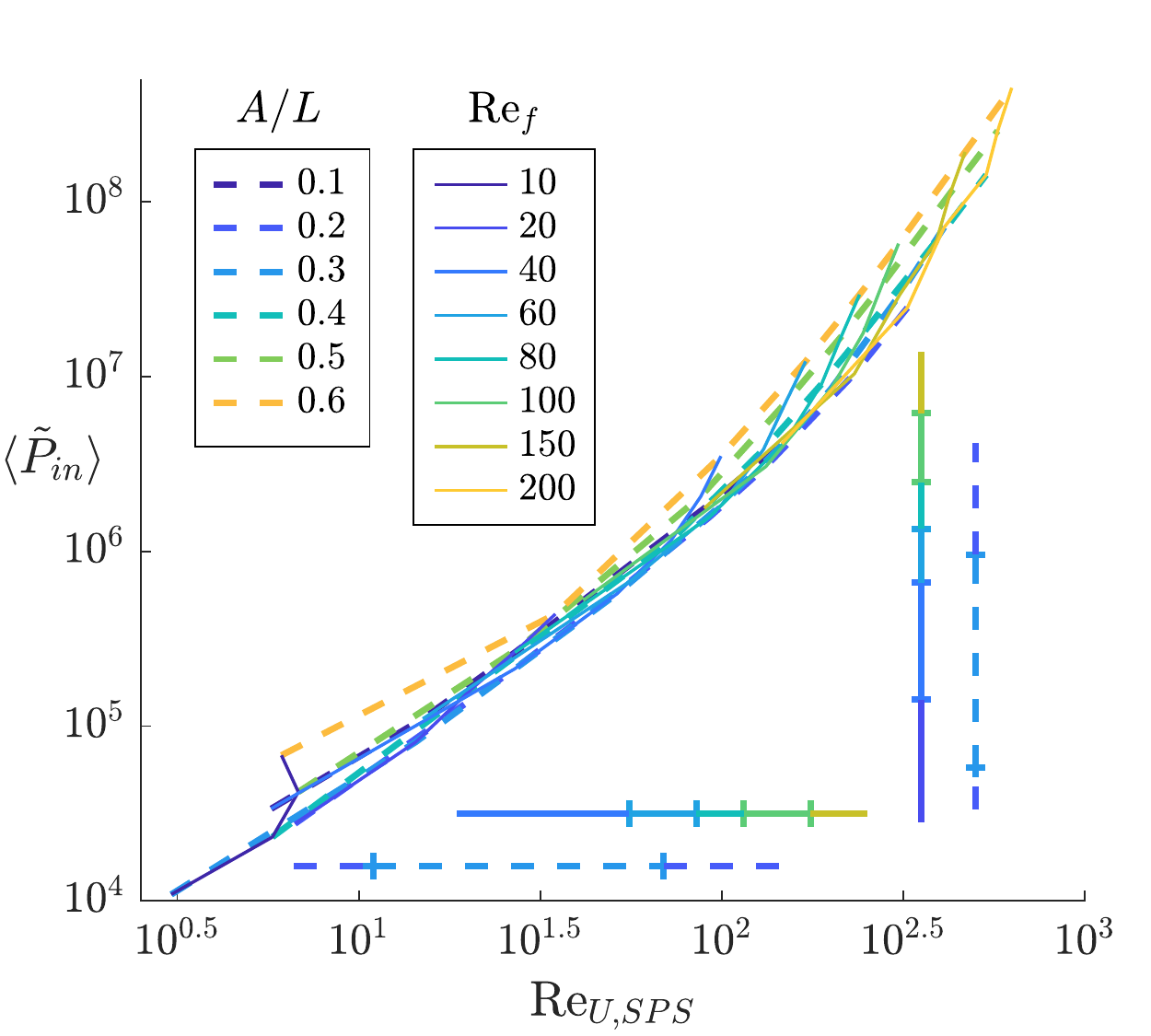}\\
           \vspace{-.25in}
           \end{tabular}
          \caption{\footnotesize Flapping states in the space of self-propelled speed (Re$_{U,SPS}$ = $LU_{SPS}/\nu$) (horizontal axis)
and average input power $\langle {\tilde P}_{in}\rangle$ (vertical axis). Solid lines denote states with a given flapping frequency (Re$_f$) and dashed lines are states with a given flapping amplitude $A/L$.
 \label{fig:ParetoOneBodyFig}}
           \end{center}
         \vspace{-.10in}
        \end{figure}

The Froude efficiency is perhaps the most common measure of efficiency in flapping-foil studies, but it is not the only way to study optimal motions. One can also consider the state that maximizes a desired output (mean thrust, or self-propelled speed, say) for various values of the input power \cite{wu1971hsp,van2015optimal}. In figure \ref{fig:ParetoOneBodyFig}
we map the two-dimensional space of flapping states with frequencies Re$_f \in [10, 200]$ and amplitudes
$A/L \in [0.1, 0.6]$ to the space of self-propelled speed (Re$_{U,SPS} = LU_{SPS}/\nu$) (horizontal axis)
and average input power $\langle {\tilde P}_{in}\rangle$ (vertical axis). The net of lines in the central portion of the figure is
the image of a rectangle in (Re$_f, A/L$) space; each solid line is a set of points with constant Re$_f$ and each dashed line
has constant $A/L$. The lines follow a common trend from lower left to upper right, showing that generally increased Re$_{U,SPS}$ correlates with increased $\langle {\tilde P}_{in}\rangle$. There is a smaller spread in the transverse direction. The Pareto frontier
is the lower right boundary of the region: the set of points that maximize Re$_{U,SPS}$ for a given $\langle {\tilde P}_{in}\rangle$, or
that minimize $\langle {\tilde P}_{in}\rangle$ for a given Re$_{U,SPS}$ \cite{van2015optimal}.  This set provides an alternative definition of maximum efficiency, that
provides different optima for a range of $\langle {\tilde P}_{in}\rangle$, and moving in the direction transverse to the Pareto frontier, we have increasing or decreasing optimality of states.
One could also replace the desired output Re$_{U,SPS}$ with Froude efficiency, $\eta_{Fr}$, and allow $U$ to vary as a third input. We discuss this alternative later in the paper. The vertical lines at the far right of the figure show the
(Re$_f, A/L$) values along the Pareto frontier for the ranges of $\langle {\tilde P}_{in}\rangle$ covered by these lines. The horizontal lines at the bottom show the (Re$_f, A/L)$ values along the Pareto frontier for the ranges of Re$_{U,SPS}$ covered by these lines. The vertical and horizontal lines show that the preferred flapping amplitude remains in the vicinity of 0.2-0.3 along the frontier, while the preferred frequency varies monotonically. In other words, to change speeds, it is most efficient to vary the frequency of the plate but keep its
amplitude roughly fixed. A similar trend has been observed for the tail beats of various fish species as they vary their swimming speed \cite{saadat2017rules}.

\begin{figure} [h]
           \begin{center}
           \begin{tabular}{c}
               \includegraphics[width=5in]{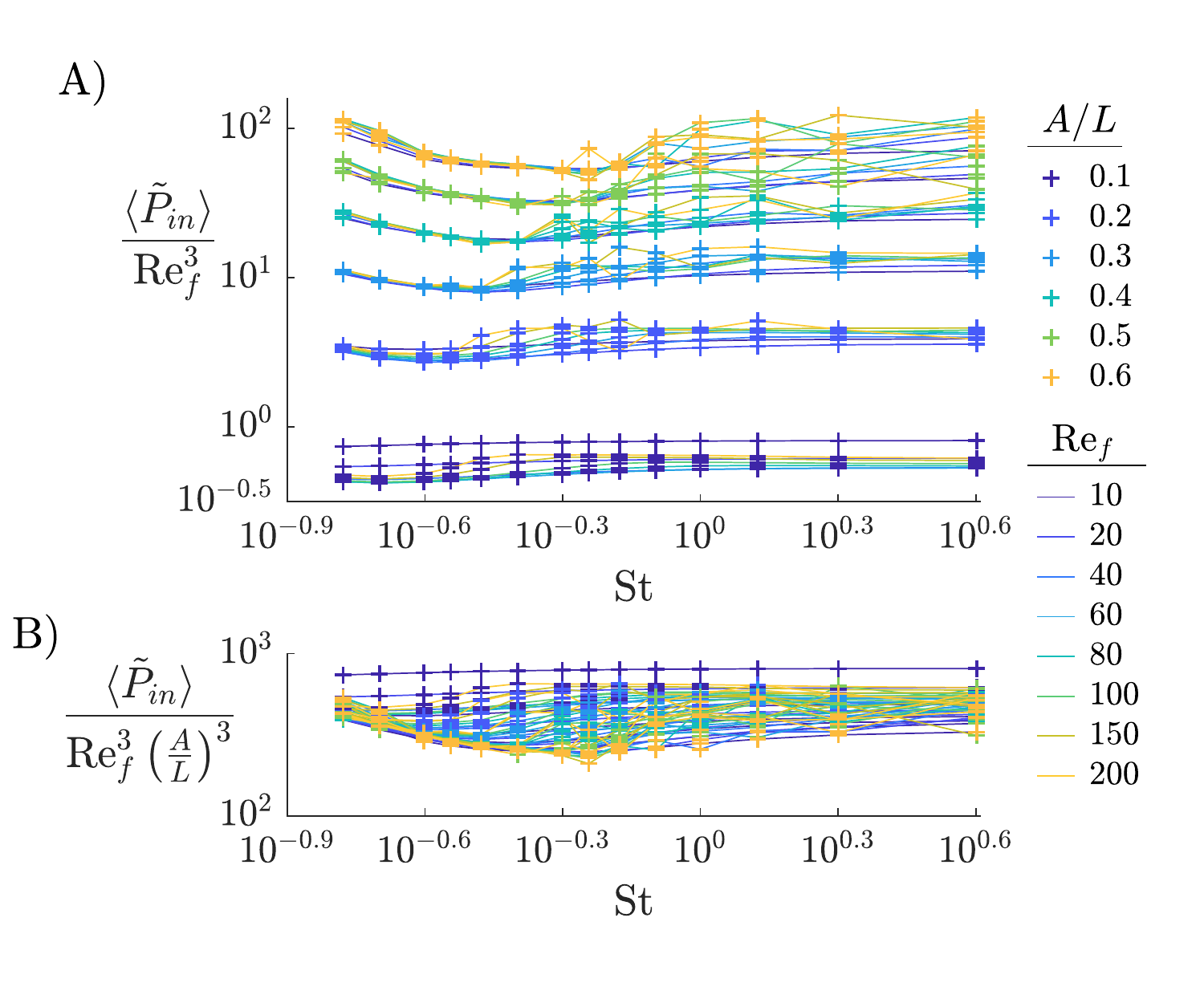}\\
           \vspace{-.25in}
           \end{tabular}
          \caption{\footnotesize For an isolated flapping body, the dependence of average input power $\langle {\tilde P}_{in}\rangle$ on frequency Re$_f$, amplitude $A/L$, and swimming speed St.
 \label{fig:PInScalingFig}}
           \end{center}
         \vspace{-.10in}
        \end{figure}

For an isolated flapping body, the average horizontal force is sensitive to the oncoming flow speed (i.e.~minus the swimming speed) as it increases from zero. The average force is zero (or close to zero) at zero oncoming flow speed, then becomes thrust, then decreases back to zero at the self-propelled state, then becomes drag as swimming speed increases. These changes reflect changes in the vortex wake structure in the vicinity of the reverse von K\'{a}rm\'{a}n
street, and the horizontal force has a subtle dependence on Re$_f$, $A/L$, and St. The input power, by contrast, has a simpler dependence on the parameters, as shown in figure \ref{fig:PInScalingFig}. Panel A shows that $\langle P_{in}\rangle$/Re$_f^3$ is
roughly constant with respect to St, with Re$_f$ varying over a factor of 20  (values listed at bottom right), but depends strongly on $A/L$ (values at right). The dependence on $A/L$ is approximately scaled out by dividing by $(A/L)^3$, as shown by the collapse of lines in panel B, particularly at larger Re$_f$.
Since the vertical plate velocity scales as $(A/L)$Re$_f$, $\langle {\tilde P}_{in} \rangle$ scales as vertical velocity cubed, a typical high-Reynolds-number scaling for a bluff body. Unlike the horizontal force, $\langle {\tilde P}_{in} \rangle$ is relatively insensitive to the oncoming flow
speed (i.e.~St) and the changes in vortex wake patterns shown in figures \ref{fig:OneBodyVorticity2BMPFig} and \ref{fig:OneBodyVorticity4BMPFig}.

\section{Doubly periodic lattices of plates \label{sec:DoublyPeriodicArrays}}

Having established some of the main properties of an isolated flapping plate in an oncoming flow, we now discuss the much larger
space of doubly periodic lattices of flapping plates. We consider two types of lattices, rectangular and rhombic (diamond), shown in figure \ref{fig:SchematicPlatesFig}, also considered by \cite{weihs1975some,hemelrijk2015increased,daghooghi2015hydrodynamic,park2018hydrodynamics,oza2019lattices}. Between these two lattices is the full range of 2D (oblique) lattices. We focus only on the two endpoint lattice types
(rectangular and rhombic) because the remaining parameter space is already quite large (five-dimensional). For the rectangular lattice, we solve the flow in a single unit cell (figure \ref{fig:SchematicPlatesFig}A, dashed line box) with periodic boundary conditions. For the rhombic case, we solve the flow in a domain consisting of two unit cells (figure \ref{fig:SchematicPlatesFig}B, dashed line box), to observe when flow modes arise that are periodic on length scales longer than a single unit cell.
We focus on time-periodic dynamics for the most part, because such states
are generally reached within 5-30 flapping periods. Dynamics usually appear to be nonperiodic at larger Re values, and may therefore require much longer run times to compute long-time averages with high accuracy. We generally avoid presenting time-averaged values for these cases except where noted explicitly in the text (e.g.~for the average input power).

\begin{figure} [h]
           \begin{center}
           \begin{tabular}{c}
               \includegraphics[width=6in]{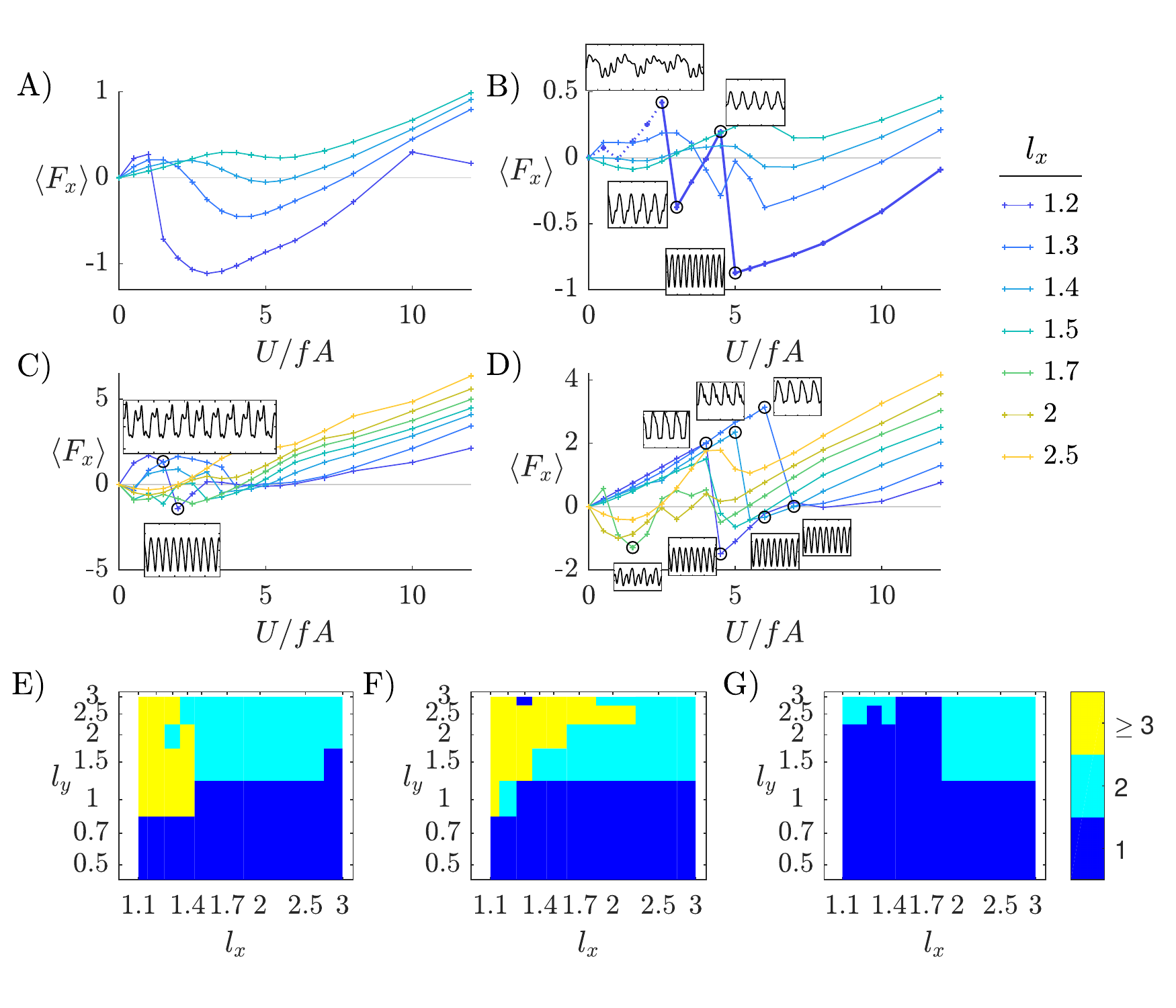}\\
           \vspace{-.25in}
           \end{tabular}
          \caption{\footnotesize Average horizontal force $\langle F_x(t) \rangle$ versus normalized horizontal flow speed $U/fA$ = 2/St for Re = 20 and various $l_x$ and $l_y$ values. In panels A-D, each line plots $\langle F_x(t) \rangle$ versus $U/fA$ for various $l_x$, listed at right. Specific values of $A/L$ and $l_y$ are chosen for each of the panels A-D: A) $A/L$ =  0.2, $l_y = 1$, B) $A/L$ =  0.2, $l_y = 1.5$, C) $A/L$ =  0.5, $l_y = 2$, D) $A/L$ =  0.5, $l_y = 3$. Bottom row: for various ($l_x$, $l_y$) pairs, the number of equilibrium states ($U/fA$ with $\langle F_x(t) \rangle$ = 0), for $A/L$ = 0.2 (E), 0.5 (F), and 0.8 (G).
 \label{fig:StateDiagramFig}}
           \end{center}
         \vspace{-.10in}
        \end{figure}

Figure \ref{fig:StateDiagramFig}A--D shows the average horizontal force $\langle F_x(t) \rangle$ versus normalized horizontal flow speed $U/fA$ = 2/St for a rectangular lattice of plates at Re = 20 and various $l_x$ and $l_y$ values.
In the single-body case, the sidewall and upstream boundary conditions may cause numerical instabilities
when vortices collide with these boundaries, i.e.~when the oncoming flow speed is too small to advect vortices to the downstream boundary. This issue does not arise with doubly-periodic boundary conditions,
and the flow computations remain stable with small oncoming flow speeds, so unlike 
in figure \ref{fig:StSPSFig}A, in \ref{fig:StateDiagramFig}A-D the curves can be computed down 
to zero $U/fA$. Panel A shows four curves with $A/L = 0.2$, $l_y = 1$, and $l_x$ ranging from 1.2 to 1.5 (labeled at right).
The vertical gap is one plate length, but the horizontal gap is smaller, 0.2 to 0.5. $\langle F_x \rangle$ initially increases with $U/fA$, so unlike a single flapping plate at this Re, zero velocity is a stable equilibrium here. After reaching a peak, each curve drops (sharply for $l_x = 1.2$, then more smoothly as $l_x$ increases), and then adopts a U-shape
somewhat similar to that in the isolated body case. For $l_x$ = 1.2 to 1.4, there are three zero crossings (counting $U/fA = 0$), corresponding to three equilibria, two stable and one unstable. Panel B shows
the same data with $l_y$ increased to 1.5. The darkest blue curve ($l_x = 1.2$) now has two sharp
drops, between which the curve increases with $U/fA$. Near zero $U/fA$, the curve is dotted, indicating
that the dynamics are nonperiodic in this region, so in the averages on the dotted line there is some uncertainty (that we do not quantify here). At the last of these nonperiodic cases (highest circled point), the time trace of $F_x(t)$ is shown in a small inset panel, with tick marks every flapping period along the horizontal axis. The graph of $\langle F_x(t) \rangle$ then drops sharply to the next point, also circled, at which the time trace becomes periodic with period 1. The dynamics remain 1-periodic as $\langle F_x(t) \rangle$ increases to the next circled point. Then $\langle F_x(t) \rangle$ drops sharply again, to a state that is 1/2-periodic, the fourth and final circled point on this curve. The curve then increases smoothly with further increases in $U/fA$. In this case, the sharp drops in the curve correspond to changes in periodicity, from nonperiodic, to 1-periodic, to 1/2-periodic. We will discuss the corresponding flow structures below. The remaining curves in this panel, for $l_x$ = 1.3 to 1.5, become increasingly smooth as $l_x$ increases, eventually resembling those in panel A, but with an additional equilibrium for $l_x$ = 1.4 and 1.5; zero velocity is unstable for these cases. Panel C shows the same quantities for $A/L$ increased to 0.5 and $l_y$  increased to 2, and a wider range of $l_x$ (labeled at right). The two inset panels show another example of the change in dynamics (from 4-periodic to 1/2-periodic) that accompany a sharp drop in $\langle F_x \rangle$ at a particular $U/fA$. Panel D ($l_y$ = 3) shows three more examples of changes from 1-periodic to 1/2-periodic dynamics that occur at sharp drops in $\langle F_x \rangle$. Panels C and D indicate a transition with respect to $l_x$ as well. For $l_x$ near 1 (blue curves), the plates experience drag at small $U/fA$. At larger $l_x$ (green and yellow curves), more complex variation of $\langle F_x \rangle$ is seen at small $U/fA$ including thrust. By counting the numbers of zero crossings of these curves (including $U/fA$ = 0), we
obtain the number of equilibrium states, and show the totals as colored patches in the bottom row, for $A/L$ = 0.2 (E), 0.5 (F) and 0.8 (G). In the dark blue regions  $U/fA$ = 0 is the only equilibrium, and there is no net locomotion. This is the case at smaller $l_y$ in most cases, and some larger $l_y$ values at the largest $A/L$ (panel G). The close vertical stacking of adjacent bodies tends to suppress vortex formation and thrust generation, as we will illustrate later. In the light blue regions, there are two equilibria: $U/fA$ = 0 is unstable and there is a stable self-propelled state, as in the isolated-body case. Examples are given by the yellow lines in panels C and D, which represent the closest approximation to the isolated body among these cases ($l_x$ and $l_y$ are largest). However, the body might not be well approximated as isolated in some cases; there can be significant flow interactions across the periodic unit cell, particularly at $A/L$ = 0.5 and 0.8. The yellow regions in the bottom row of panels have three or more equilibria, and these generally correspond to small $l_x$ and large $l_y$. The interactions between adjacent bodies' edges are strongest here, and lead to a variety of flow modes (and dynamics, indicated by the insets we have discussed) that are sensitive to small changes in $U/fA$ and the other parameters. At the largest $A/L$ (panel G), these states are suppressed by the larger amplitude of motion, which tends to suppress interactions between vortices shed by horizontally adjacent plates. 

\begin{figure} [h]
           \begin{center}
           \begin{tabular}{c}
               \includegraphics[width=6in]{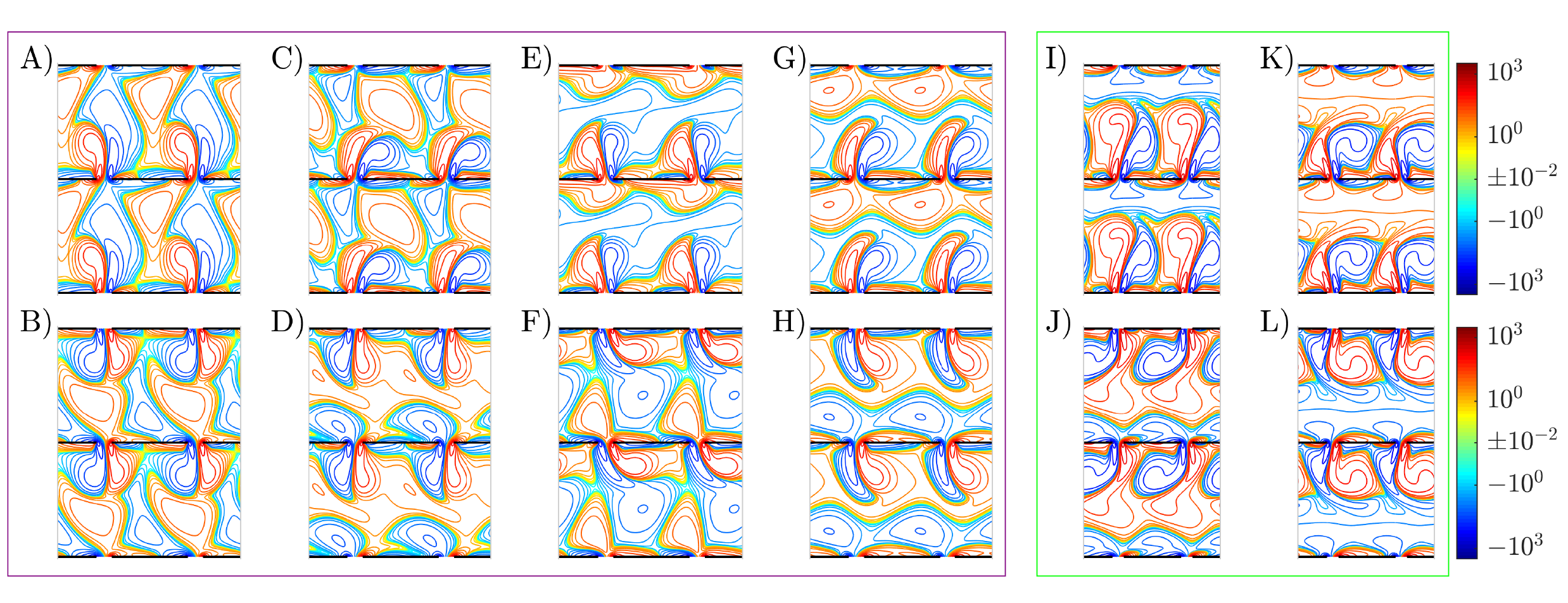}\\
           \vspace{-.25in}
           \end{tabular}
          \caption{\footnotesize Flow states that accompany the sudden drops in $\langle F_x \rangle$ highlighted in figure \ref{fig:StateDiagramFig}B (purple box) and C (green box). Panels A-B and C-D show vorticity snapshots
1/2-period apart, before and after the first sudden drop in $\langle F_x \rangle$ shown by circles in figure \ref{fig:StateDiagramFig}B. Panels E-F and G-H correspond to the second drop in figure \ref{fig:StateDiagramFig}B. The green box shows the flow state transition for the drop in
figure \ref{fig:StateDiagramFig}C. \label{fig:StateDiagramPanelsBCVorticity400}}
           \end{center}
         \vspace{-.10in}
        \end{figure}

Figure \ref{fig:StateDiagramPanelsBCVorticity400} shows examples of the flows near the sudden drops in $\langle F_x \rangle$. In figure \ref{fig:StateDiagramFig}B, four circled data points are shown, bracketing
two sudden drops in $\langle F_x \rangle$. Corresponding flows, at two instants spaced 1/2 of a flapping period apart, are shown by the four pairs of panels in the purple box of figure \ref{fig:StateDiagramPanelsBCVorticity400}. Panels A-B show the flow at $U/fA$ slightly below the first circled data point, in a quasi-periodic state giving drag. In panel A, the upward flow through the thin gap between adjacent plate edges produces an asymmetric vortex dipole. In panel B, a 1/2-period later, the downward flow produces a similar asymmetric dipole. In both cases, although the net flow is rightward, the vortices on the leftward sides of the dipoles are larger. Panels C and D show the corresponding flows at the second circled point in figure \ref{fig:StateDiagramFig}B, after the first sudden drop in $\langle F_x \rangle$, to a state of net thrust. Again, two vortex dipoles are produced on each half-cycle, but now the upward dipole curves rightward (downstream), and the rightward (blue) vortex is larger. However, the downward dipole is still roughly symmetric. Panels E and F show the corresponding flows at the third circled point in figure \ref{fig:StateDiagramFig}B. The downstream flow is larger, and net drag is obtained. In panel E, the upward dipole is more symmetric, similar to the downward dipole in panel D, and the downward dipole in panel F is curved downstream, like that in panel C. Panel G and H show the flows at the fourth circled point in figure \ref{fig:StateDiagramFig}B, after the second sudden transition from a state of drag to a state of thrust. Now both dipoles are curved rightward. Panels G and H also show a flow state that is up-down symmetric after a 1/2-period. Consequently, $F_x(t)$ (inset next to fourth circle in figure \ref{fig:StateDiagramFig}B) is 1/2-periodic---the horizontal force is the same on the up and down strokes. By contrast, the first, second, and third flow states were not up-down symmetric, and $F_x(t)$ was 1-periodic in each case. A similar phenomenon occurs
at the sudden drop in $\langle F_x \rangle$ accompanied by the insets in figure \ref{fig:StateDiagramFig}C.  $F_x(t)$ transitions from
4-periodic to 1/2-periodic in the insets. Flow snapshots are shown in the green box of figure \ref{fig:StateDiagramPanelsBCVorticity400}; panels I-J for
the first inset of figure \ref{fig:StateDiagramFig}C, and panels K-L for the second.
In panels I-J, the vortex dipoles are not up-down symmetric after a 1/2-period, and the upstream member of each vortex pair 
is larger. In panels K-L, the dipoles are up-down symmetric, and the downstream vortices are larger. A similar phenomenon also
occurs at each of the three sudden drops in $\langle F_x \rangle$ highlighted by circles in figure \ref{fig:StateDiagramFig}D.
The corresponding flow transitions, from up-down asymmetric to symmetric, are shown in 
figure \ref{fig:StateDiagramPanelDVorticity400} in appendix \ref{FxFlowTransitions}. The general phenomenon then is that sudden changes from drag to thurst can occur when $l_x$ is close to 1, when
the dipole jets  on each half stroke switch from upstream to downstream orientations. As $l_x$ increases to larger values, the curves may becomed smoothed versions of those with sharp drops, e.g.~in figure \ref{fig:StateDiagramFig}A. Eventually, at large enough $l_x$, the vortex shedding pattern changes qualitatively from a dipole between adjacent leading and trailing edges, to a single dominant vortex that
interacts with previously shed vortices in the wake, e.g.~the reverse von K\'{a}rm\'{a}n street.

\begin{figure} [h]
           \begin{center}
           \begin{tabular}{c}
               \includegraphics[width=6.8in]{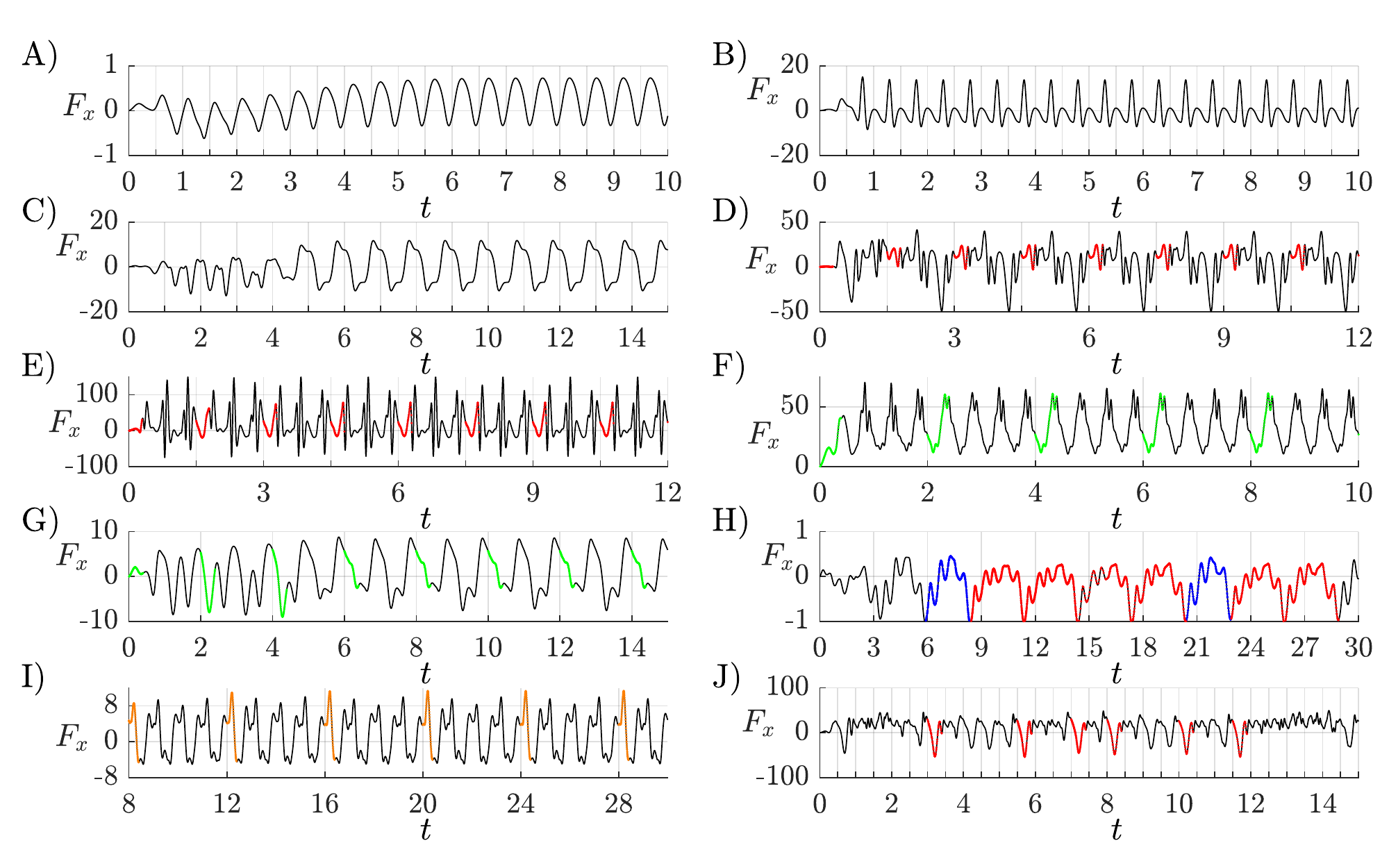}\\
           \vspace{-.25in}
           \end{tabular}
          \caption{\footnotesize Examples of $F_x(t)$ exhibiting various types of periodicity when
$l_x$ is close to unity.
All occur for rectangular lattices with Re = 20. A) $A/L = 0.2$, $l_x = 1.2$, $l_y = 1$, $U/fA = 1$;
B)  $A/L = 0.5$, $l_x = 1.2$, $l_y = 1$, $U/fA = 1$; C) $A/L = 0.5$, $l_x = 1.2$, $l_y = 3$, $U/fA = 1$;
D) $A/L = 0.8$, $l_x = 1.1$, $l_y = 2$, $U/fA = 0.5$; E) $A/L = 0.8$, $l_x = 1.1$, $l_y = 1$, $U/fA = 4.5$; F) $A/L = 0.8$, $l_x = 1.2$, $l_y = 1$, $U/fA = 12$; G) $A/L = 0.5$, $l_x = 1.3$, $l_y = 2$, $U/fA = 3.5$; H) $A/L = 0.2$, $l_x = 1.2$, $l_y = 2$, $U/fA = 1$; I) $A/L = 0.5$, $l_x = 1.2$, $l_y = 2$, $U/fA = 1$; J) $A/L = 0.8$, $l_x = 1.1$, $l_y = 2$, $U/fA = 10$. 
 \label{fig:PeriodicFxExamplesRe20Fig}}
           \end{center}
         \vspace{-.10in}
        \end{figure}

Figures \ref{fig:StateDiagramFig}, \ref{fig:StateDiagramPanelsBCVorticity400}, and
\ref{fig:StateDiagramPanelDVorticity400} have shown that for $l_x$ near 1, $F_x(t)$ can sharply change from 1-periodic (or 4-periodic)
to 1/2-periodic at certain oncoming flow speeds. Figure \ref{fig:PeriodicFxExamplesRe20Fig}
shows further examples of the diversity of periodic $F_x(t)$ that can occur when $l_x$ is near 1.
Panels A and B show 1/2-periodic $F_x(t)$, the first roughly sinusoidal, the second far from it.
Panel C shows a 1-periodic state. Panels D and E show 1.5-periodic states, with repeated features
highlighted in red. Panel E can be regarded as a perturbation of a 1/2-periodic state. Panels F and G
are 2-periodic states with repeated features highlighted in green; panel F is nearly 1/2-periodic, while
panel G is nearly 1-periodic. Panel H shows that the dynamics can switch between different
nearly-periodic states over long periods of time. The blue regions last for 2.5 periods, while the red-regions are nearly 3-periodic, and their recurrences (with slight changes) do not follow a simple
pattern up to $t = 30$. Panel I shows the 4-periodic state of figure \ref{fig:StateDiagramFig}C, top inset, with repeated features highlighted in orange; the state is nearly 1-periodic. Panel J shows
a nonperiodic state that nonetheless has recurrent downward spikes near certain times
that are spaced by multiples of 0.5: $t =$ 3.2, 5.7, 7.2, 8.2, 10.2, and 11.7.

\begin{figure} [h]
           \begin{center}
           \begin{tabular}{c}
               \includegraphics[width=6in]{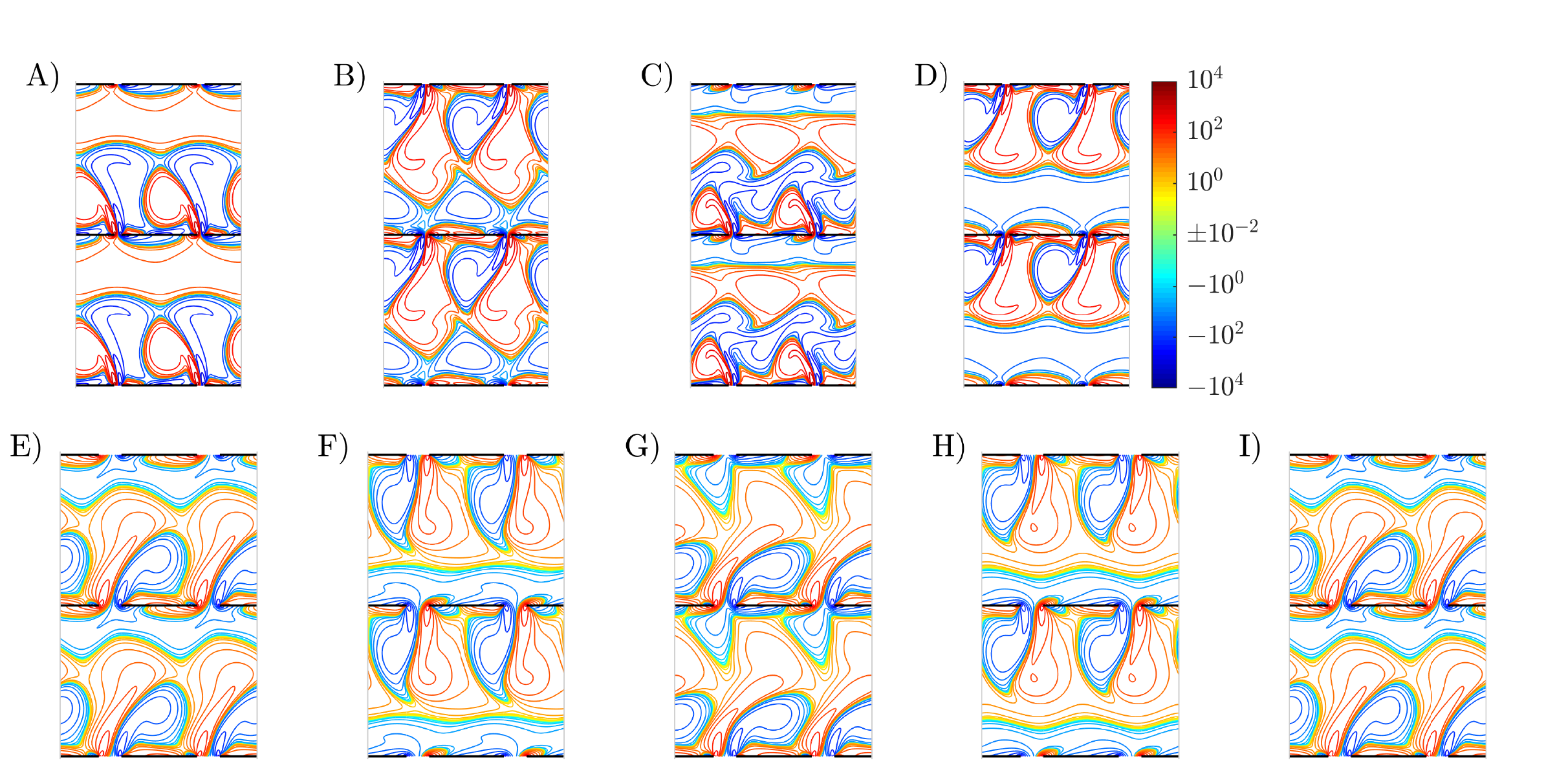}\\
           \vspace{-.25in}
           \end{tabular}
          \caption{\footnotesize Flows exhibiting different periodicities. Panels A-D: Snapshots spaced by 0.5 corresponding to $F_x(t)$ in figure \ref{fig:PeriodicFxExamplesRe20Fig}D. Panel D is essentially a mirror image of panel A. Panels E-I: Snapshots spaced by 0.5 for a 2-periodic flow (Re = 20, $A/L = 0.5$, $l_x = 1.3$, $l_y = 2$, and $U/fA = 3$). 
 \label{fig:PeriodicityFigFlows400}}
           \end{center}
         \vspace{-.10in}
        \end{figure}

Figure \ref{fig:PeriodicityFigFlows400} shows examples of flows for which $F_x(t)$ has a period larger than unity. Panels A-D show snapshots, a 1/2-period apart, that correspond to figure \ref{fig:PeriodicFxExamplesRe20Fig}D. Panel A shows an upward dipole, followed in panel B by a downward dipole. Panel C shows a smaller upward dipole, and then panel D is a mirror image of panel A. The next two snapshots (not shown) would be mirror images of panels B and C, followed by a return to panel A. Thus $F_x(t)$ has period 1.5 and the flow has period 3. Panels E-I show a flow with
period 2, and with $F_x(t)$ of period 2. The upward, rightward curving dipole in panel A is followed by a straighter downward dipole in panel F. The dipole in panel G is more curved than panel E, while that in panel H is similar to that in panel F. With panel I, the flow returns to panel E. This flow
is a more slightly perturbed version of a 1-periodic flow than the flow in panels A-D, as are many of the $n$-periodic flows we have observed.

\begin{figure} [h]
           \begin{center}
           \begin{tabular}{c}
               \includegraphics[width=6.8in]{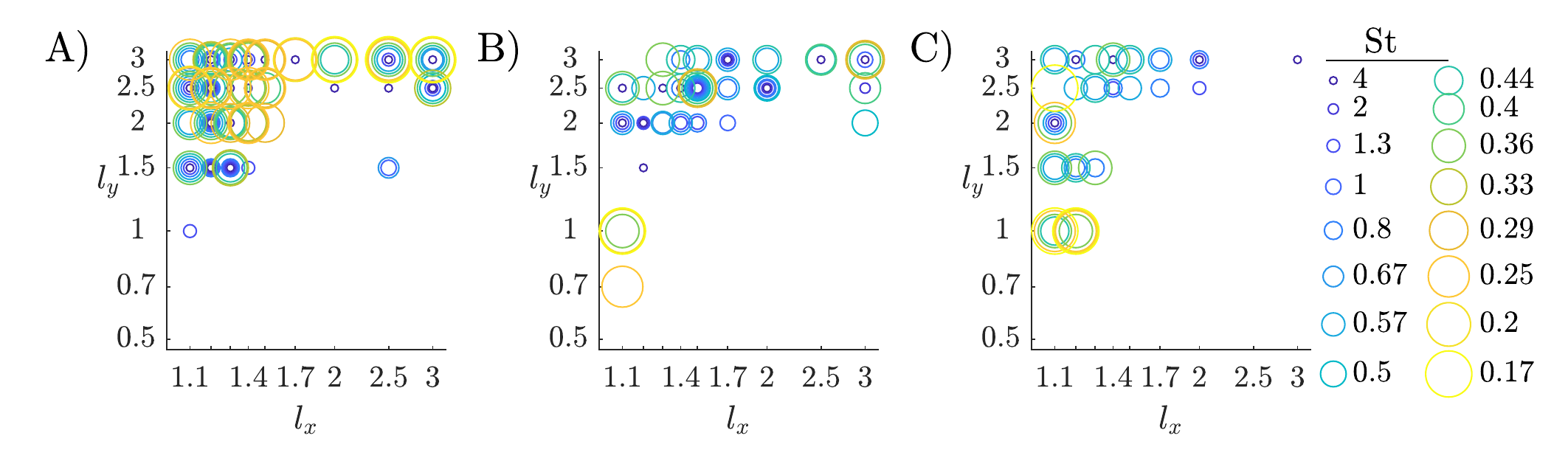}\\
           \vspace{-.25in}
           \end{tabular}
          \caption{\footnotesize Circles show parameter values where $F_x(t)$ falls below a threshold 0.01 (described in text)
for having period 1, for a rectangular lattice of plates with Re = 20. Values of $A/L$ are 0.2 (A), 0.5 (B), and 0.8 (C). Values of St (note $U/fA$ = 2/St) are labeled by circle size and color (key listed at right). Circles are centered at the corresponding values of $l_x$ and $l_y$. 
 \label{fig:DegreeOfPeriodicityRe20Fig}}
           \end{center}
         \vspace{-.10in}
        \end{figure}

We have studied examples of $F_x(t)$ and flows with different periods, mainly for $l_x$ close to 1. More broadly, there is a gradual trend towards nonperiodicity at certain parameter values. 
For Re = 20, and three different $A/L$ (0.2 (A), 0.5 (B), and 0.8 (C)), we plot circles in figure \ref{fig:DegreeOfPeriodicityRe20Fig} where $F_x(t)$
deviates from 1-periodicity by the following measure. We compute the averages of $F(t)$ over the
last eight half-periods of $V(t)$, during $t = 11$ to 15. We split the eight values into two sets of four,
one for the first half-period of $V(t)$ and the other for the second half-period. We sum the standard
deviations over the two sets and normalize by $\frac{1}{4}\int_{11}^{15} |F_x(t)| dt$ (an average magnitude of $F_x$). Where the resulting value is greater than 0.01, we plot circles in figure \ref{fig:DegreeOfPeriodicityRe20Fig}. The 0.01 threshold is somewhat arbitrary, but is chosen with certain considerations in mind. The flapping motion imposes a strong 1-periodic component in all flows, so a threshold of 0.1, say, would classify some nonperiodic $F_x(t)$ as periodic. A
threshold much smaller than 0.01 would miss some $F_x(t)$ that have almost but not completely converged to periodic near $t = 15$. The circles occur predominantly at large $l_y$, and more often at small $l_x$, though they are also found at large $l_x$. The reason is that the close spacing of plates at small $l_y$ tends to suppress complex vortical structures, and leads to more laminar, periodic flows. As we have seen, small $l_x$ leads to formation of thin dipole jets with sharp concentration of vorticity, and complicated nonperiodic dynamics can result. Many of these cases occur at small oncoming velocity (large St),
reflected in the larger number of small blue circles across the panels. As we have seen for the isolated body, above a certain flow speed, a reverse von K\'{a}rm\'{a}n street tends to form, in many cases due to merging or other regular interactions between the leading and trailing edge vortices. The larger number of yellow circles at $A/L$ = 0.2 (panel A) is perhaps because at a given $U/fA$, $U/fL$ is smaller in panel A, so in a given flapping period, vortices do not move as far downstream relative to the plate length in this case, leading to more complicated dynamics. Also, Re is constant (20) in all three panels,
so Re$_f$ decreases from left to right. To the extent that viscous regularizing effects are more controlled by Re$_f$, they are increased moving from left to right.
The corresponding data for the rhombic lattice are shown in figure \ref{fig:DegreeOfPeriodicityDiamondRe20Fig} in
the appendix, and shows similar trends but with additional nonperiodic states at smaller $l_y$.

\begin{figure} [h]
           \begin{center}
           \begin{tabular}{c}
               \includegraphics[width=6.8in]{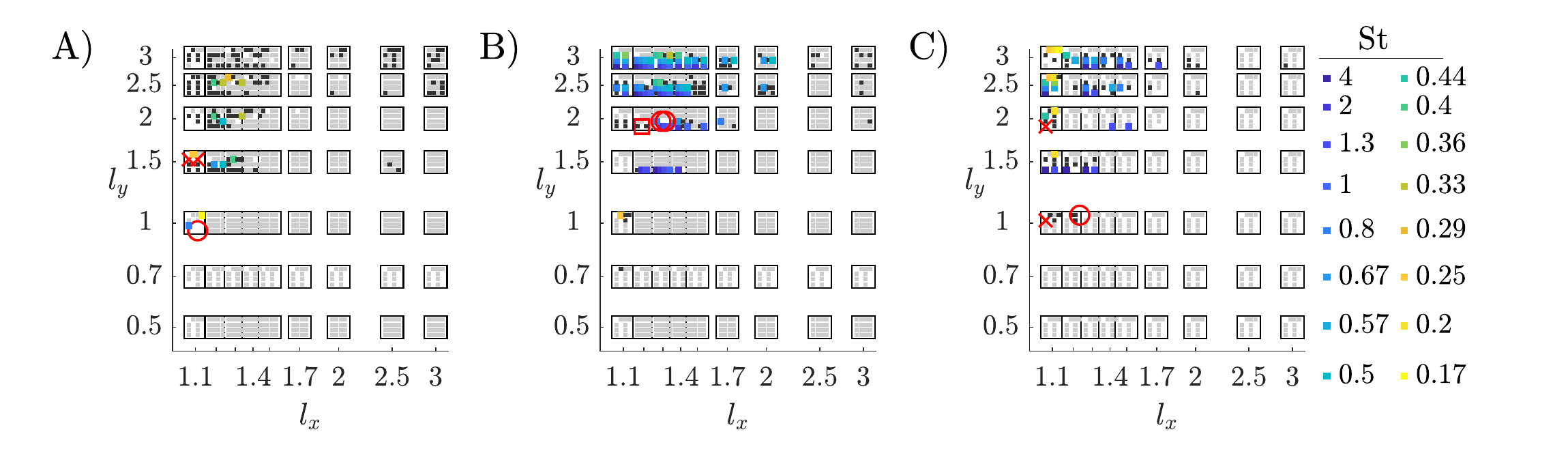}\\
           \vspace{-.25in}
           \end{tabular}
          \caption{\footnotesize Flows at Re = 20 for a rectangular lattice, classified by type of periodicity, with $A/L$ = 0.2 (A), 0.5 (B), and 0.8 (C). Each box shows data for a certain ($l_x$, $l_y$) pair that is located at the box center, and contains a set of smaller squares, each for a different value of St ranging from 0.17 to 4 (listed at right, note $U/fA$ = 2/St). The gray boxes correspond to $F_x(t)$ with period 0.5. The colored boxes correspond to period 1 but not period 0.5, i.e.~flows that are not up-down symmetric. For these flows only, 
we use the colors to label the St value. The black boxes denote nonperiodic $F_x(t)$.  
 \label{fig:PeriodicStatesRe20Fig}}
           \end{center}
         \vspace{-.10in}
        \end{figure}

Figure \ref{fig:PeriodicStatesRe20Fig} shows information about the types of periodic states
that occur in parameter space, some corresponding to the $F_x(t)$ and flows shown in
figures \ref{fig:StateDiagramFig}-\ref{fig:PeriodicityFigFlows400}. The panels again show states at
$A/L = 0.2$ (A), 0.5 (B), and 0.8 (C). Within each panel are a set of boxes (black outlines), each centered at the corresponding $(l_x, l_y)$ pair. Each box contains a set of smaller squares, each for a different value of St ranging from 0.17 to 4 (listed at right, note $U/fA$ = 2/St). 
The gray squares correspond to $F_x(t)$ with period 0.5. The colored squares correspond to period 1 but not period 0.5, i.e.~flows that are not up-down symmetric. For these flows only, 
we use the colors to label the St value. The black squares are for nonperiodic $F_x(t)$, the cases shown by circles in figure \ref{fig:DegreeOfPeriodicityRe20Fig}. White spaces lie within some of the boxes because not all parameter combinations were computed, but the overall pattern is not altered by these omissions. At Re = 20, most squares are gray, so most flows are up-down symmetric. The colored squares (1-periodic states) occur mainly at smaller $l_x$ and larger $l_y$. They are most prevalent at the intermediate $A/L$ (panel B), and there they occur at large St, i.e.~smaller $U/fA$. We have also noted a few cases of $F_x(t)$ with longer periods: 1.5 (red crosses), 2 (red circles), and 4 (red square), all of which occur at small $l_x$. In general, the period-1 and longer-period states occur near the nonperiodic states (black squares), so the former may be intermediaries in the transition from 1/2-periodic to nonperiodic states as the parameters are varied to allow more disordered flows.

\begin{figure} [h]
           \begin{center}
           \begin{tabular}{c}
               \includegraphics[width=6.8in]{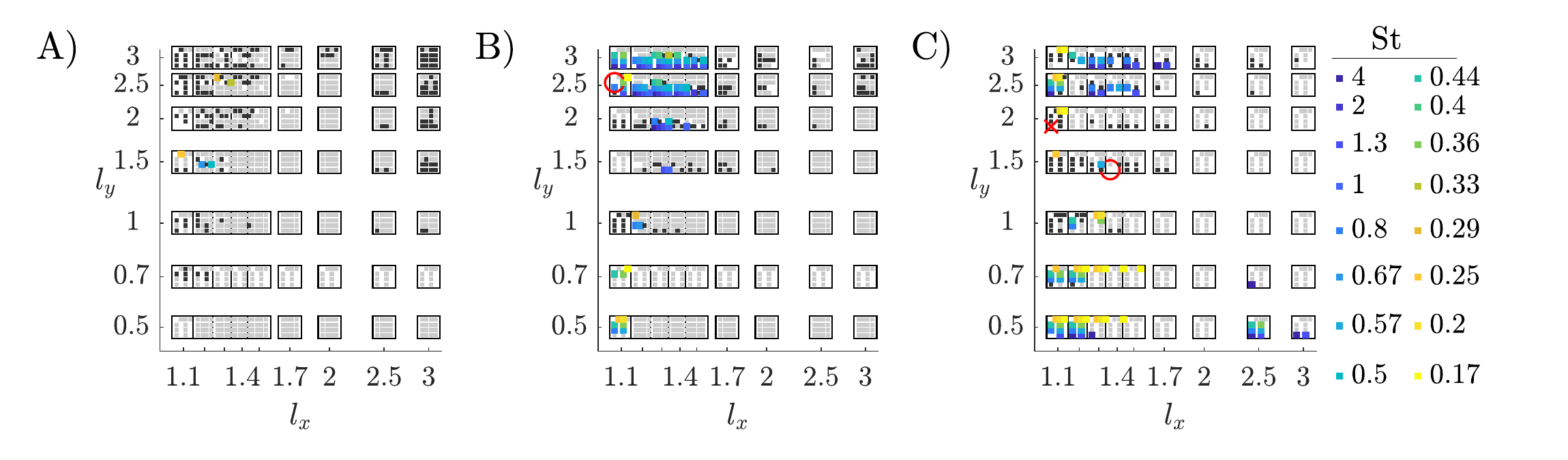}\\
           \vspace{-.25in}
           \end{tabular}
          \caption{\footnotesize Flows at Re = 20 for a rhombic lattice, classified by type of periodicity, with $A/L$ = 0.2 (A), 0.5 (B), and 0.8 (C). Each box shows data for a certain ($l_x$, $l_y$) pair (at the box center), and contains a set of smaller squares, each for a different value of St ranging from 0.17 to 4 (listed at right, note $U/fA$ = 2/St). The gray boxes correspond to $F_x(t)$ with period 0.5. The colored boxes correspond to periodicity unity but not periodicity 0.5, i.e.~flows that are not up-down symmetric. For these flows only, 
we use the colors to label the St value. The black boxes denote nonperiodic $F_x(t)$.
 \label{fig:PeriodicStatesDiamondRe20Fig}}
           \end{center}
         \vspace{-.10in}
        \end{figure}

Figure \ref{fig:PeriodicStatesDiamondRe20Fig} shows the same quantities when the
lattice is changed from rectangular to rhombic. The general trend of $F_x(t)$ with
increasing period or non-periodic at smaller $l_x$ and larger $l_y$ is basically preserved. Compared to
the rectangular lattice data, there are more 1-periodic and nonperiodic $F_x(t)$ at small $l_y$. This is perhaps because there is more $y$-distance between adjacent bodies for the rhombic lattice than for a rectangular lattice at the same $l_y$, allowing for more complex flows.

\begin{figure} [h]
           \begin{center}
           \begin{tabular}{c}
               \includegraphics[width=6in]{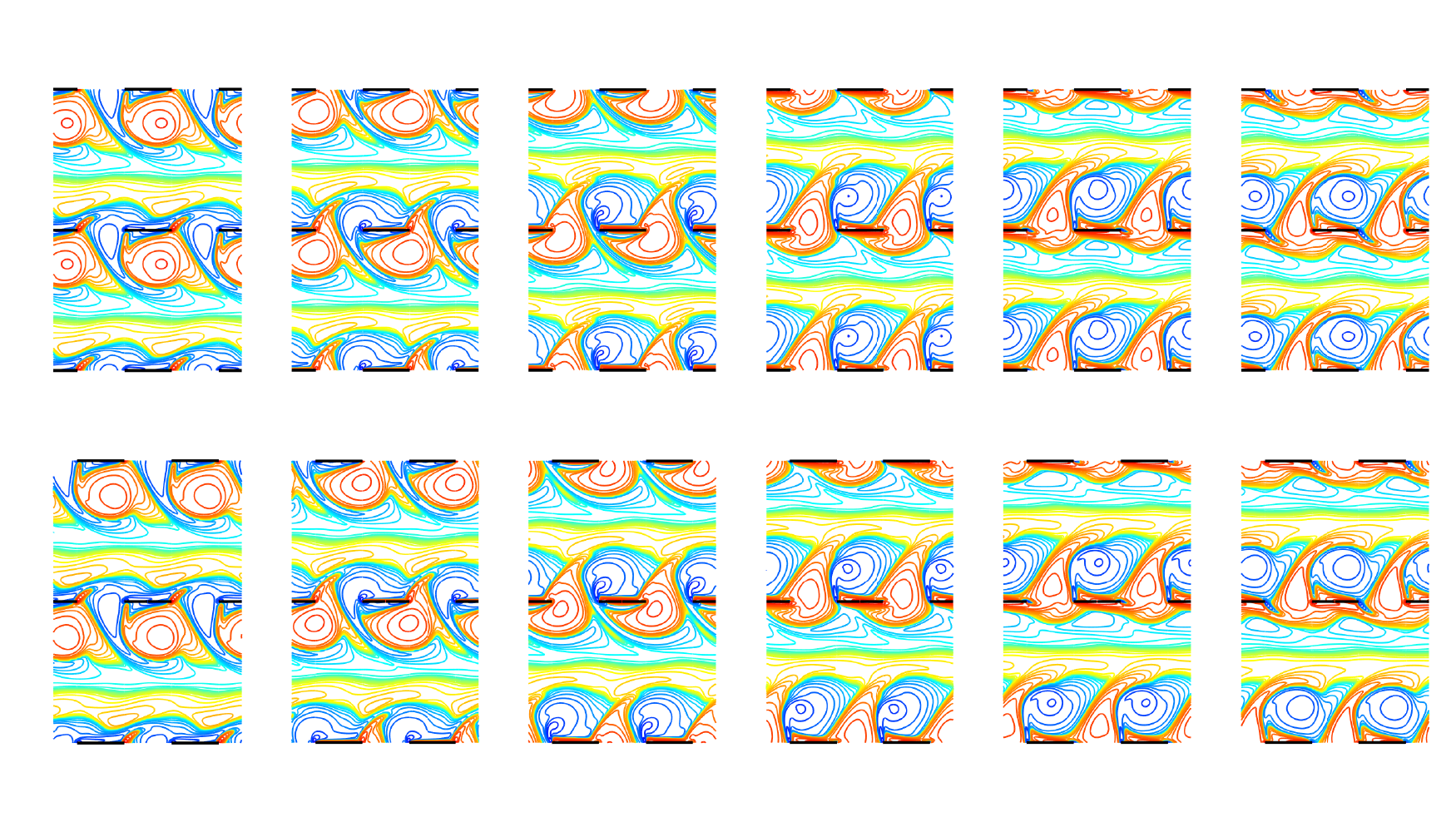}\\
           \vspace{-.25in}
           \end{tabular}
          \caption{\footnotesize Six vorticity snapshots spaced by 0.1 in $t$ during the half-period of upward flow, for rectangular (top row) and rhombic (bottom) lattices. The parameters are $A/L = 0.5$, $l_x = 2$, $l_y = 3$, Re = 70, and $U/fA = 7$.
 \label{fig:SameFlowsLargeLy400}}
           \end{center}
         \vspace{-.10in}
        \end{figure}

One important limiting case is the large-$l_y$ limit. Here the configuration consists of well-separated 1D arrays of flapping plates. The flows around the 1D arrays are essentially the same for the rectangular and rhombic lattices. Figure \ref{fig:SameFlowsLargeLy400} presents an example of such a flow, at
$l_x$ = 2 and $l_y$ = 3. The top row is a sequence of vorticity snapshots during the half-period of upward flow relative to the plates, for a rectangular lattice. The six snapshots are spaced
apart by 0.1 in $t$.  The positive (orange) vortices below the plates in the first panel, shed from the leading edge on the previous half-period, now merge with positive vorticity shed from the trailing edge during this half period. This is a case of relatively high Froude efficiency. The main difference from an isolated plate is the close interaction between the vortices shed at the leading and trailing edges of adjacent bodies, and the blue vortices are above the orange vortices, the opposite of the reverse von K\'{a}rm\'{a}n streets of figures \ref{fig:OneBodyVorticity2BMPFig} and \ref{fig:OneBodyVorticity4BMPFig}, and more similar to a regular von K\'{a}rm\'{a}n street (albeit with plates among the vortices). The lower row shows the same flow for a rhombic lattice of plates.
The flow is almost the same, because there is little interaction between vertically adjacent rows. A double layer of weak vorticity (light blue, green, and yellow) separates the vortex arrays of each row. In this region, the flow velocity is approximately uniform, with vertical flow speed equal to the average vertical flow speed, and horizontal flow speed about twice the average horizontal flow speed. The horizontal flow speed in the plate/vortex array is both positive and negative, and much smaller in magnitude.

\begin{figure} [h]
           \begin{center}
           \begin{tabular}{c}
               \includegraphics[width=6in]{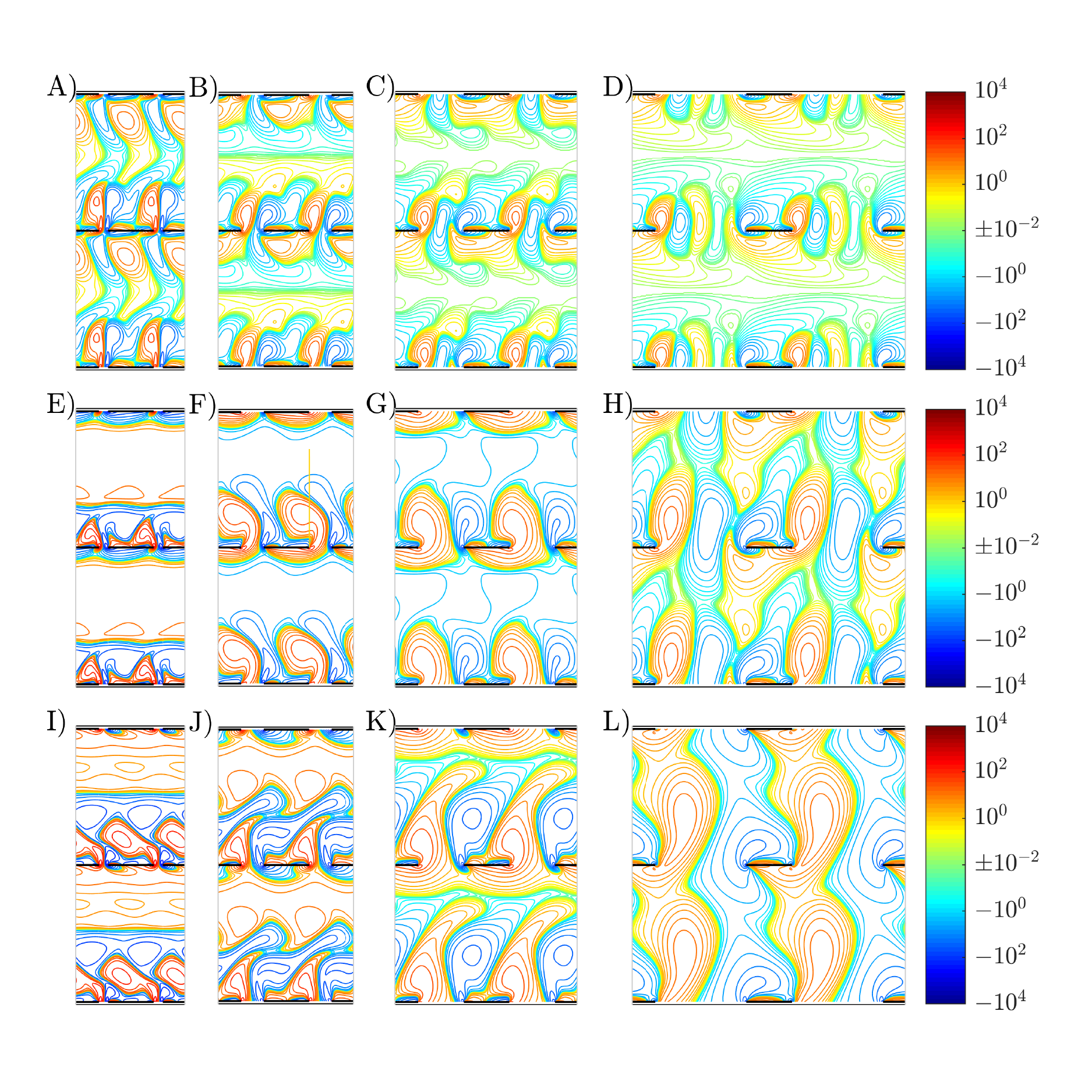}\\
           \vspace{-.25in}
           \end{tabular}
          \caption{\footnotesize Flow snapshots with $l_x$ = 1.2, 1.5, 2, and 3 (left to right columns) and $A/L$ = 0.2, 0.5, and 0.8 (top to bottom rows). The other parameters are: Re = 20, $U/fA = 3$, $l_y = 3$.
 \label{fig:LargeLyFlows400}}
           \end{center}
         \vspace{-.10in}
        \end{figure}

We now show examples of how the large-$l_y$ flows change as $l_x$ and $A/L$ are varied, in figure \ref{fig:LargeLyFlows400}. We choose Re = 20 and $U/fA = 3$ (large enough that regular vortex arrays may be generated, and small enough that thrust may occur). In the top row, $A/L = 0.2$. Moving left to right, as the gap between adjacent plates increases, the flow changes from a dipole to a vortex street. At far right, the trailing edge vortices interact more with the previously shed trailing edge vortex than with that shed at the leading edge of the adjacent plate. There is mean thrust for the flows in the first two columns of the top row, approximately zero thrust for the third, and small net drag for the fourth. In the second row, $A/L = 0.5$, there is again a transition from dipole jets to a vortex street, with larger vortices now. Now there is also nontrivial coupling between adjacent rows of plates. Instead of uniform flow, in the first three panels there are bands of nearly constant vorticity above and below the vortex arrays. These are shear flows with $u$ nearly linear with respect to $y$ and $v$ nearly constant. They differ from those that would be seen with an isolated flapping body, and therefore, unlike the flows in the first row, they would be altered for larger $l_y$ values. In the second row, only the third column corresponds to a state of mean thrust. In the third row, $A/L = 0.8$, only the first column is a state of mean thrust. It is not obvious from the form of the vortex dipole why mean thrust occurs, but there are noticeable differences with the orientation of the dipole in the second panel. Again there are linear shear zones above and below the dipole jets. In the fourth column, vertical bands of same-signed vorticity form. Now the plates are coupled strongly to both vertical and horizontal neighbors through the flow.

\begin{figure} [h]
           \begin{center}
           \begin{tabular}{c}
               \includegraphics[width=6.5in]{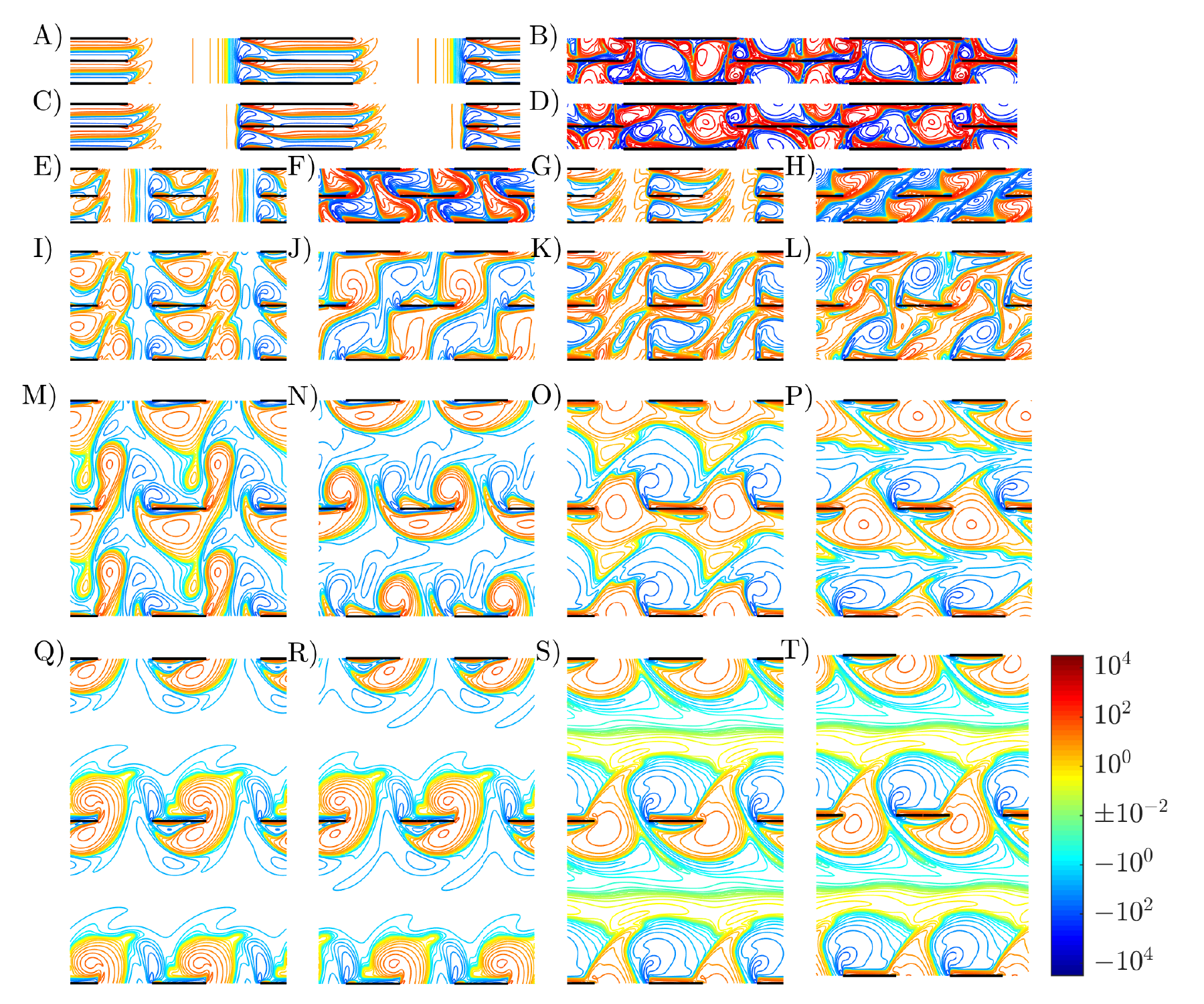}\\
           \vspace{-.25in}
           \end{tabular}
          \caption{\footnotesize Flow snapshots with $l_x$ = 2 and various $l_y$: 0.2 (A-D), 0.5 (E-H), 1 (I-L), 2 (M-P), and 3 (Q-T). Within each group of four consecutive panels, the first two are
for $U/fA$ = 3 (A-B, E-F, I-J, etc.) and the second are for $U/fA$ = 7. Within each of these two pairs of panels, the first is a rectangular lattice (A, C, E, G, etc.) and the second is a rhombic lattice. In all cases, Re = 70 and $A/L = 0.5$. The values of vorticity on the contours are labeled on the colorbar at lower right.
 \label{fig:CompareSmalltoLargeLyFlows310}}
           \end{center}
         \vspace{-.10in}
        \end{figure}

We have shown examples of flows with $l_y$ fixed and $l_x$ from small to large. We now reverse
the parameters, i.e.~hold $l_x$ fixed at 2 (an intermediate value), and vary $l_y$ from small
to large. Figure \ref{fig:CompareSmalltoLargeLyFlows310} shows examples of vorticity fields at instants of upward (and as usual, rightward) flow. 
Since $l_y$ may be small, there can be significant differences between rectangular and rhombic lattice flows and we show both. The differences are most pronounced at the smallest $l_y$. In panels
A and C ($l_y = 0.2$ and $U/fA$ = 3 and 7, respectively), the flow through the rectangular lattice is approximately horizontal Poiseuille flow between the vertical neighbors (away from the plate end regions), and unidirectional vertical flow in the space between horizontal neighbors. Most of the vertical flux occurs in these vertical channels, and the flow past the plate edges is relatively weak. These small-$l_y$ flows result in a net drag force, almost constant in time, that of the Poiseuille flow between the plates. The corresponding rhombic lattice flows (B and D) are much more complex. The plates now cover the full horizontal extent of the flow field, so the entire flow is forced through the small gaps between the interleaving plate edges. Consequently, much stronger vorticity is generated at the plate edges. This flow results in a net thrust force, is not temporally periodic but $F_x(t)$ has a strong 1/2-periodic component, and is also not spatially periodic on the scale of a single unit cell (i.e.~containing a single plate), but of course is periodic on the
the scale of a double unit cell by the definition of the periodic boundaries (the dashed box in figure \ref{fig:SchematicPlatesFig}B). This can be seen by examining the flows below the right edges of the plates. There is a small blue vortex below the top plates' right edges, but a larger (B) or smaller (D) blue region below the middle plates' right edges. Increasing $l_y$ to 0.5, the rectangular lattice flows (E and G) deviate more from Poiseuille flow, while the rhombic lattice flows (F and H) become
smoother. Panel F, at lower $U/fA$, is still a temporally nonperiodic flow, but closer to spatially periodic on the scale of the unit cell than panels B or D. The flow in panel H is both temporally periodic and spatially
periodic on the unit cell scale. Again, both rectangular lattice flows generate net drag, while the rhombic lattice flows generate net thrust. Increasing $l_y$ to 1, both rectangular flows (I and K) again generate drag, though I, at lower $U/fA$, is close to zero net drag. Of the rhombic lattice flows (J and L), only J generates thrust, but with relatively high Froude efficiency (0.04), much
higher than in panel F due both to increased thrust and decreased input power. Unlike the rectangular lattice, the rhombic lattice geometry allows an oblique vortex dipole to form at this $l_y$ (panel J), involving a positive (orange) vortex at the right edges of the middle plates and a negative vortex (blue) at the left edges of the bottom plates, which probably underlies efficient thrust generation. Increasing $l_y$ to 2, the rectangular lattice has the first occurrence of a state of net thrust in this figure, at low
$U/fA$ (M) but not high $U/fA$ (O). The rhombic lattice has a state of net thrust at both speeds (N and P). At the largest $l_y = 3$, the rectangular (Q, S) and rhombic (R, T) flows are similar as discussed below figure \ref{fig:SameFlowsLargeLy400}; both pairs generate net thrust with only small differences in their magnitudes and the corresponding Froude efficiencies.  To summarize, for the rectangular lattice, only net drag occurs below a moderate $l_y$. For the rhombic lattice, by contrast, thrust can occur at very small $l_y$, though the flows are nonperiodic and the Froude
efficiency is low. We can also see that the rhombic lattice flows converge to spatially periodic on the unit cell scale as $l_y$ becomes large, which correlates with the convergence to temporal periodicity. By computing the relative error in unit-cell-periodicity for a number of other
flows (76 in all) at various parameters, we find that unit-cell-periodicity correlates with temporal periodicity in general. Both are more prevalent when Re is small, and when $l_x - 1$ is not very small. However, there are examples of one without the other and vice versa. Therefore, in a large lattice of plates of plates at sufficiently large Re, the flow would be expected to deviate from the
periodic lattice model with unit cell periodicity. However, in most cases considered in this paper, the deviation is not very large. 

\begin{figure} [h]
           \begin{center}
           \begin{tabular}{c}
               \includegraphics[width=6.8in]{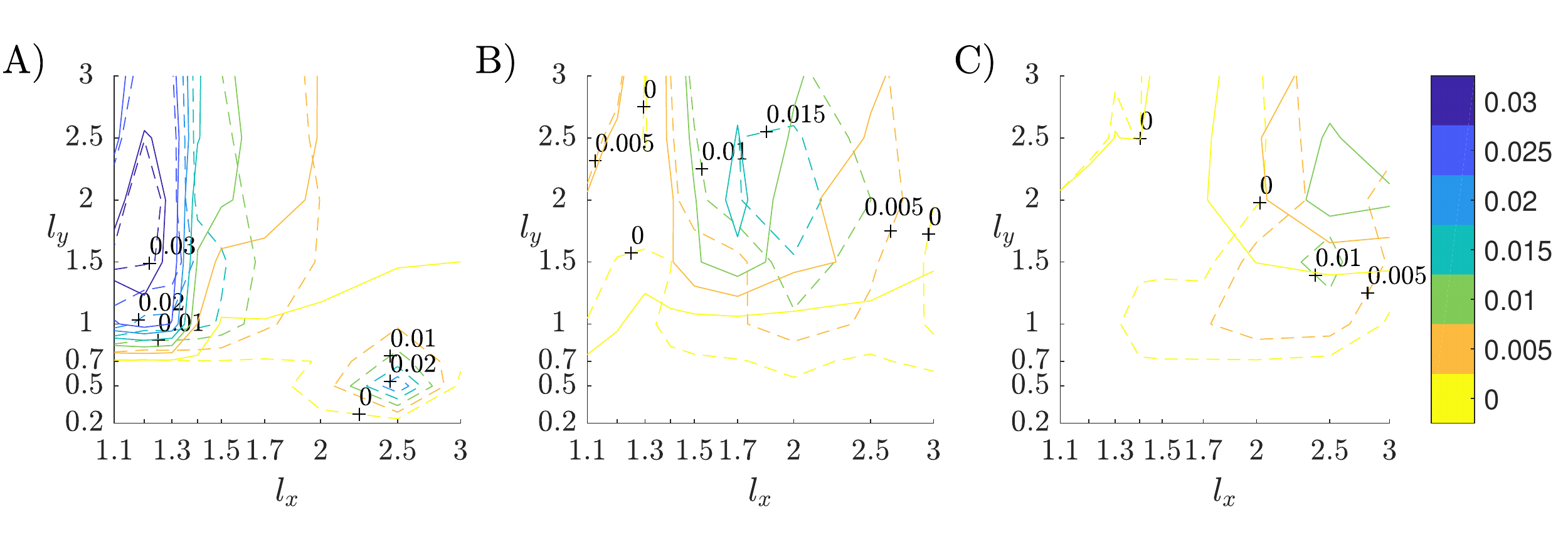}\\
           \vspace{-.25in}
           \end{tabular}
          \caption{\footnotesize Contours of Froude efficiency at $A/L$ = 0.2 (A), 0.5 (B), and 0.8 (C), for rectangular (solid lines) and rhombic (dashed lines).
 \label{fig:CompareRectDiamRe20FroudeFig}}
           \end{center}
         \vspace{-.10in}
        \end{figure}

We have discussed some examples of plate-plate interactions that lead to thrust generation in
certain one-dimensional slices through parameter space: at moderate and large $l_y$ for different $l_x$ (figures \ref{fig:StateDiagramPanelsBCVorticity400}, \ref{fig:LargeLyFlows400}, and  \ref{fig:StateDiagramPanelDVorticity400}), and
for intermediate $l_x$ at large $l_y$ (figure \ref{fig:CompareSmalltoLargeLyFlows310}).
In the $l_x$-$l_y$ plane, we present in figure \ref{fig:CompareRectDiamRe20FroudeFig} contour plots of Froude efficiency at smaller Re (= 20), where most flows are time-periodic. Contours are plotted for $A/L$ = 0.2 (A), 0.5 (B), and 0.8 (C), for both rectangular (solid lines)
and rhombic (dashed lines) lattices. In panel A, the dashed and solid lines are close for $l_y$ above 2, with a very slight advantage for the rectangular lattice. At this $A/L$ and large $l_y$, there is little interaction between different horizontal rows of plates, so the lattice type makes little difference, as in figure \ref{fig:SameFlowsLargeLy400}. Below $l_y$ = 2, the solid contour lines bend sharply leftward, and efficient thrust is only obtained for $l_x$ near 1, if at all. Here the rhombic lattice is more efficient, and yields net thrust down to $l_y = 0.2$ (yellow dashed line), as noted previously for Re = 70. Moving to panel B,
there is an overall decrease in peak Froude efficiency by about a factor of 2. The solid and dashed lines deviate more at larger $l_y$, as the now increased $A/L$ results in more interaction between different horizontal rows of plates. For both lattice types, the peak Froude efficiency moves to much larger $l_x$. For larger $A/L$, the vorticity has a larger vertical extent. For an isolated plate,
figure \ref{fig:StFrFig}B shows that $U/fA$ should be kept roughly constant as $A/L$ increases, for peak efficiency. This means that $U/fL$ increases. Thus the vortices are more spread out horizontally, and it is reasonable that the plates should be more spread out to interact with vortices efficiently. The trend continues in panel C: a further reduction in peak Froude efficiency, that
occurs at still larger $l_x$. In all cases, net thrust is obtained down to lower $l_y$ by the rhombic lattice.

\begin{figure} [h]
           \begin{center}
           \begin{tabular}{c}
               \includegraphics[width=6.8in]{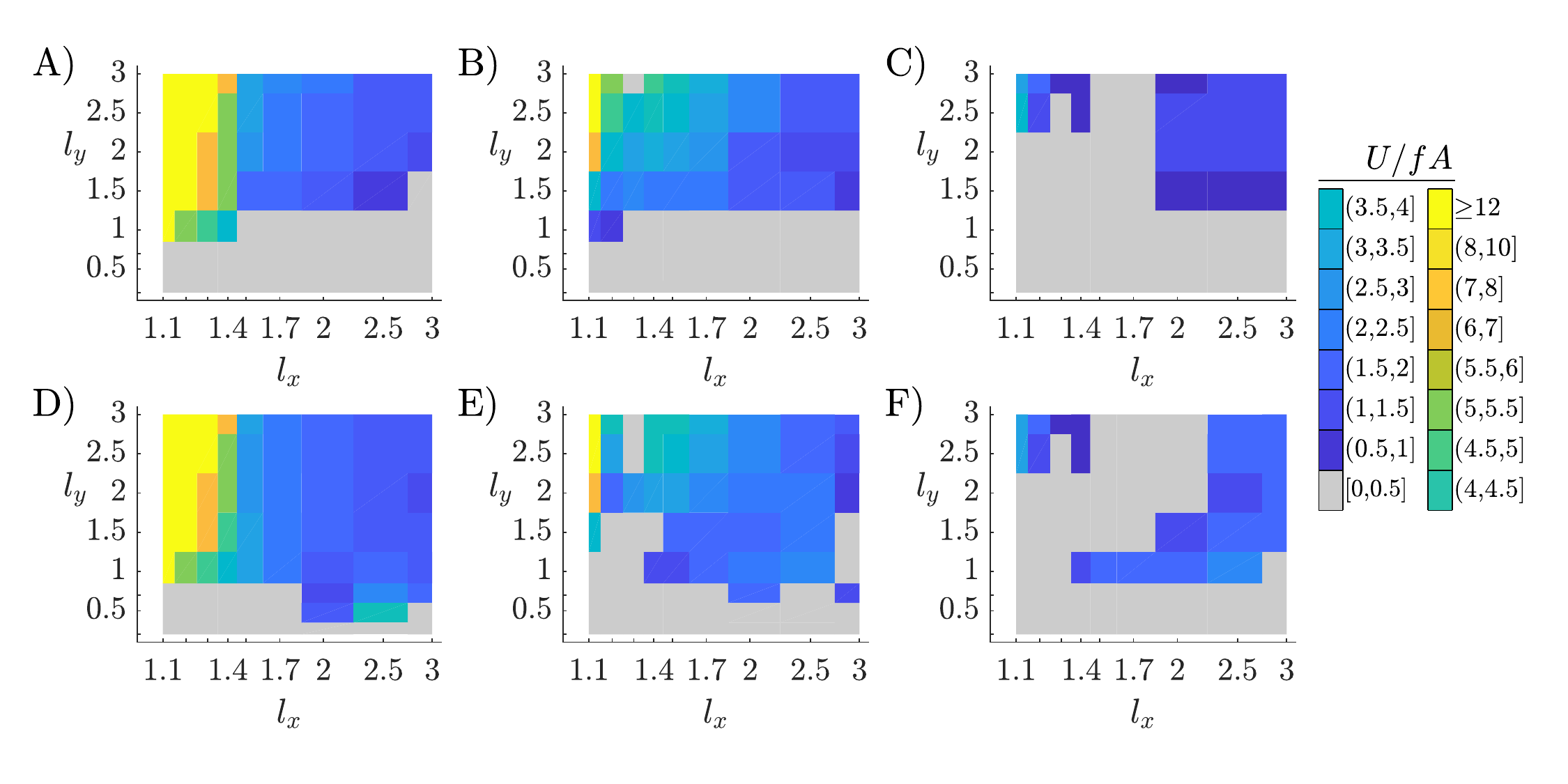}\\
           \vspace{-.25in}
           \end{tabular}
          \caption{\footnotesize Comparison of self-propelled speeds for rectangular (A-C) and rhombic (D-F) lattices at Re = 20 and $A/L$ = 0.2 (A, D), 0.5 (B, E), and 0.8 (C, F).
 \label{fig:CompareRectDiamRe20SpeedFig}}
           \end{center}
         \vspace{-.10in}
        \end{figure}

Another measure of performance is the maximum self-propelled speed $U_{SPS}/fA$ achieved by the lattice.
These data are presented in figure \ref{fig:CompareRectDiamRe20SpeedFig} for the rectangular (top row) and rhombic (bottom row) lattices, at $A/L$ = 0.2, 0.5, and 0.8 (left to right columns).  In all panels, the highest speeds are obtained at the smallest
$l_x$, where vortex dipole formation leads to strong forces on the plates. The much smaller speeds at larger $l_x$ may be partly due to the relatively small Re, which leads
to substantial diffusion of vortices as they move over a larger lattice length scale. As in figure \ref{fig:CompareRectDiamRe20FroudeFig}, the two lattices' speeds agree better at larger $l_y$; the rhombic lattice achieves propulsion at smaller $l_y$ than the rectangular lattice, with moderate to large $l_x$. An example is the local maximum in panel D at $l_x = 2.5$ and $l_y = 0.5$, close to the parameters for the flow in figure \ref{fig:CompareSmalltoLargeLyFlows310}F but at lower Re. In panels B and E, there is a small band where $l_x$ = 1.3 and $l_y = 2.5$ and 3 where the speed is greatly reduced or zero. The reason is not obvious, but may reflect a particular aspect of dipole formation at this value of Re. 

\begin{figure} [h]
           \begin{center}
           \begin{tabular}{c}
               \includegraphics[width=6.8in]{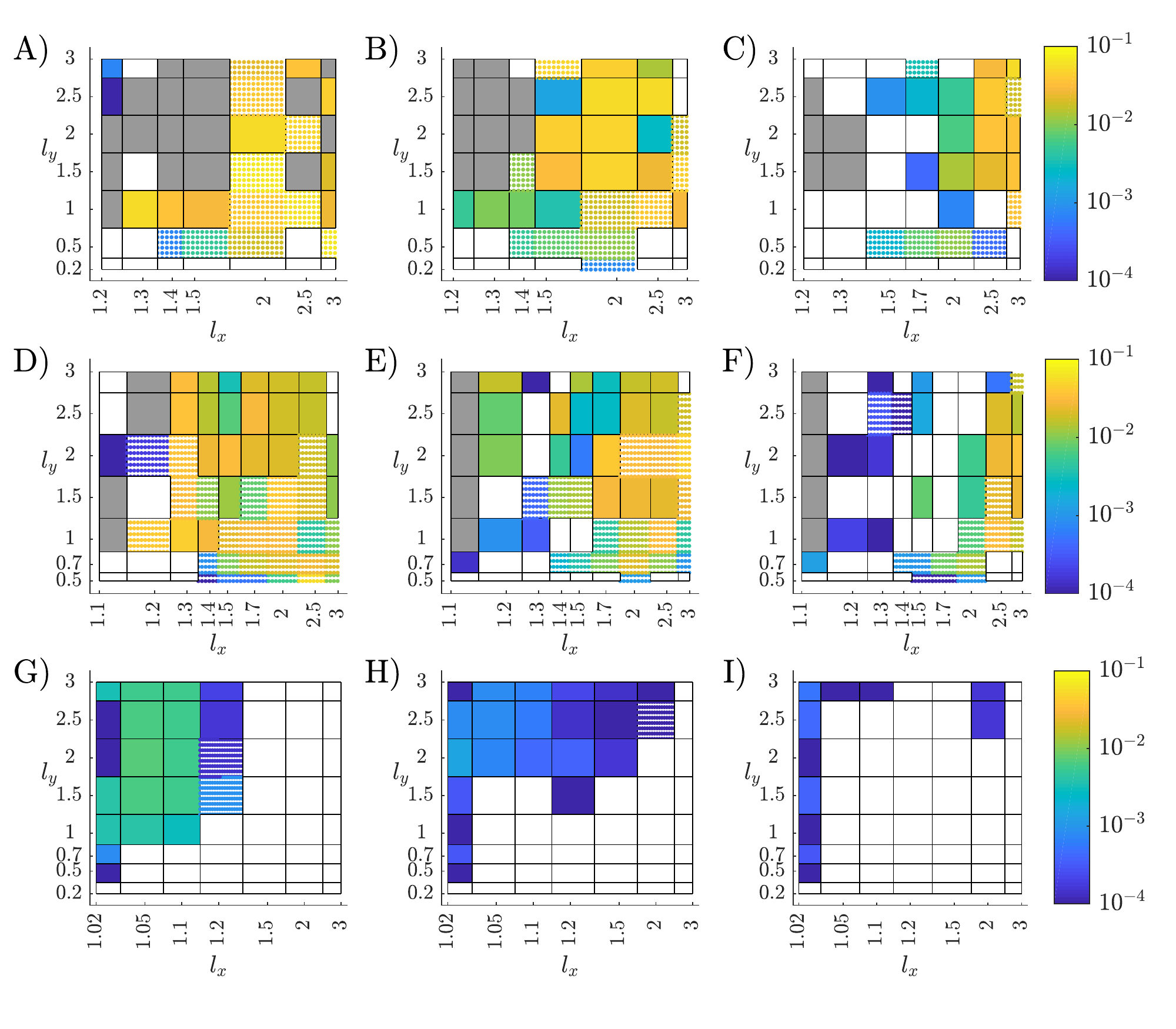}\\
           \vspace{-.25in}
           \end{tabular}
          \caption{\footnotesize Peak Froude efficiencies at Re = 70 (A-C), 40 (D-F), and
10 (G-I), for $A/L$ = 0.2 (A,D,G), 0.5 (B,E,H), and 0.8 (C,F,I). White boxes denote only net drag is obtained. Gray boxes indicate that flows were nonperiodic across $U/fA$. White boxes indicate that flows were periodic at some $U/fA$, but these only produced net drag. Colored boxes indicate peak Froude efficiency magnitude (shown by color bars at right) where periodic flow(s) that produced net thrust occurred. The color field is solid or dotted if the peak was obtained with a rectangular or rhombic lattice, respectively. Each box is centered at the corresponding $(l_x, l_y)$ value, except the boundary boxes,
where the value is given at the corresponding boundary edge. 
 \label{fig:CompareRectDiam3ReFig}}
           \end{center}
         \vspace{-.10in}
        \end{figure}

When we increase Re much above 20, more flows are nonperiodic up to $t = 30$, and a contour plot of Froude efficiency like figure \ref{fig:CompareRectDiamRe20FroudeFig} would require much longer simulations to achieve reliable long-time averages. Therefore,
for higher Re, we present data only for cases that meet a threshold for periodicity. We use
the same criterion described in figure \ref{fig:DegreeOfPeriodicityRe20Fig}, but increase
the threshold from 0.01 to 0.08. Even at this larger threshold, $F_x(t)$ is close to periodic. The larger threshold allows us to present more values, and observe the general trends more easily. These trends are not very sensitive to the particular threshold chosen (0.08). Figure \ref{fig:CompareRectDiam3ReFig} expands upon figure
\ref{fig:CompareRectDiamRe20SpeedFig} by presenting peak Froude efficiency values
for three Re: 70 (top row), 40 (middle row) and 10 (bottom row), with the same $A/L$, 0.2, 0.5, and 0.8 (left to right columns). We do not use contours now because in many cases, $F_x(t)$ is nonperiodic for all $U/fA$, and these are
shown by gray boxes. If at a given choice of parameters there are $U/fA$ such that $F_x(t)$ meets the criterion for periodicity, the color of the corresponding
box denotes the peak Froude efficiency among the periodic cases, and the values
are shown by the color bars at right. The box coloring is solid or dotted if the peak value is obtained with a rectangular or rhombic lattice, respectively. The white boxes are cases
where there are periodic flows, but all such flows yield net drag.  One basic trend is the increasing nonperiodicity of flows with Re: there are no gray boxes at Re = 10, some at
Re = 40, and more at Re = 70. Nonperiodicity is most common at small $l_x$, where the most intense vortex dipoles are created. At Re = 70, nonperiodicity is more common at smaller $A/L$, where Re$_f$ is larger. At Re = 10  (bottom row), the highest efficiency states are found
at the smallest $l_x$, which are taken as small as 1.02; periodic flows are obtained nonetheless due to the small Re.  Consistent with figure \ref{fig:CompareRectDiamRe20FroudeFig},
efficiency drops dramatically as $A/L$ increases (from left to right). At Re = 40 (middle row), the most efficient periodic flows are obtained at much larger $l_x$. This is partly because many flows at smaller $l_x$ are nonperiodic, however. At Re = 40 and 70 (top row), the basic trends of figure \ref{fig:CompareRectDiamRe20FroudeFig} are mostly preserved: rhombic lattice flows are most efficient at small $l_y$, rectangular lattice flows are usually more efficient at large $l_y$, and peak efficiencies move to larger $l_x$ as $A/L$ becomes larger (moving rightward).
In general, as Re increases, so does peak Froude efficiency, from about 0.007 at Re = 10 to 0.033 at Re = 20 (from data in figure \ref{fig:CompareRectDiamRe20FroudeFig}) to
0.055 at Re = 40, to 0.07 at Re = 70. As Re increases above 10, fewer and fewer states are periodic, so these should be regarded as conservative lower bounds for the true peak Froude efficiencies, ones that include long-time averages of nonperiodic flows.

\begin{figure} [h]
           \begin{center}
           \begin{tabular}{c}
               \includegraphics[width=6.5in]{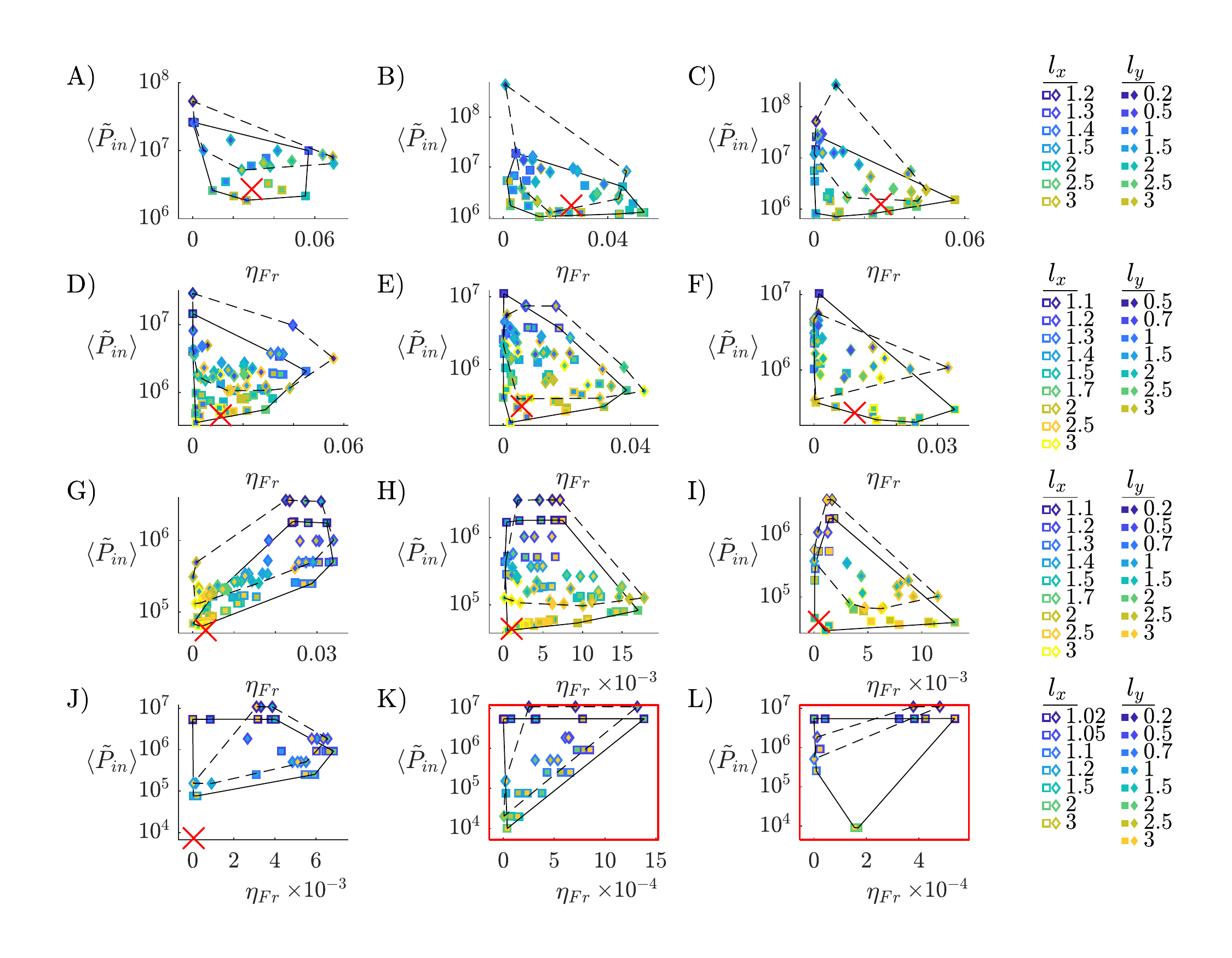}\\
           \vspace{-.25in}
           \end{tabular}
          \caption{\footnotesize  Peak values of $\eta_{Fr}$ versus $\langle {\tilde P}_{in} \rangle$ for rectangular (colored squares) and rhombic lattices (colored diamonds), at various $l_x$ and $l_y$ (colors at right). From top
to bottom row, Re = 70, 40, 20, and 10. From left to right column, $A/L$ = 0.2, 0.5, and
0.8.  
 \label{fig:ParetoFrFig}}
           \end{center}
         \vspace{-.10in}
        \end{figure}

We have shown in figure \ref{fig:CompareRectDiam3ReFig} how the maximum of the Froude efficiency over $U/fA$ varies
with respect to Re, $A/L$, $l_x$, and $l_y$ for flows that are periodic or almost periodic.
In figure \ref{fig:ParetoFrFig} we replot the same data in figure \ref{fig:CompareRectDiam3ReFig}, adding
the Re = 20 case, and showing both $\langle {\tilde P}_{in} \rangle$ and $\eta_{Fr}$ simultaneously. Thus, the four rows from top to bottom correspond to Re = 70, 40, 20, and 10, while the three columns from left to right have $A/L$ = 0.2, 0.5, and 0.8. This figure allows us to identify which configurations are effective at different
input power budgets, and describe the Pareto frontier for this set of data. In each panel we plot the data for rectangular lattices with small squares, with convex hull shown as a solid black line. The data for rhombic lattices are small diamonds with convex hull shown as a dashed black line. For each data point, the outline color gives the value of $l_x$ and the interior color gives the value of $l_y$ (listed at right). In each panel we also plot, at the panel's values of Re and $A/L$, $\langle {\tilde P}_{in} \rangle$ and $\eta_{Fr}$ for an isolated body. This is shown with a red cross, or if there is no thrust for the isolated body
at any of the $U/fA$ tested, the panel has a red outline. At the lowest two Re (bottom two rows), $\eta_{Fr}$ for the isolated body is zero or much smaller than for the lattices. $\langle {\tilde P}_{in} \rangle$ for the isolated body is at the lower end of the range of lattice values, near the values for the largest $l_x$ and $l_y$. Panel J is a somewhat special case,
as the large-$l_x$ lattice values did not yield thrust, but the isolated body did yield a very small net thrust---with $\eta_{Fr}$ and $\langle {\tilde P}_{in} \rangle$ both much smaller than for the lattice values shown. For Re increased to 40 (second row from the top), there are lattice values with $\langle {\tilde P}_{in} \rangle$ lower than that of the isolated body, and some of these---rectangular lattices with large $l_x$ and $l_y$---yield much higher efficiency. Moving to the top row (Re = 70), the isolated body's performance is relatively improved, lying near the middle of the range of lattice values. That is, the best periodic lattice flows have efficiencies about twice that of the isolated body. However, this should be regarded as only a lower bound on the efficiency advantage of the lattice flows. Many lattice flows were not counted due to nonperiodicity. These plots also show that input power for the rhombic lattices is generally larger than for the rectangular lattices, because the rhombic lattices cover more of the flow domain horizontally, forcing the fluid through smaller constrictions between the plates. Nonetheless, the largest $\eta_{Fr}$ values here, in panel A, are for rhombic lattices. We also see that $\langle {\tilde P}_{in} \rangle$ is generally largest for smaller $l_x$ and $l_y$ (blue symbols) due to increased flow constriction, except that many of these cases are omitted because they have nonperiodic flows (especially in the top row).

\begin{figure} [h]
           \begin{center}
           \begin{tabular}{c}
               \includegraphics[width=6.5in]{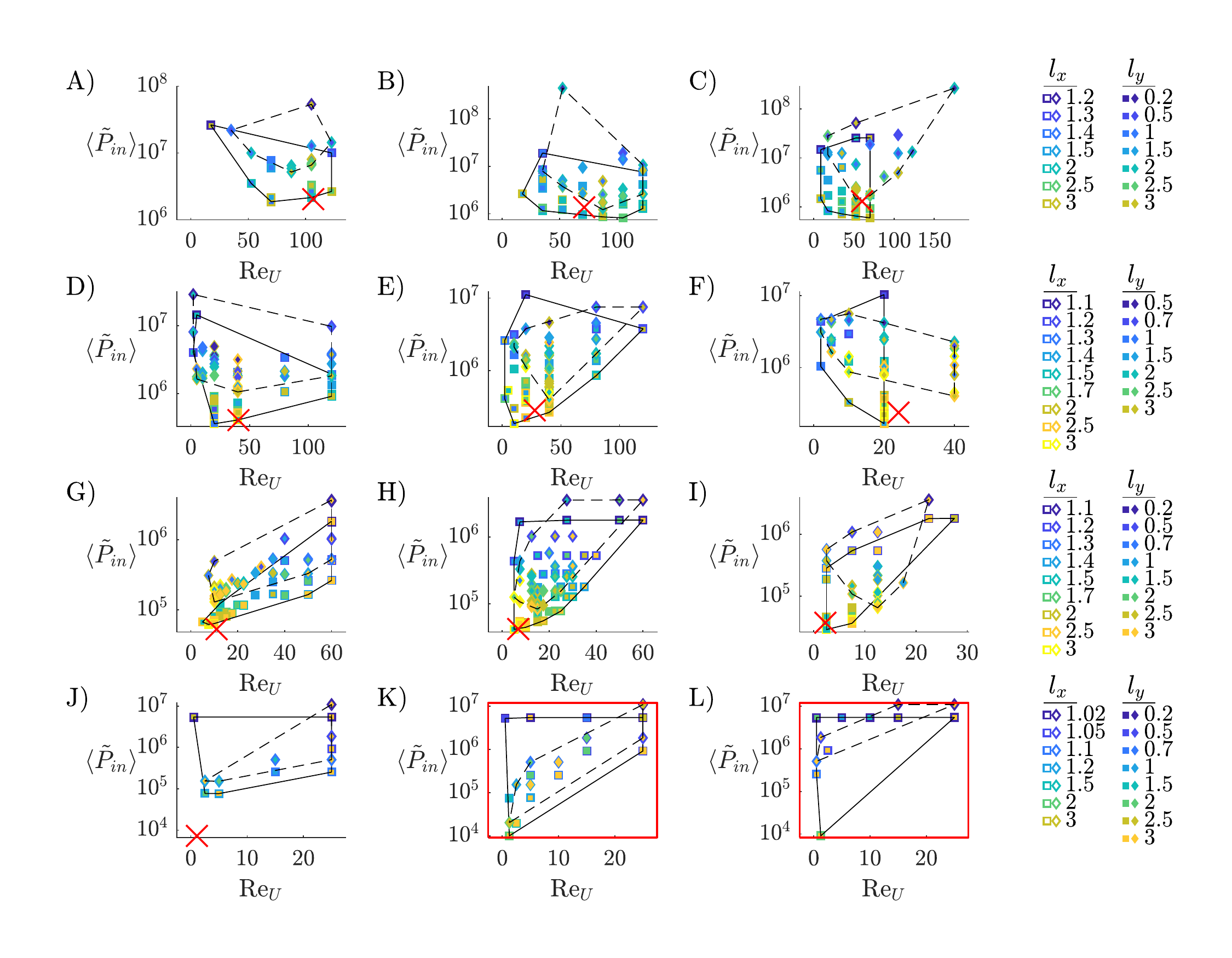}\\
           \vspace{-.25in}
           \end{tabular}
          \caption{\footnotesize Values of Re$_{U, SPS}$ versus $\langle {\tilde P}_{in} \rangle$ for rectangular (colored squares) and rhombic lattices (colored diamonds), at various $l_x$ and $l_y$ (colors at right). From top
to bottom row, Re = 70, 40, 20, and 10. From left to right column, $A/L$ = 0.2, 0.5, and
0.8.
 \label{fig:ParetoSpeedFig}}
           \end{center}
         \vspace{-.10in}
        \end{figure}

Figure \ref{fig:ParetoSpeedFig} shows the analogous plots when self-propelled speed (the maximum if there are multiple values) is substituted for Froude efficiency as the measure of performance on the horizontal axes. The self-propelled speed of the isolated body is less, but sometimes not much less, than the peak speeds of the lattices in the same panel. In some cases (panels A, F, G, and J), there is not a lattice flow that has larger Re$_{U,SPS}$ and smaller $\langle {\tilde P}_{in} \rangle$ simultaneously.  However, this could easily change if temporally nonperiodic lattice flows were included.


\section{Mean input power \label{sec:InputPower}}

The Froude efficiency and self-propelled speed depend on how the mean horizontal force
varies with oncoming flow speed. For a single flapping foil, this behavior depends on the physics of vortex creation and shedding due to large amplitude flapping at a given Reynolds number. Optimal vortex creation for thrust occurs when the foil moves a certain
angle of attack in the flow; this motion can be computed but is difficult to describe with
a simple analytical formula \cite{wang2000vortex}. The same phenomenon underlies the prevalence and optimality of St = 0.2--0.4 for flapping locomotion at high Re \cite{TNR_Nature_2003,eloy2012optimal}. For a lattice of flapping bodies, the process is further complicated by the additional length scales of separation between bodies, and the effects of vortices colliding with downstream bodies.

\begin{figure} [h]
           \begin{center}
           \begin{tabular}{c}
               \includegraphics[width=6.8in]{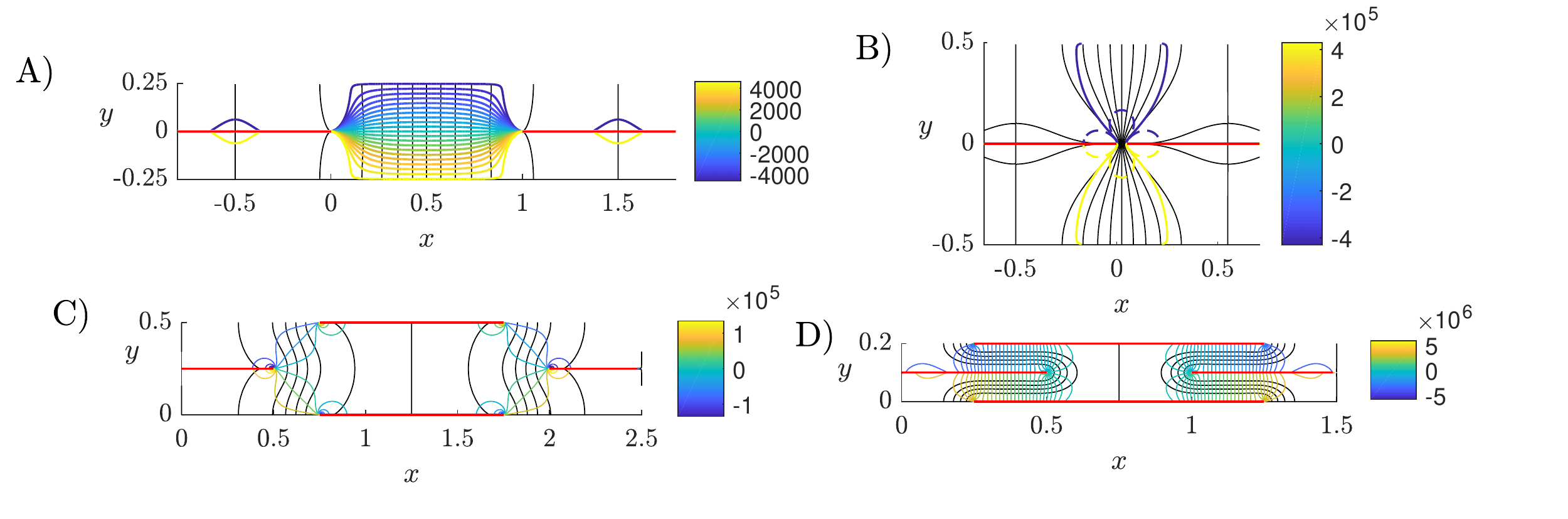}\\
           \vspace{-.25in}
           \end{tabular}
          \caption{\footnotesize Steady flows through plate lattices at Re$_V$ = 0.001. Plates are red, streamlines are black, and isopressure lines are blue, green, and yellow (values in colorbars at right). A) Flow through rectangular lattice with $l_x = 2$ and $l_y = 0.5$. B) Flow through rectangular lattice with $l_x = 1.05$ and $l_y = 1$. C) Flow through rhombic lattice with $l_x = 2.5$ and $l_y = 0.25$. D) Flow through rhombic lattice with $l_x = 1.5$ and $l_y = 0.1$.
 \label{fig:SteadyFlowsFig}}
           \end{center}
         \vspace{-.10in}
        \end{figure}

One ingredient of Froude efficiency that is easier to address analytically is the time-averaged input power. We have seen in figure \ref{fig:PInScalingFig} for an isolated flapping plate that the mean input power scales as flapping amplitude and frequency cubed, and has a weaker dependence on the oncoming flow speed. This indicates perhaps that the dominant ingredient in the resistance of the fluid is the component of the plate's motion perpendicular to the plate. A simple model problem is steady flow through a lattice of plates at a given Reynolds number. For steady vertical flows, we nondimensionalize time by $L/V$, based on the (steady) spatial average of the vertical flow $V$
(because there is no flapping frequency), giving a new definition of Reynolds number:
\begin{align}
\mbox{Re}_V = VL/\nu.
\end{align}
Figure \ref{fig:SteadyFlowsFig} shows steady vertical flows at Re$_V$ = 0.001 through different types of lattices. For the rectangular lattice, there are two limiting regimes: $l_y/(l_x-1) \ll 1$ (the ratio is 1/2 in panel A) and $l_y/(l_x-1) \gg 1$ (the ratio is 20 in panel B). The rhombic lattice has three limiting regimes. One is the same as panel B, $l_y/(l_x-1) \gg 1$. Here the streamlines would be altered from those in B away from the gap between the plates, but would become the same near the gap. The second is $l_y/(l_x/2-1) \ll 1$ with $l_x > 2$ (panel C has $l_y/(l_x/2-1) = 1$ and $l_x$ = 2.5), and the third is $l_y/(1-l_x/2) \ll 1$ with $l_x < 2$ (panel D has $l_y/(1-l_x/2) = 0.4$ and $l_x$ = 1.5). Only panel B is firmly in the asymptotic regime, while the other panels are at moderate ratios. In all cases, the flows resemble the limiting flows, even at moderate ratios. 

The steady versions of equations (\ref{NS1})-(\ref{NS2}) are
\begin{align}
\mathbf{u}\cdot\nabla\mathbf{u} = -\nabla p
+\frac{1}{\mbox{Re}_V} \nabla^2 \mathbf{u},  \label{NS1s} \\
\nabla \cdot \mathbf{u} = 0. \label{NS2s}
\end{align}
\nn Integrating the $y$-component of (\ref{NS1s}) over a periodic unit cell,
the left side vanishes because there is zero net vertical (or horizontal) outflow. So does the viscous term, because the inverse square root singularity in the shear stress diverges too slowly to produce a vertical resultant in the limit of zero plate thickness. Hence the integral of $-\partial_y p$ over the unit cell is zero. The integral has two contributions: one from the vertical change in $p$ across the unit cell, and the other from the jump in $p$ across the plate,
resulting in a force applied by the plate to the fluid. Hence
\begin{align}
0 = -\int_{plate} [p]^+_- dx - l_x l_y \frac{\Delta p_y}{l_y},  \label{vbal} 
\end{align}
\nn where $\Delta p_y$ is the change in pressure across the unit cell in the $y$
direction. For the rhombic lattice, with two plates in a double unit cell (figure \ref{fig:SchematicPlatesFig}B), the integral in (\ref{vbal}) includes both plates, and $\Delta p_y$ is the change in pressure across the double unit cell in the $y$ direction.
 For steady flow with dimensionless mean $y$-velocity 1,
\begin{align}
\langle P_{in} \rangle = P_{in} = \int_{plate} [p]^+_- dx = - l_x \Delta p_y,  \label{Pinbal} 
\end{align}
\nn which relates the input power to the pressure difference across a unit cell in the $y$-direction. The latter can be determined analytically for the limiting cases of the flows in figure \ref{fig:SteadyFlowsFig}.

In the limit $l_y/(l_x-1) \ll 1$, the flow in panel A becomes Poiseuille flow in the $x$-interval between the plate edges (0 $\leq x \leq$ 1 in panel A), and zero flow
in the rest of the flow field. Poiseuille flow is a good approximation to panel A even though $l_y/(l_x-1) = 1/2$, not very small. The lack of streamlines above and below the plates indicates
slow flow there (the local density of streamlines is proportional to flow speed). In the
interval 0 $\leq x \leq$ 1 the flow is nearly unidirectional with a parabolic profile for $v(x)$, and the isopressure contours are nearly equally spaced, corresponding to the constant pressure gradient of Poiseuille flow. Using Poiseuille flow to relate the pressure gradient $\Delta p_y/l_y$  to the net fluid flux through the unit cell, $Q = 1\cdot l_x$, 
\begin{align}
P_{in} = -l_x l_y \frac{\Delta p_y}{l_y} = l_x l_y \frac{12 Q}{(l_x - 1)^3 \mbox{Re}_V} =
   \frac{12 l_x^2}{(l_x - 1)^2 \mbox{Re}_V} \frac{l_y}{l_x - 1}. \label{PoisPin}
\end{align}
\nn The last expression in (\ref{PoisPin}) is written in terms of $l_y/(l_x-1)$, the
parameter that sets the validity of the approximation.

The limit $l_y/(l_x-1) \gg 1$ is exemplified by figure \ref{fig:SteadyFlowsFig}B. This
is the Stokes flow through small gaps in a periodic array of plates. The black lines are again the streamlines. They show flow converging toward the gap, and a small recirculation region centered at the midpoints of the plates. To determine $\Delta p_y$ and therefore $P_{in}$ in (\ref{Pinbal}), we can use the Stokes flow solution for a single gap in an infinite wall, derived by \cite{hasimoto1958flow}.
In the region that is much farther from the gap than its width, the pressure is approximately constant, one constant above the wall and a different constant below the wall. In figure \ref{fig:SteadyFlowsFig}B this is shown by the colored lines, isopressure contours (values at right). The solid line contours (dark blue and yellow) have pressures close to the values at distance 0.5 above and below the gaps (i.e.~far from the gap). The dashed lines (dark blue and yellow) have pressures 1\% above and below those of the solid lines. Hence, in panel B above and below the ``cloverleaf" regions near the gap bounded by the dashed lines, the pressure varies by less than 2\%. Therefore, the pressure field is essentially the same as that far from a single gap in an infinite wall. We approximate $\Delta p_y$ in (\ref{Pinbal}) as the difference between the far-field pressure constants for the infinite-wall case with net flux $Q$ through the gap, from \cite{hasimoto1958flow}: $\Delta p_y/Q = -32$Re$_V/(l_x-1)^2\pi$ . Then we have
\begin{align}
P_{in} = -l_x \Delta p_y = \frac{32 l_x^2}{\pi (l_x - 1)^2 \mbox{Re}_V}. \label{SmallGapPin}
\end{align}

\begin{figure} [h]
           \begin{center}
           \begin{tabular}{c}
               \includegraphics[width=6.8in]{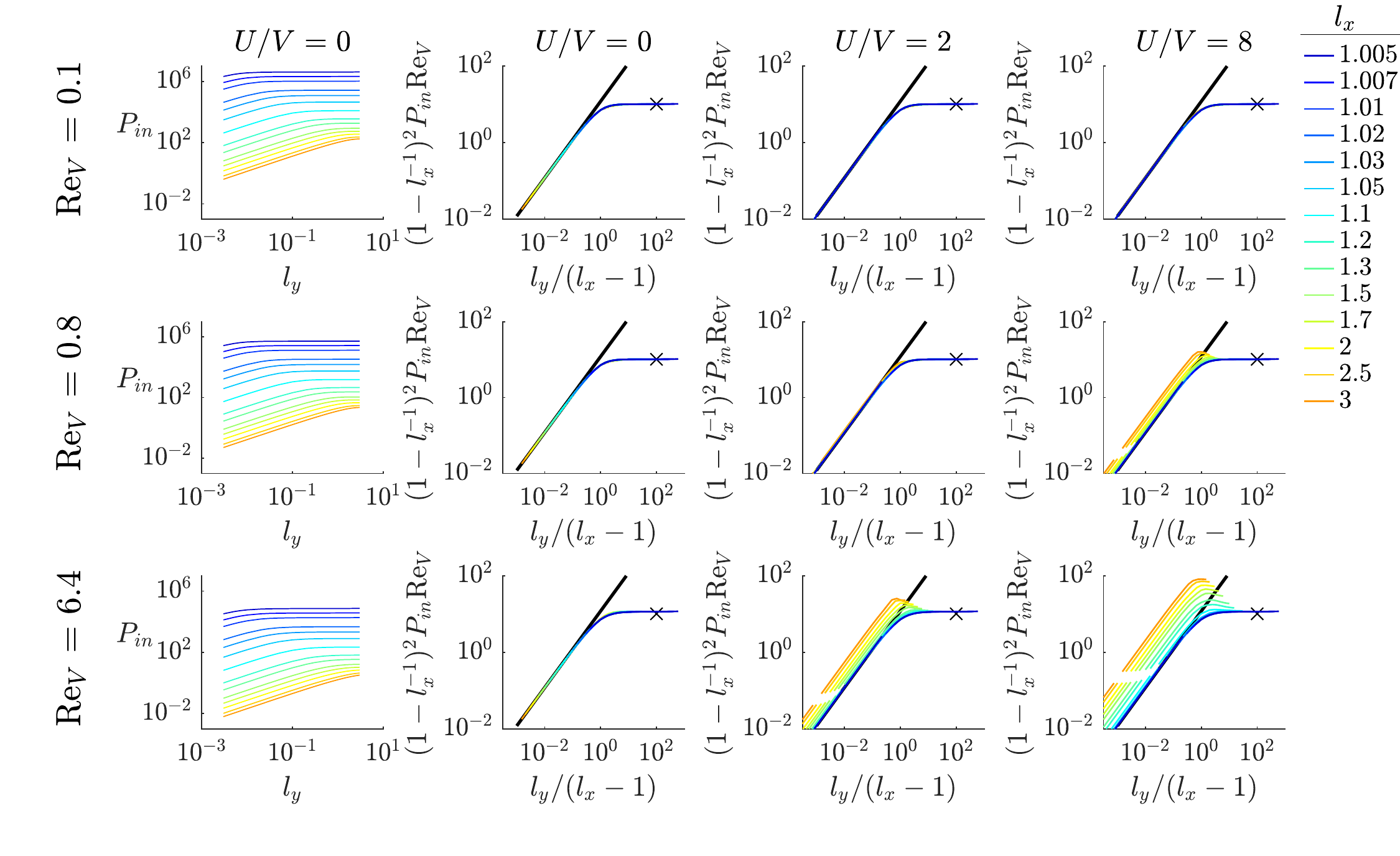}\\
           \vspace{-.25in}
           \end{tabular}
          \caption{\footnotesize For steady flow through a rectangular lattice, $P_{in}$ versus $l_y$ (left column) and in rescaled
variables (second through fourth columns) for different values of Re$_V$ (labeled at left) and $U/V$ (labeled at top).
 \label{fig:PInSteadyScalingRectangleFig}}
           \end{center}
         \vspace{-.10in}
        \end{figure}

In figure \ref{fig:PInSteadyScalingRectangleFig} we plot $P_{in}$ for steady
flows through rectangular lattices. Each row shows data for a different value of Re$_V$, up to 6.4, close to the threshold at which the steady flow state becomes unstable for some choices of $l_x$ close to 1. The first column shows $P_{in}$ versus $l_y$ for different choices of $l_x$ (colored lines, values listed at right), at $U/V = 0$. The second column shows the same data in rescaled variables. $P_{in}$ is divided by
$l_x^2/(l_x -1)^2$Re$_V$, a factor that appears in both (\ref{PoisPin}) and (\ref{SmallGapPin}). On the horizontal axis, $l_y/(l_x-1)$ is used as the dependent
variable. The black line shows the Poiseuille flow scaling (\ref{PoisPin}), while
the black cross shows the value 32/$\pi$ given by (\ref{SmallGapPin}). The agreement is almost exact. The third and fourth columns show the same data for $U/V$ = 2 and 8, respectively. Here the Poiseuille flow result (\ref{PoisPin}) can be modified by including a dimensionless crossflow $U/V$ in the flow equations \cite{Batchelor1967}, resulting
in a $v(x)$ that is linear plus exponential. In place of (\ref{PoisPin}) we obtain:
\begin{align}
 P_{in} = \frac{U}{V}l_x^2l_y\left[\frac{l_x-1}{\mbox{Re}_U} - \left(\frac{1}{2} +\frac{1}{e^{\mbox{Re}_U (l_x-1)} -1}\right)(l_x-1)^2\right]^{-1} \label{PinCross}
\end{align}
with Re$_U = UL/\nu$. $P_{in}$ in (\ref{PinCross}) tends to (\ref{PoisPin}) as Re$_U \to 0$, i.e.~if either Re$_V$ or $U/V$ $\to 0$, so the crossflow has no effect in Stokes flow (as can also be seen by linearity).
So only in the three panels near the lower right corner of figure \ref{fig:PInSteadyScalingRectangleFig}---i.e the cases (Re$_V, U/V$) = (0.8,8), (6.4, 2), and (6.4, 8)---do the colored lines deviate noticeably from the black line; the deviation depends on
$l_x$.  The upper portions of the colored lines are the data computed from steady Navier-Stokes solutions and the lower portions (separated by a gap from the upper portions) are the Poiseuille-plus-crossflow approximations (\ref{PinCross}), different for each $l_x$, and which are linear in $l_y$. These line up almost exactly with the computed data. In summary, at small Re$_V$, $P_{in}$ grows linearly with $l_y/(l_x-1)$
when the ratio is small. The linear growth is because the rate of 
viscous energy dissipation per plate in the channel flows are
proportional to the $y$-spacing between the plates, $l_y$. At large $l_y/(l_x-1)$, 
$P_{in}$ is independent of $l_y$ because the small-gap flow, and the corresponding rate of viscous energy dissipation, becomes independent of $l_y$.

\begin{figure} [h]
           \begin{center}
           \begin{tabular}{c}
               \includegraphics[width=6.8in]{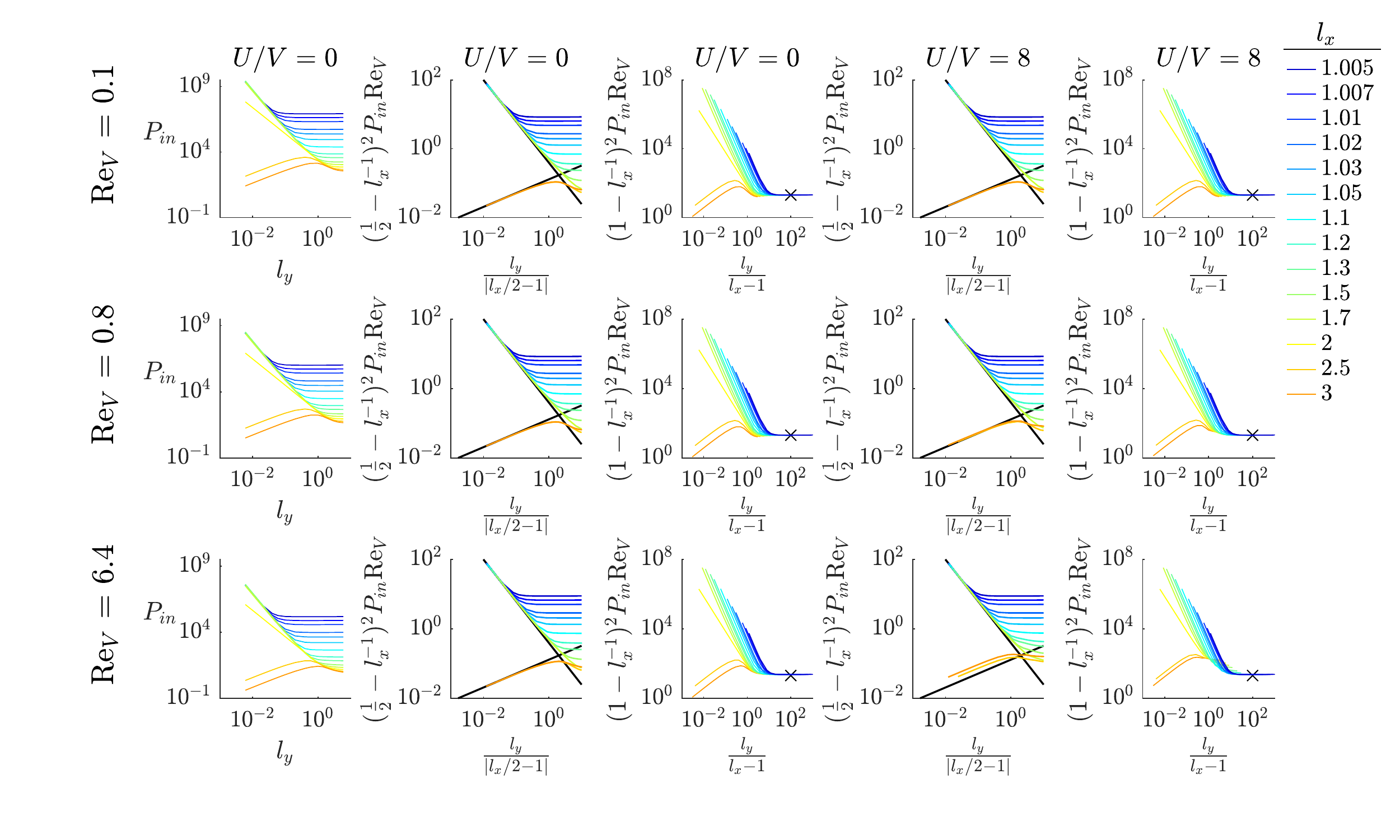}\\
           \vspace{-.25in}
           \end{tabular}
          \caption{\footnotesize For steady flow through a rhombic lattice, $P_{in}$ versus $l_y$ (left column) and in rescaled
variables (second through fifth columns) for different values of Re$_V$ (labeled at left) and $U/V$ (labeled at top).
 \label{fig:PInSteadyScalingDiamondFig}}
           \end{center}
         \vspace{-.10in}
        \end{figure}

We now consider analytical models for the rhombic lattice, with examples of flows
in figure \ref{fig:SteadyFlowsFig}C and D. The flow in panel C tends to two Poiseuille flows, each in a channel of width $l_x/2-1$, as $l_y/(l_x/2-1)$ becomes small. The flow
in panel D tends to four Poiseuille flows, oriented horizontally, each in a channel of width $l_y$ and length $1-l_x/2$ (the horizontal overlap between the plates), as 
$l_y/(1-l_x/2)$ becomes small. The pressure drop between the ends of the channels is half the total pressure drop over the double unit cell. The special case $l_x = 2$ and $l_y \to 0$,
at the boundary between these flows, is more complicated and we do not address it here.
Using these approximations, we obtain for the limit $l_y/|l_x/2-1| \ll 1$,
\begin{align}
P_{in} = \left.
\begin{cases}
   \displaystyle \frac{6 l_x^2}{(l_x/2 - 1)^2 \mbox{Re}_V} \frac{l_y}{l_x /2- 1}, & l_x > 2 \\   \displaystyle \frac{96 l_x^2}{(1-l_x/2)^2 \mbox{Re}_V} \left(\frac{1-l_x /2}{l_y}\right)^3, & l_x < 2
\end{cases}
\right.
\label{RhombicPoisPin}
\end{align}
\nn using the appropriate expressions for $\Delta p_y$. The case of large $l_y$ is essentially the same as figure \ref{fig:SteadyFlowsFig}B, so for the double unit cell of the rhombic lattice, we have twice the $P_{in}$ of (\ref{SmallGapPin}). We compare these results with the computed steady Navier-Stokes results in figure \ref{fig:PInSteadyScalingDiamondFig}.
We use only two values of $U/V$ now, 0 and 8, and the same Re$_V$ as in figure
\ref{fig:PInSteadyScalingRectangleFig}. The first column shows the unscaled data at $U= 0$. Unlike for the rectangular lattice, the scaling of $l_y$ changes from $l_y/|l_x/2-1|$ to $l_y/(l_x-1)$ at small and large $l_y$ respectively, so additional columns are needed to show the data with these separate scalings. The second and fourth columns show the small-$l_y$ scalings, with black lines showing the relationships in (\ref{RhombicPoisPin}).
In the bottom panel of the fourth column ((Re$_V$, $U/V)$ = (6.4,8)), the colored lines with $l_x > 2$ are shifted at nonzero Re$_U$, due to the crossflow effect discussed for the rectangular lattice (not rederived here). There is no visible shift for the $l_x < 2$ lines, because the effect of the crossflow cancels for the four horizontal channel flows,
two in each direction, in figure \ref{fig:SteadyFlowsFig}D. The black crosses (third and fifth columns) again
show the large $l_y/(l_x-1)$ values of $P_{in}$. The main difference from the rectangular lattice is the large growth in $P_{in}$ at small $l_y$ when $l_x < 2$. Propulsion occurs in many of these cases, e.g.~at $l_y$ = 0.5 and 0.7 (top and middle rows of figure \ref{fig:CompareRectDiam3ReFig}). However, the peak Froude efficiency decreases noticeably when $l_x$ drops below 2 and when $l_y$ decreases in most of these cases.

\begin{figure} [h]
           \begin{center}
           \begin{tabular}{l}
               \includegraphics[width=7.5in]{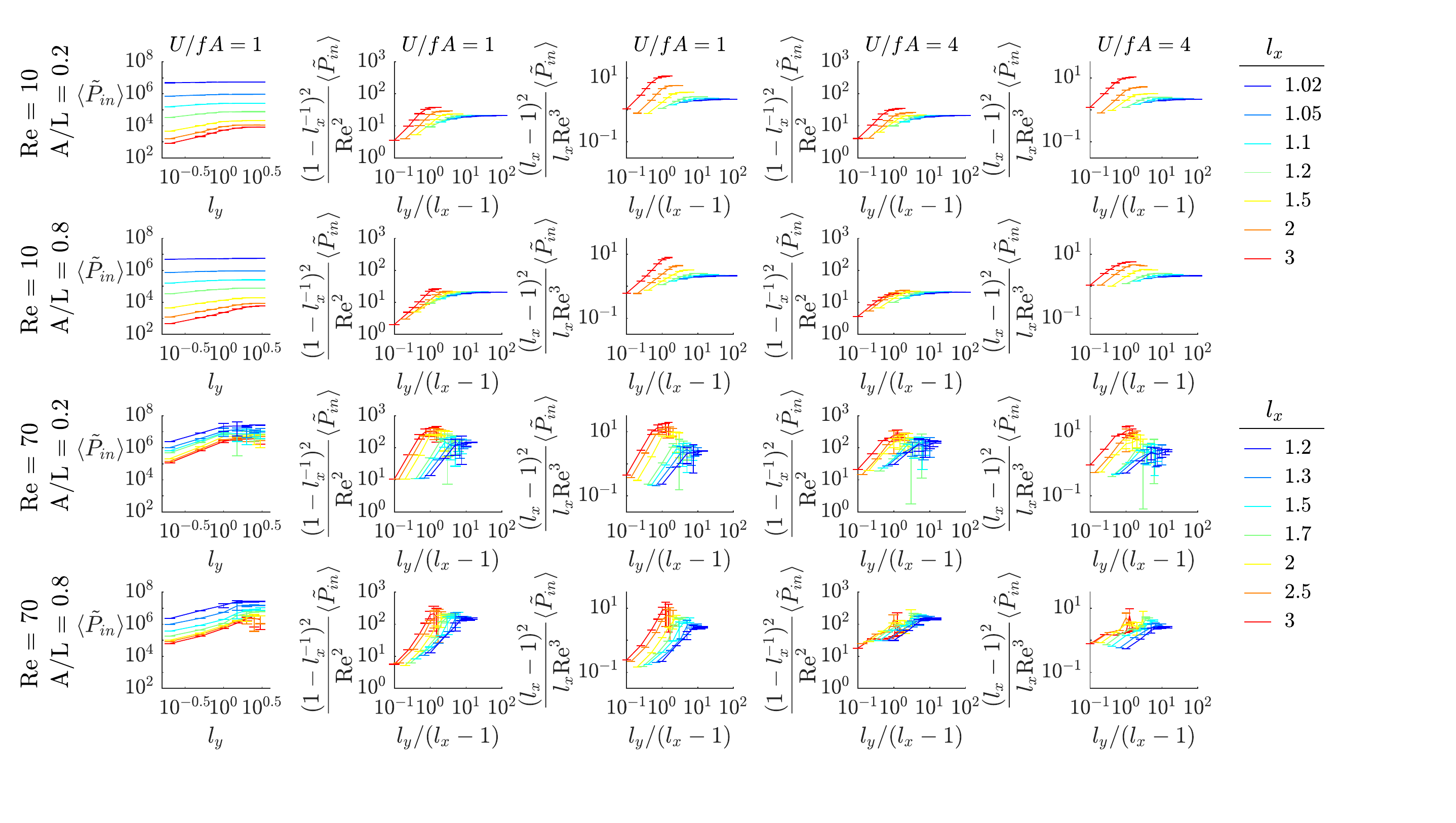}\\
           \vspace{-.25in}
           \end{tabular}
          \caption{\footnotesize Time-averaged input power $\langle {\tilde P}_{in} \rangle$ for flapping rectangular lattices at $U/fA$ = 1 (first to third columns) and 4 (fourth and fifth columns), at Re = 10 and 70 and $A/L$ = 0.2 and 0.8 (values labeled at left), for various $l_x$ (listed separately for each Re at right). The error bars show the range of values within one standard deviation of $\langle {\tilde P}_{in} \rangle$. The standard deviation is computed using the last five period-averages of ${\tilde P}_{in}(t)$.
 \label{fig:PInUnsteadyRectangleFig}}
           \end{center}
         \vspace{-.10in}
        \end{figure}

Now we consider the input power in the {\it unsteady}, fully nonlinear flapping problem, at Re = 10--70.
For the steady problem, the flows were plotted at much lower Reynolds number (Re$_V$ = 0.001) in figure \ref{fig:SteadyFlowsFig}, mainly because an analytical solution is available in this limit for panel B, the small-gap case. However, the Poiseuille flows with a crossflow or without (as in panels A, C, and D) remain valid at larger Re$_V$, until they become unstable (at Re$_V$ = O(10$^3$) \cite{schmid2002stability}, above the Reynolds numbers in the present study). Nondimensionalizing (\ref{PoisPin}) using $\nu/L$ in place of $V$, consistent with the unsteady $\langle{\tilde P}_{in}\rangle$ results in this paper, we have
\begin{align}
{\tilde P}_{in} = 
   \frac{12 l_x^2}{(l_x - 1)^2}\mbox{Re}_V^2 \frac{l_y}{l_x - 1}. \label{PoisTildePin}
\end{align}
\nn The small-gap flow in panel B changes to a jet flow (e.g.~figure \ref{fig:StateDiagramFig}) as the Reynolds number rises to 20. For Re = 10--70 the flow is intermediate between viscous-dominated
(resulting in (\ref{SmallGapPin})) and inertia-dominated. In the latter case, the momentum theorem can be used to calculate ${\tilde P}_{in}$ in the case of steady inertia-dominated (high-Re) flow. The calculation is the same as for 
the steady drag on an infinite, periodically perforated plate, given in 
\cite{Batchelor1967}, section 5.15. In the small-gap limit $l_y/(l_x - 1) \gg 1$,
the pressure drop through a unit cell of our periodic lattice becomes the same as that
through the periodically perforated plate, which in our notation (and nondimensionalized by
$\rho_f V^2$) is
\begin{align}
\Delta p_y = -\frac{1}{2(l_x-1)^2}.
\end{align}
\nn The pressure is assumed to follow Bernoulli's law on the upstream sides of the plates, while on the downstream sides, where separated flow occurs, it is assumed constant and equal to that in the gap (see \cite{Batchelor1967} for details). The input power follows from (\ref{Pinbal}):
\begin{align}
P_{in} = \frac{l_x}{2(l_x-1)^2} \; ; {\tilde P}_{in} = \frac{l_x}{2(l_x-1)^2}\mbox{Re}_V^3. \label{sep}
\end{align}
\nn where again, ${\tilde P}_{in}$ has been nondimensionalized using $\nu/L$ in place of $V$, consistent with the unsteady results in this paper. For both Stokes flow (\ref{SmallGapPin}) and separated flow (\ref{sep}) with small
$l_x -1$, $P_{in}$ diverges like $(l_x -1)^{-2}$, though with different prefactors.

 Time-averaged input power for the rectangular lattice is plotted in figure
\ref{fig:PInUnsteadyRectangleFig} at two values of Re and $A/L$ (labeled at left) and $U/fA$ (labeled at top) in the ranges already considered. The unscaled $\langle {\tilde P}_{in} \rangle$ data are shown in the first column. $\langle {\tilde P}_{in} \rangle$ is rescaled according to the small-$l_y$ Poiseuille flow scaling (\ref{PoisTildePin}) in the second and fourth columns, and according to the large-$l_y$ separated flow scaling (\ref{sep}) in the third and fifth columns, with Re (from (\ref{params2})) in place of Re$_V$.  
The error bars show
the range of values within one standard deviation of $\langle {\tilde P}_{in} \rangle$, computed using the last five period-averages of ${\tilde P}_{in}(t)$. In many cases (i.e.~Re = 10, small $l_y$), ${\tilde P}_{in}(t)$ is periodic, so the error bar has almost zero height, and the upper and lower horizontal hash marks overlap, appearing as a single hash mark. At Re = 70 and larger $l_y$, the vertical extent of the error bar is noticeable, and gives a measure of the nonperiodicity of the data. 

Although the data are more scattered in the unsteady case, they are qualitatively similar to the steady case
(figure \ref{fig:PInSteadyScalingRectangleFig}), including
the shift at increasing $l_x$ in the Poiseuille flow (linear growth regime). The steady problem neglected the effects of unsteadiness (an oscillating instead of steady vertical flow) and nonlinearity in the Navier-Stokes equations. We extended the steady model to the case of an unsteady but linearized model,  
\begin{align}
\partial_t v + U\partial_x v = -P_y e^{2\pi i t}
+\frac{1}{\mbox{Re}_f} \partial_{xx} v,  \label{NS1v}
\end{align}
\nn for a harmonically oscillating channel flow $v(x,t) = V_0(x)e^{2\pi i t}$ with crossflow. Analytical solutions are again possible, but more complicated than the steady case because we are back in the five-dimensional parameter space. We did not analyze the results in detail but they seemed to agree qualitatively with the small $l_y$-results in
figures \ref{fig:PInSteadyScalingRectangleFig} and \ref{fig:PInUnsteadyRectangleFig}.
Despite the complications due to vortex shedding and nonperiodicity,
the steady models and the fully nonlinear simulations agree qualitatively in the behavior of 
$\langle {\tilde P}_{in} \rangle$---linear growth at small $l_y/(l_x-1)$ in the second and fourth columns of figure
\ref{fig:PInUnsteadyRectangleFig}, saturation at large $l_y/(l_x-1)$ in the third and fifth columns. The main discrepancy is that the Re = 10 data have a larger magnitude than the Re = 70 data. A better fit is provided by the relation 
$\langle {\tilde P}_{in} \rangle \sim \mbox{Re}^3$ (typical for pressure losses at high Re, as in the third and fifth columns), rather than the $\sim \mbox{Re}^2$ scaling for steady Poiseuille flow used in the second and fourth columns of figure \ref{fig:PInUnsteadyRectangleFig}.

\begin{figure} [h]
           \begin{center}
           \begin{tabular}{l}
             \includegraphics[width=7.5in]{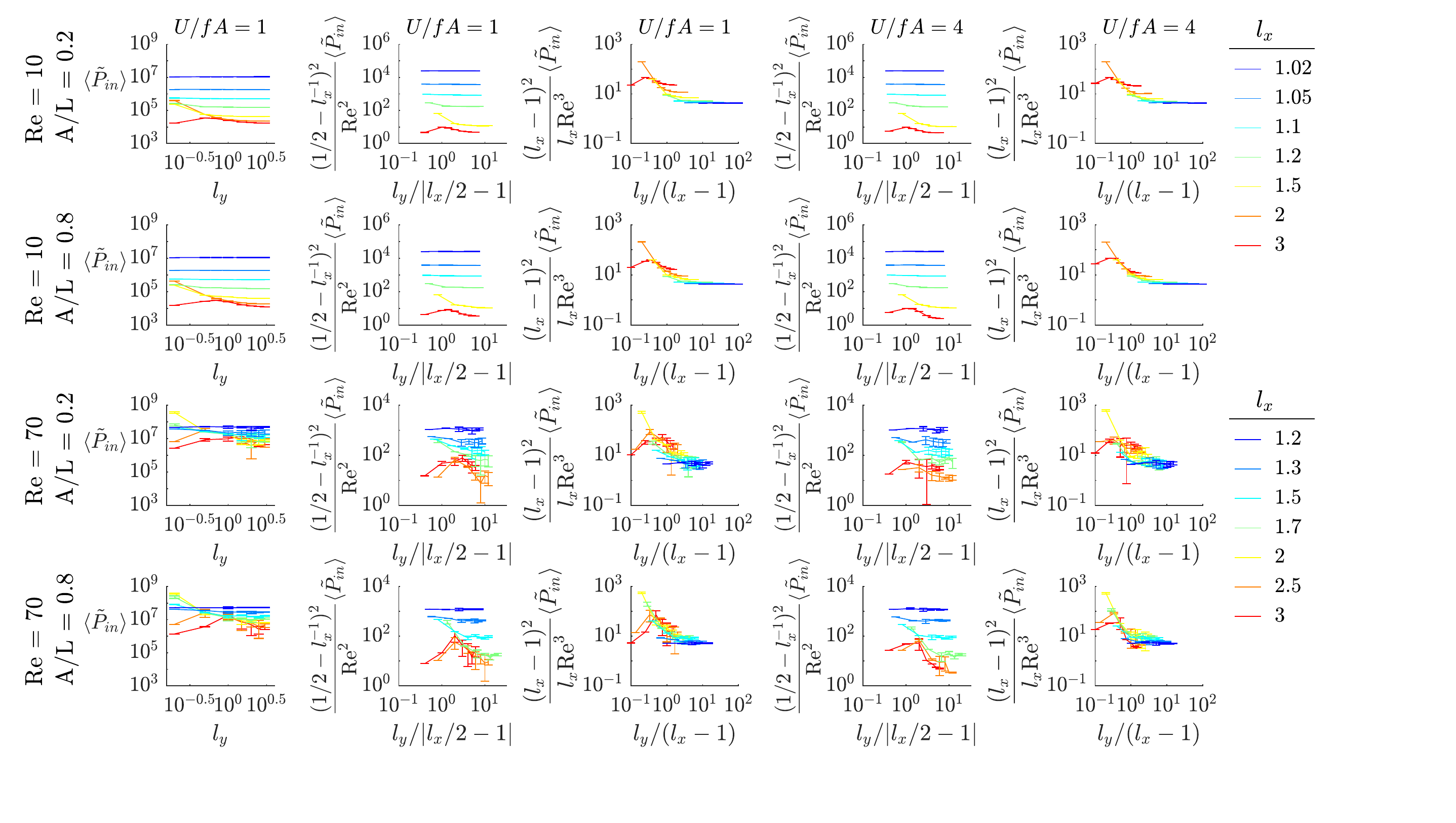}\\ 
           \vspace{-.25in}
           \end{tabular}
          \caption{\footnotesize Time-averaged input power $\langle {\tilde P}_{in} \rangle$ for flapping rhombic lattices at $U/fA$ = 1 (first to third columns) and 4 (fourth and fifth columns), at Re = 10 and 70 and $A/L$ = 0.2 and 0.8 (values labeled at left), for various $l_x$ (listed separately for each Re at right). The error bars show the range of values within one standard deviation of $\langle {\tilde P}_{in} \rangle$. The standard deviation is computed using the last five period-averages of ${\tilde P}_{in}(t)$.
 \label{fig:PInUnsteadyDiamondFig}}
           \end{center}
         \vspace{-.10in}
        \end{figure}

The fully nonlinear $\langle {\tilde P}_{in} \rangle$ data for the rhombic lattices are presented
in figure \ref{fig:PInUnsteadyDiamondFig}. As for the rectangular lattices, the data are presented with different scalings at small $l_y$ (second and fourth columns) and large $l_y$ (third and fifth columns). In spite of the effects of nonperiodicity, there is good qualitative agreement between the unsteady and steady (figure \ref{fig:PInSteadyScalingDiamondFig}) cases. In the second and fourth columns, there is a clear divergence at small $l_y/|l_x/2-1|$ between the lines with $l_x >2$ (red and orange) and the remaining lines. We did not compute cases
at $l_y/|l_x/2-1|$ as small as in figure \ref{fig:PInSteadyScalingDiamondFig}, however, because they are not useful for locomotion, and in the unsteady case, the parameter space is larger and each simulation requires much more computing time. Again, the Re = 10 data are generally larger than the Re = 70 data in the second and fourth columns, 
perhaps indicating the importance of pressure losses beyond those of Poiseuille flow at higher Re.
In the third and fifth columns, the lines at various $l_x$ seem to agree reasonably well at large $l_y/(l_x-1)$. 

%

\section{Summary and conclusions \label{sec:Conclusions}}

We have studied propulsive properties of flapping lattices of plates, at Re = 10--100,
where the flows often become time-periodic within 5--30 flapping periods. This Re range is typical for submillimeter- to centimeter-scale flying and swimming organisms \cite{childress2004transition,miller2004vsa,miller2009flexible,jones2015lift,santhanakrishnan2018flow,skipper2019characterization},
and is relevant to the increasing number of robotic flying and swimming vehicles that inhabit this size range \cite{chen2017biologically,zhang2018untethered,hu2018small,ren2019multi,chen2019controlled}.
Froude efficiency is typically much lower in this Re range than in the higher Re range typical of most fish and birds \cite{SBL_PAeroSci_1999,TTY_AnnRevFluidMech_2000,fish2006passive}, so collective locomotion may be relatively more important for achieving locomotion (efficiently, or at all, at very low Re where it is no longer possible for an isolated flapping body).

In our simulations, we used a rectilinear grid with grid points concentrated near the singularities at the plates' edges. The condition number of the discrete Laplacian remains many orders of magnitude below 10$^{16}$ for the grids used in this paper, so round-off error is not a major obstacle. For a Laplace equation with similar singular behavior, we found that the method gives better than 1\% accuracy in the solution and the integral of its gradient along the plate for modest mesh sizes, and converges at 3/2 order.

We first used the method to determine the propulsive properties of an isolated flapping plate in
this specific context (a plate with zero thickness flapping in a flow with fixed velocity upstream).
The solutions show many properties that resemble those of previous flapping-foil studies: a reversed von K\'{a}rm\'{a}n street at certain oncoming flow speeds, more complicated vorticity fields at lower speeds, and a single stable self-propelled speed. The Strouhal numbers for maximal Froude efficiency increase from about 0.4 to $\gg 1$ as Re$_f$ decreases from 200 to 10, while the Strouhal numbers corresponding to the self-propelled speeds increase from 0.25 to $\gg 1$ in the same range of Re$_f$.
The maximum Froude efficiency for an isolated plate decreases from 0.06 at Re$_f$ = 200 to less than 0.003 at Re$_f$ = 10. These values are much lower than for higher-Re swimming fish and robots. This is consistent with the fact that flapping locomotion is not possible when Re decreases below a critical value close to 1, and there the maximum Froude efficiency becomes zero. The Pareto-optimal flapping amplitudes for maximizing speed at a given mean input power stay nearly fixed at 0.2--0.3. By contrast, the optimal flapping frequency increases with increasing self-propelled speed and/or input power.  The input power scales as flapping amplitude and frequency, both raised to the third power, and depends only weakly on the oncoming flow velocity.

We then studied the propulsive properties of rectangular and rhombic lattices of flapping plates at low to moderate Reynolds numbers. Not surprisingly, there is a much wider range of flows than for an isolated body (which can also be regarded as the solution in the limit of large lattice spacing). When the plates are closely spaced in the streamwise direction ($l_x$ close to 1), there are sharp transitions from drag to thrust with slight increases in the oncoming flow speed ($U/fA$). These correspond to changes in flow modes, characterized by vortex dipoles switching from upstream to downstream directions. Some of these flows have $F_x(t)$ with periods that are various integer multiples of half the flapping period, and others are nonperiodic. In general, nonperiodicity is more common at large Re, small $l_x$, large $l_y$, and small $U/fA$, for both rectangular and rhombic lattices, with more nonperiodicity at small $l_y$ in the rhombic case. 

As $l_x$ increases beyond the vicinity of 1 with large $l_y$ (akin to a 1D tandem array), the vortex dipoles transition to vortex-street wakes like those of isolated bodies, but which collide with the leading edges of plates downstream. Varying $l_y$ from small to large with intermediate $l_x$ (fixed at 2), the flows in rectangular and rhombic lattices are initially very different, with drag-producing Poiseuille-type flows in the rectangular case, and less periodic, thrust-producing flows in the rhombic case. Interactions between vorticity at laterally-adjacent leading and trailing edges in the rhombic lattices produced thrust efficiently at small $l_y$. As $l_y$ increased, the rectangular lattice flows eventually shed discrete vortices and produced thrust, with somewhat higher efficiencies than the rhombic lattice in many cases.

We produced maps of maximum Froude efficiency and self-propelled speeds in different portions of the four-dimensional (Re, $A/L$, $l_x$, $l_y$) parameter space where dynamics are close to periodic in time. At fixed Re, Froude efficiency is higher
at $A/L$ = 0.2 than at 0.5 and 0.8, and the peak occurs at gradually increasing $l_x$ (and moderately large $l_y, \approx$ 2) as
$A/L$ increases. The rhombic lattice is more efficient at small $l_y$, and the rectangular lattice is slightly more efficient in most cases at large $l_y$. The highest self-propelled speeds occurred at small $l_x$ and large $l_y$ for both lattice types. As Re was increased from 10 to 70, the peak Froude efficiency increased from 0.007 to 0.07. The isolated flapping body had a much lower Froude efficiency at Re = 10, eventually rising to about half that of the optimal lattices at Re = 70. However, many lattice flows are nonperiodic at Re = 70, and including these would increase the advantage over the isolated flapping body. The lattices showed similarly-sized advantages over the isolated bodies in the maximum self-propelled speeds. 

The advantage in Froude efficiency occurs even with the much larger input power required for the lattices (due to the confinement of flow between the plates). In a steady model, we found that good analytical approximations to the input power could be derived for both lattice types. These relate to the flow through a small
gap at small $l_x$ and moderate $l_y$ and Poiseuille flows with crossflow at small $l_y$ and moderate $l_x$. The unsteady flows showed scaling behaviors that agreed qualitatively with those of the steady model flows.

To limit the number of parameters under consideration and the computational complexity, we have assumed all of the plates are moved together in phase (as in \cite{becker2015hydrodynamic,peng2018hydrodynamic} but not \cite{lin2019phase}, for a small group of plates). Even with this restriction, varying the spacing between the plates gives us a degree of control of the phase between shed vortices and the motions of downstream bodies with which the vortices collide. In our study, the spacing between them is not allowed to vary with time (e.g.~under fluid forces \cite{becker2015hydrodynamic}). Incorporating these effects would greatly expand the parameter spaces under consideration but would be natural areas for future work.

\begin{acknowledgments}
This research was supported by the NSF Mathematical Biology program under
award number DMS-1811889 (S.A.). 
\end{acknowledgments}

\appendix
\section{Appendix \label{FxFlowTransitions}}

\begin{figure} [h]
           \begin{center}
           \begin{tabular}{c}
               \includegraphics[width=6in]{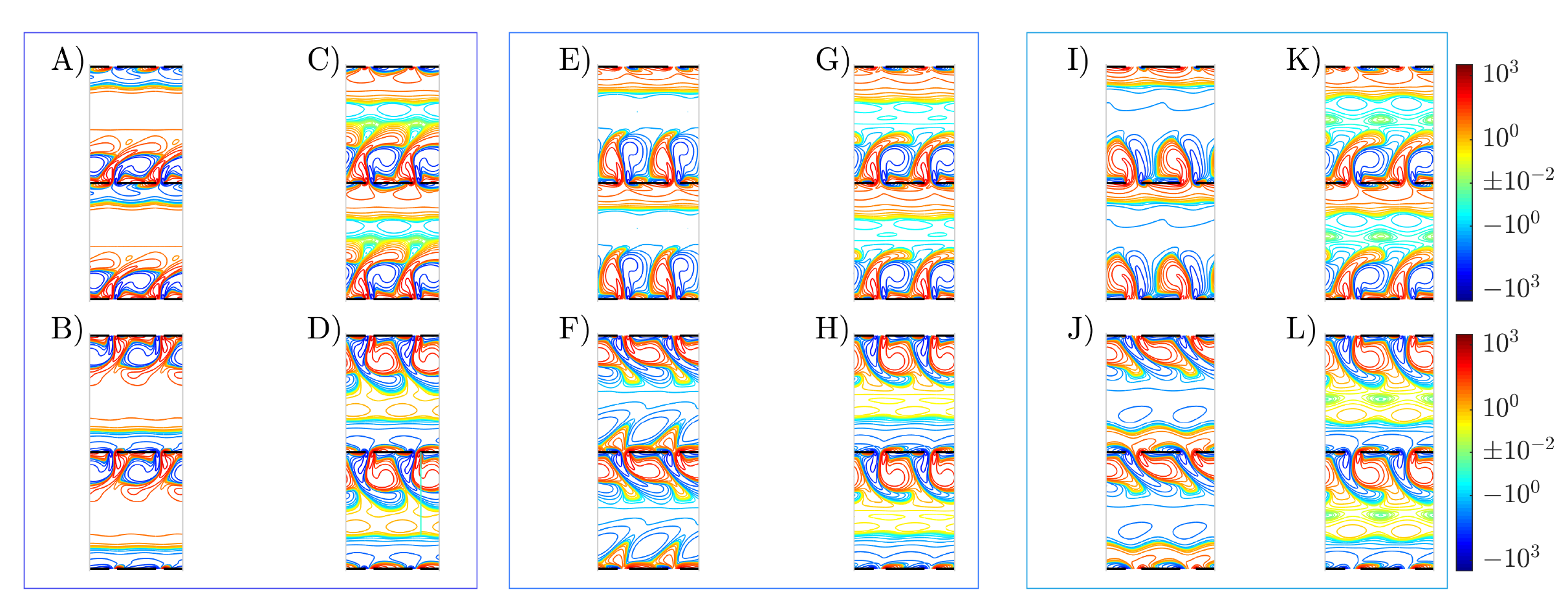}\\
           \vspace{-.25in}
           \end{tabular}
          \caption{\footnotesize Flow states that accompany each of the three sudden drops in $\langle F_x \rangle$ shown in figure \ref{fig:StateDiagramFig}D. In each case the flows transition from up-down asymmetric (panels A-B, E-F, and I-J) to symmetric (panels C-D, G-H, and K-L), with larger vortices downstream (rightward half of vortex dipoles). The first, second, and third boxes correspond to
$l_x$ = 1.2, 1.3, and 1.4, respectively.
 \label{fig:StateDiagramPanelDVorticity400}}
           \end{center}
         \vspace{-.10in}
        \end{figure}

Figure \ref{fig:StateDiagramPanelDVorticity400} shows examples of the vorticity fields at transitions in flows
corresponding to the three sudden drops in $\langle F_x \rangle$ in figure \ref{fig:StateDiagramFig}D, as $U/fA$ is increased slightly. In each case the vortex dipoles emitted from the gaps between the plates are oriented more
downstream after the transition.

\begin{figure} [h]
           \begin{center}
           \begin{tabular}{c}
               \includegraphics[width=6.8in]{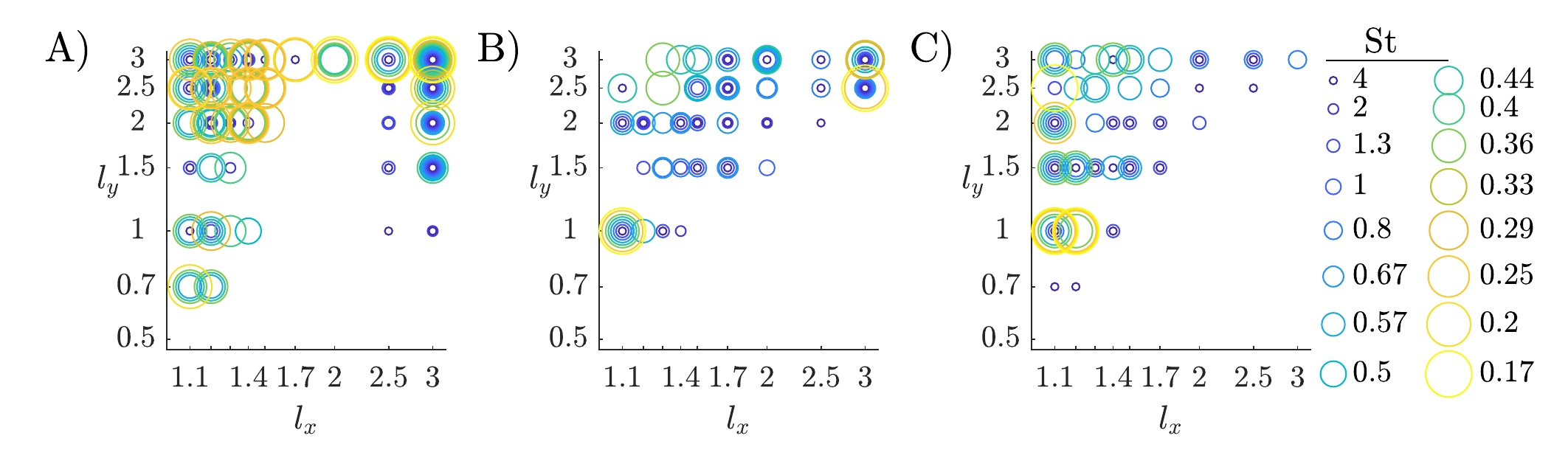}\\
           \vspace{-.25in}
           \end{tabular}
          \caption{\footnotesize Circles show parameter values where $F_x(t)$ falls below a threshold 0.01 (described in text adjacent to figure \ref{fig:DegreeOfPeriodicityRe20Fig})
for having period 1, for a rhombic lattice of plates with Re = 20. Values of $A/L$ are 0.2 (A), 0.5 (B), and 0.8 (C). Values of St are labeled by circle size and color (key listed at right). Circles are centered at the corresponding values of $l_x$ and $l_y$.
 \label{fig:DegreeOfPeriodicityDiamondRe20Fig}}
           \end{center}
         \vspace{-.10in}
        \end{figure}

Figure \ref{fig:DegreeOfPeriodicityDiamondRe20Fig} shows parameters where $F_x(t)$ falls below a threshold for periodicity in flows through rhombic lattices at Re = 20.  Nonperiodicity is associated with small $l_x$, large $l_y$, and
large St (small $U/fA$).

\bibliographystyle{unsrt}
\bibliography{DoublyPeriodicPlates}

\end{document}